\documentclass[preprints,article,accept,moreauthors,pdftex]{Definitions/mdpi}

\firstpage{1}
\pubvolume{1}
\issuenum{1}
\articlenumber{0}
\pubyear{2023}
\copyrightyear{2023}
\datereceived{ }
\daterevised{ } 
\dateaccepted{ }
\datepublished{ }
\hreflink{https://doi.org/} 

\Title{A Comparative Modelling of Essential Characteristics of Volatility: Simulation and Empirical Study}

\TitleCitation{Title}

\Author{Richard T. A. Samuel $^{1,\dagger,\ddagger}$*\orcidA{}, Charles Chimedza $^{1,\ddagger}$\orcidA{} and Caston Sigauke $^{2}$\orcidA{}}

\AuthorNames{Richard T. A. Samuel, Charles Chimedza and Caston Sigauke}

\AuthorCitation{Samuel, R. T. A.; Chimedza, C.; Sigauke, C.}

\address{%
$^{1}$ \quad School of Statistics and Actuarial Science, University of the Witwatersrand, Johannesburg, South Africa; richieeniks@gmail.com \\
$^{1}$ \quad School of Statistics and Actuarial Science, University of the Witwatersrand, Johannesburg, South Africa; Charles.Chimedza@wits.ac.za\\
$^{2}$ \quad Department of Mathematical and Computational Sciences, University of Venda, Thohoyandou, South Africa; caston.sigauke@univen.ac.za}

\corres{Correspondence: richieeniks@gmail.com}

\firstnote{Current address: School of Statistics and Actuarial Science, University of the Witwatersrand, Johannesburg, South Africa.}
\secondnote{These authors contributed equally to this work.}

\abstract{This study utilised the dynamics of five time-varying models to estimate six essential features of financial return volatility that are relevant for robust risk management.  These features include pronounced persistence, mean reversion, leverage effect or volatility asymmetry, conditional skewness, conditional fat-tailedness, and the long memory behaviour of volatility decomposition into long-term and short-term components. Both simulation and empirical evidence are provided. Through the applications of these models using the S\&P Indian index, the study shows that the market returns are characterised by these volatility features. Our findings from the long-memory behaviour revealed that although the response to shocks is greater in the short-term component, it is however short-lived. On the contrary, despite a high degree of persistence in the long-term component, market information or unexpected news arrival only has a low long-run impact on the market. Based on this, the long-run investment risks within the Indian stock market seem to be under control. Hence, our findings suggest that rational investors should try to stay calm with the arrival of unexpected news in the market because the long-run effect of such news will not be severe, and the market will eventually return to its normal state.}
\begin{document}

\section{Introduction}
Stock market indices are prone to be characterised by features of volatility such as pronounced persistence, mean reversion, leverage effect (i.e., volatility asymmetry), conditional skewness, conditional fat-tailedness, and the long memory behaviour of volatility decomposition \cite{HarveySuca2014}. Hence, these characteristics are usually exhibited by financial returns (see \cite{Sucarrat2013,EnglePatton2001}). Volatility persistence is a process where the return of today affects the future's forecast variance \cite{EnglePatton2001}. The economic implications of the degree of persistence of a shock include its influence on dynamic hedging policies, the valuation of options and the price of securities \cite{PortSumm1986}. Mean reversion implies that there is a normal volatility level to which volatility will eventually return. Asymmetry is a process in which positive and negative shocks have different impacts on volatility. Asymmetry in equity returns is sometimes known as leverage effect \cite{EnglePatton2001}. Leverage effect implies that volatility tends to be higher following negative returns \cite{Sucarrat2013}. Conditional fat-tailedness implies that the standardised conditional return is more fat-tailed than the Normal distribution, while conditional skewness denotes that the standardised return is not symmetric \cite{Sucarrat2013}. The long-memory behaviour of volatility decomposition occurs when volatility is decomposed into one short-term component and one long-term component \cite{Sucarrat2013}.

Volatility on its own is not directly observable, hence its measurement and the modeling of its evolution rely on some measurement methods. Two well documented methods in the literature are the Generalised Autoregressive Conditional Heteroscedasticity (GARCH) model and the Generalised Autoregressive Score (GAS) model. These two models use the conditional variance for measuring volatility (see \cite{Ardiaetal2019,Chou1988,TakaishiT2018}). These models can reliably capture some features of asset returns. However, it is well reported that asymmetric GARCH models perform better at modelling volatility than the simple GARCH model (see \cite{TakaishiT2018}), since the simple GARCH model is incapable of capturing asymmetries in the volatility process \cite{FengShi2017}. Hence, this study applies a robust extension of the GARCH model called the family GARCH (fGARCH) model \cite{Hentschel1995} that incorporates asymmetric volatility process. Based on this, the fGARCH and GAS models, with their extensions, are used to model the stated essential features of volatility using the daily equity returns.

To be specific, the study comparatively applies the fGARCH and GAS models to estimate the magnitude and dynamics of the persistence (with mean reversion) in the conditional volatility of the returns. The main difference between these models is that the fGARCH model uses the dynamics of the residuals to drive the conditional variance, while the GAS model uses the dynamics of the conditional score to drive the time-varying conditional variance. Furthermore, the study uses the extensions of these two models, i.e., the extensions of the (f)GARCH model that consist of the Threshold GARCH (TGARCH) and Component GARCH (CGARCH) models, and the extension of the GAS model that consists of the Beta-Skew-$t$-EGARCH models to estimate the remaining stated features of returns. Specifically, the study compares the applications of the TGARCH and CGARCH models on one side, with the multi-functions of the Beta-Skew-$t$-EGARCH model on the other side to model the asymmetry (leverage effect), skewness, fat-tails and long-memory behaviour of volatility.

Engle and Patton \cite{EnglePatton2001} used the GARCH(1,1) model to estimate various stylised facts of volatility in the Dow Jones Industrial Index returns from 1988 to 2000, and found a highly persistence volatility estimate with a half-life of about 73 days. Moreover, the tendency of volatility to persist over time is well documented in the literature (see \cite{Chou1988,PandeyKumar2017} among others). Tsay \cite{Tsay2005} used the TGARCH model to study asymmetry in the IBM daily returns. Oh and Patton \cite{OhPatton2016} used the GAS model with a factor copula model to study systemic risk. Other GAS modelling applications are in credit risk analysis \cite{Crealetal2014}, spatial econometrics \cite{BlasqKOOPetal2014,CataniaBill2017}, high-frequency data \cite{Opschooretal2018,Gorgietal2019}, among others. See \cite{Ardiaetal2019,HaddadETAL2023} for more dynamic applications of the GAS model. Studies associated with the decomposition of return volatility to its permanent-transitory components through the CGARCH model are increasingly found in the literature (see \cite{ChenShen2004,ChiangETAL2009,Ane2006,EngleLee1999}). The applications of the Beta-Skew-$t$-EGARCH model for modelling relevant features of volatility are well studied in Sucarrat \cite{Sucarrat2013}, and Harvey and Sucarrat \cite{HarveySuca2014}.

Our study should result in responses to the following questions. One, which assumed innovations are the most adequate from the fGARCH and GAS modelling to estimate the persistence of the return volatility? Two, how persistent is the market returns volatility? Three, how are the models compared in terms of performances? Four, what volatility features characterise the returns and what are the implications on investment in the market? This study used the daily S\&P Indian index data, obtained from Thomson Reuters Datastream \cite{Datastream2021} from January 4th, 2010 to June 18th, 2021 with a total of 2990 observations. We chose these periods so as to include the periods of both relative calm and the recent volatility spike (market turmoil) caused by the global COVID-19 pandemics. Moreover, with a vast population, rapidly growing economy, and admirable investment opportunities, India is one of the greatest investment destinations in the world. It is projected to have the largest population globally before 2030 \cite{HertogETAL2022}, and its financial markets are increasingly attracting many foreign and domestic investors. The rest of the paper is structured as follows. In Section \ref{CCC222}, the applied research theories and methodologies are presented. Section \ref{CCC333} presents the simulation and empirical evidence of the estimations. Section \ref{CCC444} discusses the novel findings, and Section \ref{CCC555} concludes.

\section{Materials and Methods}\label{CCC222}
\subsection{The GARCH model}\label{theGML}
The GARCH model \cite{Bollers1986} is an extension of the Autoregressive Conditional Heteroscedasticity (ARCH) model proposed by Engle \cite{Engle1982} for volatility estimation. It is usually described by its conditional mean and variance equations. The mean equation can be stated as:

\begin{equation}\label{GodsHeaven}
  r_t = \mu_t + \varepsilon_t,
\end{equation}

where $r_{t}$ denotes the returns, $\varepsilon_t=z_t\sigma_t$ is the random or unpredictable residual. The $z_t$ is a standardized residual returns $(z_t = \varepsilon_t/\sigma_t)$ which are independent, identically distributed (i.i.d.) random variable with zero mean and unit variance. The $\mu_t$ in Equation \eqref{GodsHeaven} represents the mean function and it is usually expressed as an ARMA$(m,n)$\footnote{ARMA means Autoregressive Moving Average.} process,

\begin{equation}
 \mu_t = \sum\limits_{j=1}^{m}\varsigma_jr_{t-j} +
\sum\limits_{j=1}^{n}\psi_j\varepsilon_{t-j},
\end{equation}

where $\varsigma_j$ and $\psi_j$ are unknown parameters. The conditional variance equation of the general GARCH$(p,q)$ model can be defined as:

\begin{equation}\label{11331177}
\sigma^2_t = \underbrace{
\underbrace{\omega + \sum\limits_{j=1}^{p}\alpha_j\varepsilon^2_{t-j}}_\text{ARCH} +
\sum\limits_{j=1}^{q}\beta_j\sigma^2_{t-j}}_\text{GARCH},
\end{equation}

where $\alpha_j \geq 0$ and $\beta_j \geq 0$ are the ARCH and GARCH coefficients, respectively, while $\omega > 0$ is the intercept. The first-order GARCH(1,1) is possibly the best candidate and the most widely used GARCH model for modelling volatility \citep{Harvey2013}. The rate of decay of shocks to volatility in the conditional variance of the GARCH process is measured by summing the coefficients ($\alpha$, $\beta$). This refers to the volatility persistence of the GARCH models and it indicates the speed of the decay of volatility after a shock. If the sum of the coefficients equals to one, then shocks to the volatility do not decrease over time, hence the persistence is felt forever, and the unconditional variance of the process does not exist. Such situation is called integrated-GARCH \cite{Chou1988}. Shocks to volatility shows long persistence into the future when the sum is close to one. This produces a mean-reversion system in which the variance process (volatility), be it high or low, eventually returns very slowly to the mean (normal) state. Lastly, the process of shocks to the conditional variance shows high persistence when the sum is greater than one, which implies explosive volatility forecasts.

\subsection{Persistence and Mean Reversion in Volatility}\label{FaithMercyGrace}
Volatility can be described as persistent if today’s return produces a large effect on the prediction variance many periods in the future \cite{EnglePatton2001}. Volatility clustering means that volatility comes and goes. Hence, a period of low volatility will be followed by a volatility rise, while a period of high volatility will sooner or later make way to more normal volatility \cite{EnglePatton2001}. Mean reversion in volatility implies that there is a normal volatility level to which volatility will return eventually. A familiar classical measure of volatility persistence is called the "half-life" of volatility, denoted as $h2l$ \cite{Ghalanos2018}. This can be described as the number of days it will take the volatility to revert or move halfway back towards its unconditional mean after deviating from it (see \cite{Ghalanos2018,EnglePatton2001}).

\begin{equation}\label{h2llifelife}
  h2l = \displaystyle\frac{\log_{e}\frac{1}{2}}{\log_{e}\hat{P}},
\end{equation}

where $\log_{e}$ denotes the natural logarithm, and $\hat{P}$ represents the estimate of the persistence parameter.

\subsection{The fGARCH Model}\label{thefGARCH777}
The family GARCH abbreviated fGARCH \cite{Hentschel1995} is an omnibus model that subsumes some familiar asymmetric and symmetric GARCH models as sub-classes \cite{Ghalanos2018}. These sub-classes include the standard GARCH (sGARCH) model \citep{Bollers1986}, the Threshold GARCH (TGARCH) model \citep{Zakoian1994}, the Nonlinear Asymmetric GARCH (NAGARCH) model \citep{EngleNg1993}, the Absolute Value GARCH (AVGARCH) model \citep{Taylor1986,Schwert1990}, the Nonlinear ARCH (NGARCH) model \citep{HigginsBera1992}, the Exponential GARCH (EGARCH) model \citep{Nelson1991}, the GJR GARCH (GJRGARCH) model \citep{GlostenETAL1993}, and the Asymmetric Power ARCH (apARCH) model \citep{Dingetal1993}. The fGARCH$(p,q)$ model can be stated as:

\begin{equation}\label{fGARCHEqn}
		\sigma^{\gamma}_t = \omega + \sum_{j=1}^{p} \alpha_j \sigma^{\gamma}_{t-j}(|z_{t-j} - \zeta_{2j}| - \zeta_{1j}\{z_{t-j} - \zeta_{2j}\})^{\delta} + \sum_{j=1}^{q}\beta_j \sigma^{\gamma}_{t-j}.
\end{equation}
This omnibus model allows the decomposition of the residuals to be driven by different powers for $z_{t}$ and $\sigma_t$ in the conditional variance equation. The model also allows for both rotations and shifts in the news impact curve, where the rotation drives large volatility shocks while the shift is the main origin of asymmetry for small volatility shocks. As discussed in Ghalanos \cite{Ghalanos2018}, Equation \eqref{fGARCHEqn} is related to the Box-Cox transformation of the conditional standard deviation, where the absolute value function can be transformed by the parameter $\delta$, while the shape is determined by $\gamma$. The parameter $\delta$ is subject to shifts and rotations through the $\zeta_{2j}$ and $\zeta_{1j}$, respectively. The full specification of the family GARCH model can be fitted when $\delta$ = $\gamma$ (see \cite{Ghalanos2018}). The volatility persistence of the fGARCH model can be obtained through the parameter estimate $\hat{P}$, stated as:
\begin{equation}\label{fGARCHEqnPers}
		\hat{P} = \sum_{j=1}^{q}\beta_j + \sum_{j=1}^{p} \alpha_j \varrho_j,
\end{equation}

where $\varrho_j$, as stated in Equation \eqref{fGARCHPersExt}, denotes the expectation of $z_{t}$ below the Box-Cox transformation that is associated with the absolute value asymmetry term. This study used the ``persistence()'' function from the rugarch package of R software to estimate the persistence. The unconditional variance of the fGARCH model, in relation to the persistence, is $\hat{\sigma}^{2} = \hat{\omega} / (1 - \hat{P})^{2/\gamma}$ \cite{Ghalanos2018}. Readers can refer to \citep{Ghalanos2022,Ghalanos2018,Hentschel1995} for more information on the nested models and fGARCH model.
\begin{equation}\label{fGARCHPersExt}
		\varrho_j = \verb"E"(|z_{t-j} - \zeta_{2j}| - \zeta_{1j}(z_{t-j} - \zeta_{2j}))^{\delta} = \displaystyle\int_{-\infty}^{\infty} (|z - \zeta_{2j}| - \zeta_{1j}(z - \zeta_{2j}))^{\delta}f(z, 0, 1, \ldots)dz
\end{equation}

\subsection{The Threshold GARCH Model}
The asymmetric TGARCH model was independently developed by Glosten,
Jaganathan, Runkle \cite{GJR1993} and Zakoian \citep{Zakoian1994}. The conditional variance of the
Threshold GARCH (1,1) model can be specified as:

\begin{equation}\label{(10)}
\sigma^2_t= \omega + (\alpha + \gamma
d_{t-1})\varepsilon^2_{t-1} + \beta\sigma^2_{t-1},
\end{equation}

where the coefficient $\gamma$ represents the leverage or asymmetry term, $\alpha$ and $\beta$
are unknown parameters. The dummy variable $d_{t-1}$ is a negative innovations indicator defined as $d_{t-1} = 1$ if $\varepsilon_{t-1} < 0$, and $d_{t-1} = 0$ if $\varepsilon_{t-1}\geq 0$. In this model, depending on
the $\varepsilon_{t-1}$ being above or under the threshold value (zero), good news $\varepsilon_{t-1} > 0$, and bad news $\varepsilon_{t-1} < 0$ have different effects on volatility \cite{Tsay2005}. Good news has
an impact of $\alpha$, while bad news has an impact of $\alpha + \gamma$.
If $\gamma > 0$, then volatility is increased by bad news, which shows the
existence of leverage effect. This indicates that negative shocks or bad news will impact
future volatility more than positive shocks or good news of the same size. The news impact is asymmetric if $\gamma \neq 0$.

\subsection{The Component GARCH Model}
The CGARCH model was introduced by Engle and Lee \cite{EngleLee1999} to decompose the time-varying GARCH
conditional variance into transitory and permanent components for describing the short-run and
long-run effects of volatility. The model is applied in this study to investigate the decomposition of volatility in the financial returns.

Consider the GARCH (1,1) model's conditional variance $\sigma^2_t$ as:
\begin{equation}
\sigma^2_t = \omega + \alpha(\varepsilon^2_{t-1} - \omega) +
\beta(\sigma^2_{t-1} - \omega).
\end{equation}

From the equation, the conditional variance shows mean reversion to a time invariant value $\omega$. Therefore, the effect of a past innovation decays to zero eventually as the volatility converges to this value $\omega$ with powers of $\alpha + \beta$.

By contrast, for the permanent specification, the component GARCH
model allows mean reversion to a varying level by replacing the
time-invariant mean value $\omega$ in the standard GARCH model with a time
varying component $m_t$, modeled as:
\begin{equation}
m_t = \hat{\omega} + \rho(m_{t-1} - \hat{\omega}) + \phi(\varepsilon^2_{t-1} - \sigma^2_{t-1}),
\end{equation}

where $m_t$ is the time varying long run volatility level that converges to
the long-run time-invariant volatility level $\hat{\omega}$ with the power of
$\rho$. The permanent component therefore explains the long run persistence behaviour of $\sigma^2_t$. The coefficients $\phi$ and $\alpha$ are used to measure the effects of a shock to the permanent and transitory components, respectively \cite{ChiangETAL2009}.
The value of $\rho$ measures the long-run persistence and it usually
lies close to one, implying that the long-run volatility converges or
approaches the mean at a very slow pace, which in turn makes the effect of
shocks to persist for a long time. This situation is mostly experienced when $\rho$
is between 0.99 and 1; in that case $m_t$ will approach $\hat{\omega}$ very
slowly and volatility will persist much longer. The nearer the estimated
value of $\rho$ is to zero the faster it approaches $\hat{\omega}$ and the
nearer it is to one, the slower it approaches $\hat{\omega}$.

The second aspect of the component GARCH model describes the transitory
specification, $\sigma^2_t - m_t$, as:
\begin{equation}\label{GodsThrone}
\sigma^2_t - m_t = \alpha(\varepsilon^2_{t-1} - m_{t-1}) + \beta(\sigma^2_{t-1} - m_{t-1}).
\end{equation}

As shown in Equation \eqref{GodsThrone}, the difference between the conditional variance and the permanent (or trend) component,
$\sigma^2_{t-1} - m_{t-1}$, is the transitory component of $\sigma^2_t$, with the persistence parameter $\alpha + \beta$. Hence, the transitory component converges to zero with powers of $\alpha + \beta$.

\subsection{The GAS Model}\label{beular}
An alternative method to the family GARCH model for modelling volatility can be found in a Score Driven (SD) model known as the Generalised Autoregressive Score (GAS) model, introduced by Harvey \cite{Harvey2013} and Creal et al. \cite{Crealetal2013a} (see \cite{Ardiaetal2019,Cataniaetal2020}). The model uses the score of the conditional density function to determine the time variation in the parameters. The score functions are robust to outliers, and the model is quite suitable for modelling skewed or fat-tailed time series data like financial returns \cite{Opschooretal2018,Harvey2013,AlanyaBeltran2022,HaddadETAL2023}. Moreover, like the other observation-driven\footnote{Observation-driven models are developed to model large changes (which may occur in the form of jumps or shifts) and distributional asymmetries that often exist in financial time series \cite{HaddadETAL2023}.} models, extensions to long memory behaviour, asymmetric, and other time series dynamics are possible. Furthermore, likelihood estimation using the GAS model is simple and direct \cite{Ardiaetal2019}.

\subsubsection{Model Specification}\label{beular2}
Let an $N \times 1$ vector $\bm{r}_{t}$ imply the dependent variable of interest, $\bm{\vartheta}_{t}$ the vector of time-varying parameter, $\bm{x}_{t}$
a vector of exogenous variables (i.e., the covariates), all at time $t$, and $\bm{\zeta}$ a vector of time-invariant parameters. Define $\bm{R}^{t} = \{\bm{r}_{1}, \ldots, \bm{r}_{t}\}$, $\bm{\Theta}^{t} = \{\bm{\vartheta}_{0}, \bm{\vartheta}_{1}, \ldots, \bm{\vartheta}_{t}\}$, and $\bm{X}^{t} = \{\bm{x}_{1}, \ldots, \bm{x}_{t}\}$. The information set that is available at time $t$ consists of $\{\bm{\vartheta}_{t}, \calF_{t}\}$, where

\begin{equation}\label{gasrefa}
 \calF_{t} = \{\bm{R}^{t-1}, \bm{\Theta}^{t-1}, \bm{X}^{t}\}, ~~~~\mbox{for} ~~t ~= ~1, \ldots, n.
\end{equation}

It is assumed that the generation of $\bm{r}_{t}$ is through the observation density \cite{Crealetal2012,Crealetal2013a}

\begin{equation}\label{gasrefaAA}
 \bm{r}_{t} \sim p \left(\bm{r}_{t} | \bm{\vartheta}_{t}, \calF_{t}; \bm{\zeta} \right).
\end{equation}

It is further assumed that the mechanism for updating $\bm{\vartheta}_{t}$ (i.e., the time-varying parameter) is given by the autoregressive updating equation:

\begin{equation}\label{bvngas1123}
   \bm{\vartheta}_{t + 1} = \bm{\kappa} + \sum_{i = 1}^{p} \bm{A}_{i}\bm{s}_{t - i + 1} + \sum_{j = 1}^{q} \bm{B}_{j} \bm{\vartheta}_{t - j + 1}.
\end{equation}

Equation \eqref{bvngas1123} is presented in Ardia et al. \cite{Ardiaetal2019} as:

\begin{equation}\label{bvn23}
   \bm{\vartheta}_{t + 1} \equiv \bm{\kappa} + \textbf{A} \bm{s}_{t} + \textbf{B}\bm{\vartheta}_{t},
\end{equation}

where $\bm{\kappa}$, $\textbf{A}$ and $\textbf{B}$ are matrices of coefficients with appropriate dimensions and they are functions of the static parameter $\bm{\zeta}$, while the scaled score $\bm{s}_{t}$ is a suitable function of past data, $\bm{s}_{t} = \bm{s}_{t}(\bm{r}_{t}, \bm{\vartheta}_{t}, \calF_{t}; \bm{\zeta})$ (see \cite{Crealetal2012,Crealetal2013a}). Vector $\bm{\kappa}$ controls the level of the process $\bm{\vartheta}_{t}$, the matrix of coefficients $\textbf{A}$ controls (or determines) the impact of $\bm{s}_{t}$ on $\bm{\vartheta}_{t + 1}$, while matrix $\textbf{B}$ determines the persistence of the process \cite{Ardiaetal2019,Koopmanetal2017}. In particular, $\bm{s}_{t}$ denotes the direction for updating the vector of parameters from $\bm{\vartheta}_{t}$ to $\bm{\vartheta}_{t + 1}$, hence, $\textbf{A}$ can be described as the step of the update. In other words, $\bm{s}_{t}$ acts as a steepest ascent algorithm to improve the local fit of the model given the current parameter position (see \cite{Ardiaetal2019}). To implement the GAS models\footnote{The GAS model is implemented in this study with the use of the R package GAS developed by Ardia et al. \cite{Ardiaetal2019}.}, the $\textbf{A}$ and $\textbf{B}$ matrices are constrained to exist as diagonals (see \cite{Ardiaetal2019,OhPatton2016}), e.g., for a GAS model with Student's $t$ error distribution, $\textbf{A} \equiv \mbox{diag}(a_{\mu}, a_{\sigma}, a_{\nu})$ and $\textbf{B} \equiv \mbox{diag}(b_{\mu}, b_{\sigma}, b_{\nu})$, where $\mu$, $\sigma$ and $\nu$ are location, scale and shape parameters, respectively. Hence, $b_{\mu}$ refers to the persistence of the conditional mean (location), while $b_{\sigma}$ is the persistence of the conditional variance (or scale) (see \cite{Ardiaetal2019}). This persistence parameter $b_{\sigma}$ of the GAS model coincides with the persistence parameters $\alpha + \beta$ of the standard GARCH model\footnote{The GAS model with assumed Normal distribution coincides with the standard GARCH(1,1) model of Bollerslev \cite{Bollers1986} (see \cite{Blasquesetal2014,Crealetal2013a,OhPatton2016,Ardiaetal2019}). Hence, we investigated this by comparing the estimate of the persistence $\hat{b}_{\sigma}$ from the GAS model fitted with a time-varying scale parameter, and the estimate $\hat{\alpha}_1+\hat{\beta}_1$ from the GARCH(1,1) model. Both models were fitted to the real return S\&P Indian stock data under the Normal error, and their outcomes yielded $\hat{b}_{\sigma} \equiv \hat{\alpha}_1+\hat{\beta}_1 \approx 0.97$.} of Bollerslev (see \cite{Blasquesetal2014,Crealetal2013a,OhPatton2016,Ardiaetal2019}).

The GAS approach depends on the observation density in Equation \eqref{gasrefaAA} for a given parameter $\bm{\vartheta}_{t}$. When an observation $r_{t}$ is realised, the time-varying $\bm{\vartheta}_{t}$ to the next period ${t + 1}$ can be updated using Equation \eqref{bvngas1123} with

\begin{equation}\label{vn2jvh3b}
   \bm{s}_{t} = \bm{S}_{t}\cdot \bm{\nabla}_{t}, ~~~~~~\bm{\nabla}_{t} = \frac{\partial\mbox{ln}p \left(\bm{r}_{t} | \bm{\vartheta}_{t}, \calF_{t}; \bm{\zeta} \right)}{\partial \bm{\vartheta}_{t}}, ~~~~~~ \bm{S}_{t} = S\left(t, \bm{\vartheta}_{t}, \calF_{t}; \bm{\zeta}\right),
\end{equation}

where $S(\cdot)$ represents a matrix function, $\mbox{ln}$ denotes the natural logarithm, $\bm{\nabla}_{t}$ is the score of Equation \eqref{gasrefaAA} evaluated at $\bm{\vartheta}_{t}$, and $\bm{S}_{t}$ is the scaling matrix (see \cite{Ardiaetal2019,Crealetal2012,Crealetal2013a,HaddadETAL2023}). Given the dependence of the stated driving mechanism in Equation \eqref{bvngas1123} on the scaled score vector in Equation \eqref{vn2jvh3b}, the GAS model with orders $p$ and $q$ can be defined by Equations \eqref{gasrefaAA}, \eqref{bvngas1123}  and \eqref{vn2jvh3b}. The model can be referred to as GAS ($p$, $q$) and the orders $p$ and $q$ are typically taken as $p$ = $q$ = 1 (see \cite{Crealetal2012,Crealetal2013a}). However, for details on including more lags in the GAS process, see \cite{Crealetal2013a,Blasquesetal2014}.

As reported by Ardia et al. \cite{Ardiaetal2019}, the authors Creal et al. \cite{Crealetal2013a} suggested setting $\bm{S}_{t}$ to the inverse of the information matrix ($\calI$) to a power $\gamma > 0$ of $\bm{\vartheta}_{t}$ to account for the variance of $\bm{\nabla}_{t}$. To be precise,

\begin{equation}\label{11aa}
  \bm{S}_{t} = \bm{\calI}_{t|t-1}^{-\gamma}, ~~~~~~~~~~~~ \bm{\calI}_{t|t-1} = \verb"E"_{t-1} \left[\bm{\nabla}_{t} \bm{\nabla}_{t}^{\top}\right],
\end{equation}

where the expectation $\verb"E"_{t-1}$ is taken with respect to the conditional distribution of $\bm{r}_{t}|\bm{r}_{1:t-1}$. The parameter $\gamma$ normally takes value in the set $\{0, \frac{1}{2}, 1\}$. However, other choices of $S_{t}$ are possible as well (see \cite{Blasquesetal2014}). When $\gamma$ = 0, $\bm{S}_{t}$ = $\textbf{I}$ (identity matrix), which means there is no scaling. If $\gamma$ = $\frac{1}{2}$ ($\gamma$ = 1), then the conditional score $\bm{\nabla}_{t}$ is pre-multiplied by the square root of (the inverse of) its covariance matrix $\calI_{t}$. However, whatever the choice of $\gamma$, $\bm{s}_{t}$ is a martingale difference with respect to the distribution of $\bm{r}_{t}|\bm{r}_{1:t-1}$, i.e., \verb"E"$_{t - 1}$[$\bm{s}_{t}$] = 0 for all $t$ (see \cite{Ardiaetal2019}). The GAS framework embodies many available observation-driven models in the literature for a suitable choice of the scaling matrix $\bm{S}_{t}$ \cite{Crealetal2013a,HaddadETAL2023}. Readers can refer to \cite{Ardiaetal2019,Harvey2013,CrealETAL2011,Crealetal2012,Crealetal2013a,Cataniaetal2020,Koopmanetal2017,HaddadETAL2023} for more details on the GAS model.

\subsection{The Beta-Skew-$t$-EGARCH Model}
To begin with, the Beta-\textit{t}-EGARCH as a model without skewness was
proposed by Harvey and Chakravarty \cite{HarvChakr2008}, and Harvey \cite{Harvey2013}. It is an extension
made to the EGARCH model where an equation that depends on the conditional score of the last observation drives the variance or scale \cite{CrealETAL2008,CrealETAL2011}. The Beta-\textit{t}-EGARCH model can be regarded as an unrestricted version of the GAS model of Creal et al. \cite{Crealetal2013a} (see \cite{Sucarrat2013}). In other words, it is a dynamic model driven by conditional score, which is a martingale difference. A martingale difference has a constant (or zero) conditional expectation \cite{Harvey2013}. Due to various implications of conditional skewness on asset pricing, conditional score models were extended to skew distributions \cite{HarveySuca2014}. Hence, the Beta-Skew-$t$-EGARCH model is a skewed version of the Beta-$t$-EGARCH model, where skewness can be brought in through the method introduced by \mbox{Fern$\acute{a}$ndez} and Steel \cite{FernSteel1998} (see \cite{Sucarrat2013,HarveySuca2014}).

A number of useful and attractive properties are attributed to this
skewed model. As in the Beta-$t$-EGARCH model where observations that could
be seen as outliers for a Normal distribution are down-weighted with the use
of the conditional score \cite{HarveySuca2014}, the
Beta-Skew-$t$-EGARCH model also displays robustness to outliers or jumps. In particular,
it performs quite well in modelling of key stylised facts of financial returns like fat-tails, leverage effect,
conditional skewness and the long-memory behaviour of volatility decomposition
into long-term and short-term components. The model has two versions, described below as "one-component and two-component models" .

\subsubsection{The One-Component Beta-Skew-$t$-EGARCH Model}\label{trippleA}
The three equations below illustrate the first order Beta-Skew-$t$-EGARCH model's martingale difference version \cite{HarveySuca2014}.

\begin{eqnarray}\label{GodsLIGHT}
  R_t &=& \mbox{exp}(\lambda_t)z_t = \sigma_tz_t, ~~~ z_t \sim st(0, \sigma^2_z, \nu, \eta), ~~ \nu>2, ~~ \eta\in(0, \infty), \\
  \lambda_t &=& \omega + \lambda^\dag_t, \\
  \lambda^\dag_t &=& \phi_1\lambda^\dag_{t-1} + \kappa_1 u_{t-1} + \kappa^\ast sgn(-R_{t-1})(u_{t-1} + 1), ~~~ |\phi_1|<1,
\end{eqnarray}

where $R_t$\footnote{The demeaned return $R_t$ is the same as the error term $\varepsilon_t$ (i.e., the unpredictable part of return) in Equation \eqref{GodsHeaven}, but with $z_t$ distributed as a skew Student's $t$, and it is not standardised to have a unit variance (see \cite{Sucarrat2013}).} is the demeaned return, $\sigma_t$ represents the volatility or conditional scale, and the conditional error $z_t$ is not standardised i.e., the variance is not one. Equation \eqref{GodsLIGHT} shows that $z_t$ follows the skew Student's $t$ distribution with a mean of zero, scale $\sigma^2_z$, degrees of freedom $\nu$, and skewness parameter $\eta$. The $z_t$ is defined as $z_t = z^{\ast}_t - \mu_{z^{\ast}}$, where $z^{\ast}_t$ is an uncentred\footnote{That is, the mean is not necessarily equal to zero.} skew Student's $t$ variable with $\nu$ degrees of freedom, mean $\mu_{z^{\ast}}$, and skewness parameter $\eta$. The $\omega$
denotes the log-scale intercept that is described as the long-term log-volatility, while $\phi_1$ represents the persistence parameter (the bigger the persistence, the more the clustering of volatility). The $\kappa_1$ is the ARCH parameter that indicates the response to shocks (the bigger the absolute value of the parameter, the greater the response of volatility to shocks), $\kappa^\ast$ is the leverage parameter, $u_{t}$ denotes the conditional score, that is, the derivative of the log-likelihood of $R_t$ at $t$ with respect to $\lambda_t$, and \textit{sgn} is the sign function. Details on this model are discussed in \cite{Sucarrat2013}.

\subsubsection{The Two-Component Beta-Skew-$t$-EGARCH Model}
In order to accommodate the long memory feature of financial return like a
two-component GARCH model, the two-component Beta-Skew-$t$-EGARCH model was
introduced. The model is used to decompose volatility persistence into long
term (or long-run) and short term (or short-run) components \cite{AlizadETAL2002}. The first order
two-component Beta-Skew-$t$-EGARCH model's martingale difference version
\cite{HarveySuca2014} is given as:

\begin{eqnarray}
  R_t &=& \mbox{exp}(\lambda_t)z_t = \sigma_tz_t, ~~~ z_t \sim st(0, \sigma^2_z, \nu, \eta), ~~ \nu,\eta\in(0, \infty), \\
  \lambda_t &=& \omega + \lambda^\dag_{1,t} + \lambda^\dag_{2,t}, \\
  \lambda^\dag_{1,t} &=& \phi_1\lambda^\dag_{1,t-1} + \kappa_1 u_{t-1}, ~~~ |\phi_1|<1,\\
  \lambda^\dag_{2,t} &=& \phi_2\lambda^\dag_{2,t-1} + \kappa_2 u_{t-1} + \kappa^\ast sgn(-R_{t-1})(u_{t-1} + 1), ~~~ |\phi_2|<1, ~~ \phi_1 \neq \phi_2,
\end{eqnarray}

where $\lambda_{1,t}$ and $\lambda_{2,t}$ represent the time-varying long-run and short-run components of log-volatility, respectively. Both components are driven by the conditional score $u_{t}$. The $\phi_1$ and $\phi_2$ are the long-term and short-term persistence parameters and the model is not identifiable if $\phi_1$ = $\phi_2$\footnote{To avoid explosive recursions during the model’s estimation, it is necessary to restrict $\phi_1$ and $\phi_2$ so that  $|\phi_1| < 1$ and $|\phi_2| < 1$. Hence, to ensure this, Sucarrat \cite{Sucarrat2013} suggested the use of the nlminb function as the optimizer during estimation (computation).}. The $\kappa_1$ and $\kappa_2$ are parameters that indicate the long-run and short-run responses to shocks, respectively.

\subsection{The True Parameter Recovery Measure}
Monte Carlo simulation (MCS) studies largely focus on the estimator's ability to recover the true data-generating parameter (see \cite{Chalmers2019}). Hence, the True Parameter Recovery (TPR) measure was introduced by Samuel et al. \cite{SamuelETAL2023} as a way of measuring how the MCS estimator performs at recovering the true parameter. The measure is used as a proxy for the coverage\footnote{The probability that the true parameter is contained within a confidence interval of estimates is known as the coverage probability \cite{Hilary2002}.} of the MCS experiment to calculate the level of recovery of the true parameter by the MCS estimator. It can be stated as:
\begin{equation}\label{TPRCovProb}
\mbox{TPR} = \left(K - \left[\frac{(\vartheta - \hat{\vartheta})}{\vartheta} \times K\right]\right)\%,
\end{equation}

where $\vartheta$ is the true data-generating parameter, $\hat{\vartheta}$ denotes the estimator, and $K$ = 0, 1, 2, $\ldots$, 100 represents the nominal recovery level. A TPR outcome of 90\% or 95\%, for instance, implies that the MCS estimator is able to recover 90\% or 95\% of the true parameter. The MCS estimator $\hat{\vartheta}$ will fully recover the true parameter $\vartheta > 0$ when $\hat{\vartheta}$ = $\vartheta$, such that the outcome of the TPR is equal to the specified nominal recovery level $K$ (that is, $K$\% = TPR) (see \cite{SamuelETAL2023}).

\subsection{Meta-Statistics and Quasi-Maximum Likelihood Estimation}\label{llbLondani@}
After the simulation modelling, the general simulation behaviours are summarised using meta-statistics. Meta-statistics are the metrics or performance measures for evaluating the modelling outcomes. The two meta-statistical summaries used in this study are the bias and standard error (SE). Bias is the average difference between the true (population) parameter and its estimate \cite{FengShi2017}. The optimal value of bias is 0, and the nearer the absolute value of the bias estimator is to zero, the better the outcome. The SE is used to evaluate sampling variability in the estimation (see \cite{Chalmers2019,Yuanetal2015}). The lower the SE, the higher the precision or efficiency of the MCS estimator \cite{Morrisetal2019}. The consistency of the MCS estimator occurs when SE decreases (or tends to zero) such that $\hat{\vartheta} \rightarrow \vartheta$ as the sample size $N$ increases (or tends to infinity).

The (R)MSE is also used to measure efficiency but is not used here because it is claimed to be more relevant as a performance measure when the simulation’s purpose relates to forecasting rather than estimation (see Morris et al. \cite{Morrisetal2019}). The authors further reported that RMSE is more sensitive to the chosen amount of observations when used for comparing methods than when bias or SE metric are used alone. Therefore, the SE is used as the key metric for the estimators’ precision and efficiency in this volatility modelling study. Moreover, it is reported by several authors that MSE is sensitive to the level of returns volatility and extreme observations (see \cite{Patton2011,WangYang2018}). Therefore, the optimal error assumption required to estimate the persistence of volatility using the fGARCH and GAS model will emerge from estimator with the best precision (efficiency) in terms of SE. See \cite{ChalmerAndAdkins2020,SigalChalm2016,Morrisetal2019} for details on meta-statistics.

In view of the fact that financial data are heavy-tailed and characteristically leptokurtic,
i.e., non-Normal (see \cite{Bollerslev1987}), the standard errors of the estimates
using MLE will be inconsistent if the true error distribution is not Normal.
However, this can be resolved using Quasi-Maximum Likelihood Estimation
(QMLE) with Normal error (see \cite{BollerWoold1992}). It is believed that even
if the true error distribution is not Normal, the QMLE's standard error is
asymptotically consistent \cite{FengShi2017}. Hence, the QMLE's standard errors from the fit of the fGARCH model \cite{Zivot2013,White1982,Ghalanos2018} are used in the simulation since they are robust against violation or misspecification of the assumption of the error distribution \cite{Qietal2010}.

\section{Results: Simulation and Empirical}\label{CCC333}
This section presents the Monte Carlo simulation and empirical studies of the modelling involving the (family) GARCH and GAS models with their extended (or nested) versions. All the applied models in this study can be estimated through the MLE once a distribution for the innovations is specified.

\subsection{Simulation Study through the fGARCH Model}\label{THELORD}
In this section, we present the outcomes of the Monte Carlo simulation study involving the fGARCH model. It is believed that observation-driven\footnote{Observation-driven models are developed to model large changes (which may occur in the form of jumps or shifts) and distributional asymmetries that often exist in financial time series \cite{HaddadETAL2023}.} models like the (family) GARCH can yield efficient outcomes when fitted with the true (or appropriate) innovation distribution \cite{Bollerslev1987,FengShi2017}. Hence, among ten selected innovations, we use MCS approach to determine an optimal or the most suitable assumed innovation distribution that is relevant for volatility persistence estimation through the fGARCH model. The ten selected assumed innovations are the Gaussian (or Normal), Student’s $t$, Generalised Error Distribution (GED), skew-Normal, skew-Student’s $t$, skew-GED, Johnson’s reparametrised SU (JSU) distribution, Generalised Hyperbolic (GHYP) distribution, Generalised Hyperbolic Skew-Student’s $t$ (GHST) distribution, and Normal Inverse Gaussian (NIG). See  \cite{Ghalanos2018,Eling2014,BarndorNieletal2013,LeePai2010,AshAbdham2010,Pourahmadi2007,AzzaliniCapitanio2003,BrancoDey2001,Azzalini1985} for details on the assumed innovations.

\subsubsection{Modelling Design}\label{LoveGOD}
MCS studies are computer experiments that involve data creation by pseudo-random sampling through known probability distributions \cite{Morrisetal2019}. The random data can be generated by repeated resampling (where the true data generating model is not known) or through known parametric models \cite{Morrisetal2019}. This study generates the simulated return datasets through the parametric model fGARCH for volatility persistence estimation. The general design of the simulation study is described as follows. To begin with, simulated return observations are generated from the true model, where the true model is the applied model fitted with the true distributed error \cite{FengShi2017}, e.g., fGARCH-Student’s $t$, if Student’s $t$ is the true error distribution. In other words, we fit the true model to the actual return data\footnote{The actual return data are the S\&P Indian market returns.}, and the empirical outcomes from the fit are used to generate the required simulated return data.

Next, the simulated return data are analysed, where the performance of the volatility estimator $\hat{\alpha}+\hat{\beta}$ is evaluated through meta-statistics (SE and bias) comparisons, under each of the ten assumed errors, to obtain the estimator with the best efficiency and precision. That is, we carry out the comparisons under the ten selected assumed innovations, and based on this, the estimator with the best efficiency (from comparing the meta-statistics) yields the optimal or most adequate innovation to estimate the persistence of the volatility.

\subsubsection{Presentation of the Simulation Study}
The simulation begins with the use of fGARCH(1,1) model in Equation \eqref{paper2MCSMummy} as the true data generating process (DGP) for generating the simulated returns, where the true innovation is Student's $t$ with degrees of freedom $\nu$ = 4.1.

\begin{equation}\label{paper2MCSMummy}
		\sigma^{\gamma}_t = \omega + \alpha_1 \sigma^{\gamma}_{t-1}(|z_{t-1} - \zeta_{21}| - \zeta_{11}\{z_{t-1} - \zeta_{21}\})^{\delta} + \beta_1 \sigma^{\gamma}_{t-1}.
\end{equation}

Here, the fGARCH(1,1) model is used as the true DGP to generate the data because the first lag of conditional variance can considerably capture existing volatility clustering in the dataset. That is, volatility of returns depends less on distant past events but more on recent past events \cite{JavedMant2013}. Hence, we present the applications of the fGARCH model for obtaining an optimal assumed innovation to estimate the persistence of the conditional variance, where the results of fitting the model with each of the selected ten assumed innovations are compared.

Furthermore, a Student’s $t$ with degree of freedom (or shape parameter) $\nu$ = 4.1 is used to ensure that $\verb"E"[z_t^4]$ < $\infty$, which is relevant for $\sqrt{N}$ consistent quasi-maximum likelihood (QML) estimation following the assumption of Francq and Thieu \cite{FrancqThieu2019}. Moreover, the Student’s $t$ is applied as the true distribution for the disturbances because it can adequately deal with fat-tailed or leptokurtic features \cite{HarveySuca2014} experienced in financial data \citep{Hentschel1995}. It is also assumed that financial data like the interest rates, stock prices and exchange rates appear to exhibit a distribution like that of the Student’s $t$ \citep{Heracleous2007}. Bollerslev \cite{Bollerslev1987} also claims that the fat-tailed property for some data can be approximated more accurately through a conditional Student’s $t$ distribution \cite{Chou1988}.

The true parameter values used for the simulated data generation are $\omega = 0.0518$\footnote{We approximated the presented values to four decimal places for brevity.}, $\mu = 0.0570$, $\alpha = 0.1011$, $\beta = 0.7988$, $\gamma = 2.0349$, $\zeta_{11} = -0.1815$, $\zeta_{21} = 1.2621$ and $\nu = 4.1$. These parameter values are estimates obtained by fitting the true model fGARCH(1,1)-Student’s $t$ to the Indian return data. Next, with seed value 12345, we use these true parameter values to generate simulated datasets of sample size $N$ = 22,000, replicated 1,000 times. However, to prevent the effect of initial values in the simulation process, we discard the first $N$ = \{8,000; 7,000; 6,000\} sets of observations at each stage of the generated 22,000 return observations. Hence, we use the last $N$ = \{14,000; 15,000; 16,000\} data points under each of the ten assumed innovations, as presented in Table \ref{fGARCHSTD}. These trimmings are carried out following the data generating designs of Feng and Shi \cite{FengShi2017}. The true parameter values ($\alpha$, $\beta$, $\alpha+\beta$) = (0.1011, 0.7988, 0.8999) from the fit are set a priori for the simulation process as shown in Table \ref{fGARCHSTD}.

\begin{table}[h!]
\caption{The fGARCH modelling simulation outcomes.\label{fGARCHSTD}}
	\begin{adjustwidth}{-\extralength}{0cm}
		\newcolumntype{C}{>{\centering\arraybackslash}X}
		\begin{tabularx}{\fulllength}{C|CCCCCCC}
\hline		
  \multicolumn{1}{l|}{} &  \multicolumn{7}{c}{True data-generating model: fGARCH(1,1)-Student's $t$}  \\
  \multicolumn{1}{l|}{} &  \multicolumn{7}{c}{True parameters set a priori: $\alpha+\beta$ = 0.8999, $\alpha$ = 0.1011 and $\beta$ = 0.7988}  \\ \hline
			\textbf{Assumed innovation}&\textbf{$N$}& \textbf{$\hat{\alpha}$} &\textbf{$\hat{\beta}$} &\textbf{$\hat{\alpha}+\hat{\beta}$} &\textbf{SE$_{\hat{\alpha}+\hat{\beta}}$} & \textbf{Bias$_{\hat{\alpha}+\hat{\beta}}$} &\textbf{TPR$_{\hat{\alpha}+\hat{\beta}}$ (95\%)}  \\ \hline
   & 14,000&   0.0938  &  0.8059  &  0.8997   &  0.0986&  -0.0002  &  94.98\%  \\
   		  	\multirow[m]{1}{*}{Normal}& 15,000 &  0.0966 &  0.8027  & 0.8993     &  0.0204  &  -0.0006   & 94.94\% \\
				 & 16,000 & 0.0954  &  0.8054  &  0.9008    &  0.0173  &   0.0009 &  95.10\% \\ \hline	
& 14,000&  0.0934 &  0.8060  &  0.8994    &  5.3090  &  -0.0005   &  94.95\%  \\
   		  	\multirow[m]{1}{*}{skew-Normal}& 15,000 & 0.0963  &  0.8030  &  0.8993    &  0.0207  &  -0.0006 & 94.94\% \\
		\multirow[m]{1}{*}{}	& 16,000 & 0.0952  &  0.8058  &  0.9009    &  0.0175  &  0.0011  &  95.11\% \\ \hline	
& 14,000& 0.1136  &  0.7945  &  0.9080    &  0.0100  &  0.0082  &  95.86\%  \\
   		  	\multirow[m]{1}{*}{Student's $t$}& 15,000 &  0.1137 & 0.7960   &  0.9097    &  0.0098  & 0.0098  & 96.04\% \\
				 & 16,000 &  0.1132 &  0.7971  &  0.9103    &  0.0096  &  0.0104 &  96.10\% \\ \hline	
& 14,000&   0.1135   &  0.7944  &  0.9080   & 0.0100 &  0.0081  &  95.86\%  \\
   		  	\multirow[m]{1}{*}{skew-}& 15,000 &  0.1138 &  0.7959  &  0.9098    &  0.0098  &  0.0099  & 96.04\% \\
		\multirow[m]{1}{*}{Student's $t$}	& 16,000 &  0.1132 &  0.7970  &  0.9102     &  0.0097  &  0.0103  &  96.09\% \\ \hline	
& 14,000&  0.1015 &  0.7984  &  0.8999    &  0.0121  &   0.0000  &  95.00\%  \\
   		  	\multirow[m]{1}{*}{GED}& 15,000 & 0.1032  &  0.7984  &  0.9015    &  0.0101  &  0.0017 & 95.17\% \\
				 & 16,000 &  0.1025 &  0.8001  &  0.9026    &  0.0090  &   0.0027 &  95.29\% \\ \hline	
& 14,000& 0.1015  &  0.7982  &  0.8997    &  0.0118  &   -0.0002  &  94.98\%  \\
   		  	\multirow[m]{1}{*}{skew-GED}& 15,000 & 0.1032  &  0.7981  &  0.9013   &  0.0100  &   0.0015 & 95.16\% \\
		\multirow[m]{1}{*}{}	 & 16,000 & 0.1025  &  0.7999  &  0.9025    &  0.0090  &   0.0026  &  95.27\% \\ \hline	
& 14,000& 0.1123  &  0.7945  &  0.9069    &  0.0101  &  0.0070   &  95.74\%  \\
   		  	\multirow[m]{1}{*}{GHYP}& 15,000 & 0.1126  &  0.7961  &  0.9087    &  0.0097  &   0.0089 & 95.93\% \\
				 & 16,000 &  0.1121 &  0.7972  &  0.9094    &  0.0095  &  0.0095  &  96.00\% \\ \hline	
& 14,000&  0.1064 &  0.7955  &  0.9018    & 0.0107   &  0.0020  &  95.21\%  \\
   		  	\multirow[m]{1}{*}{NIG}& 15,000 &  0.1073 &   0.7966 &  0.9040    &  0.0093  &  0.0041 & 95.43\% \\
				 & 16,000 & 0.1068  &  0.7980  &  0.9048   &  0.0089  &   0.0050 &  95.52\% \\ \hline	
& 14,000& 0.1034  &  0.7977  &  0.9011    &  0.0111  &   0.0013  &  95.13\%  \\
   		  	\multirow[m]{1}{*}{GHST}& 15,000 & 0.1035  &  0.7969  &  0.9004   &  0.0109  &  0.0005  & 95.06\% \\
				 & 16,000 &  0.1034 &  0.7973  &  0.9007   &  0.0104  &  0.0008 &  95.09\% \\ \hline	
& 14,000& 0.1094  &  0.7947  &  0.9041   &  0.0102  &   0.0042 &  95.44\%  \\
   		  	\multirow[m]{1}{*}{JSU}& 15,000 & 0.1099  & 0.7961   &  0.9060  &  0.0094  &   0.0061  & 95.65\% \\
				 & 16,000 & 0.1094  &  0.7973  &  0.9068   & 0.0092   &  0.0069 &  95.73\% \\ \hline				
		\end{tabularx}
	\end{adjustwidth}
	\noindent{\footnotesize{Monte Carlo simulations with $L = 1,000$ replications in R version 4.0.3, where $N$ is the sample size. SE is the QMLE robust standard error of Bollerslev and Wooldridge \cite{BollerWoold1992}, while TPR is the True Parameter Recovery measure with a nominal recovery level of 95\% (i.e., 0.95).}}
\end{table}

After generating the simulated returns, we fit the fGARCH(1,1) model to each simulated return dataset under the ten assumed innovations. The parsimonious ARMA(1,1) model is also used as the most appropriate candidate ARMA models (where ar = 0.2598 and ma = -0.1922)\footnote{The ar = 0.2598 and ma = -0.1922 are the true parameter values of the Autoregressive and Moving Average parts, respectively, of the ARMA model.} to remove autocorrelation in the simulated returns. However, dependence in the simulation process does not affect the achievement of consistency by the estimator \citep{Chib2015}.

Next, we use the selected bias and SE meta-statistics to assess the estimators. The most adequate assumed innovation distribution that will be used to estimate the persistence of the volatility will be obtained from the estimator with the best efficiency and precision from the meta-statistical comparisons carried out under the ten selected assumed innovations. To avoid coding errors, the SimDesign \cite{ChalmersP2020} package, with predefined meta-statistical functions, is used for the bias calculations \citep{SigalChalm2016,ChalmersAdkin2020}. The bias and SE for the $\hat{\alpha}+\hat{\beta}$ estimator are presented in Table \ref{fGARCHSTD}, but we use SE as the main metric (or performance measure) for precision and efficiency.

Next, we proceed with comparing the bias for the estimator $\hat{\alpha} + \hat{\beta}$. The results from the table show that the GED takes the lead at the start, but the GHST relatively outperforms the rest of the nine innovation assumptions in absolute value of bias as $N$ tends to the peak at 16,000. For comparisons through the main metric SE, the true innovation Student's $t$ and its skew version (the skew-Student's $t$) both take the lead at the beginning, but the NIG relatively outperforms the rest of the nine assumed innovations in efficiency and precision with the lowest values as $N$ tends to the peak.

\begin{figure}[H]
\begin{adjustwidth}{-\extralength}{0cm}
\centering
\includegraphics[height=2.25in, width=0.90\linewidth]{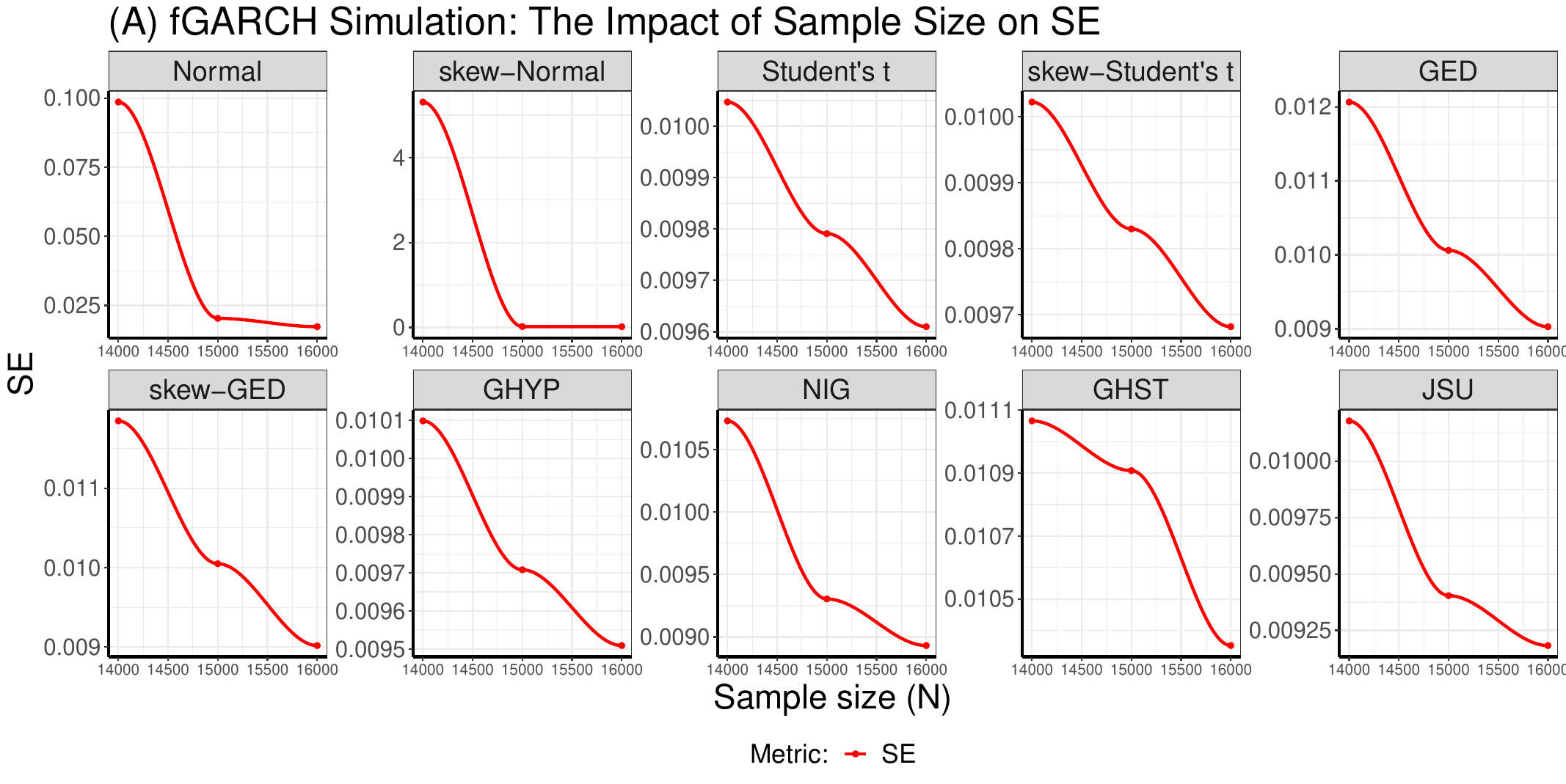}\hfil
      \includegraphics[height=2.25in, width=0.90\linewidth]{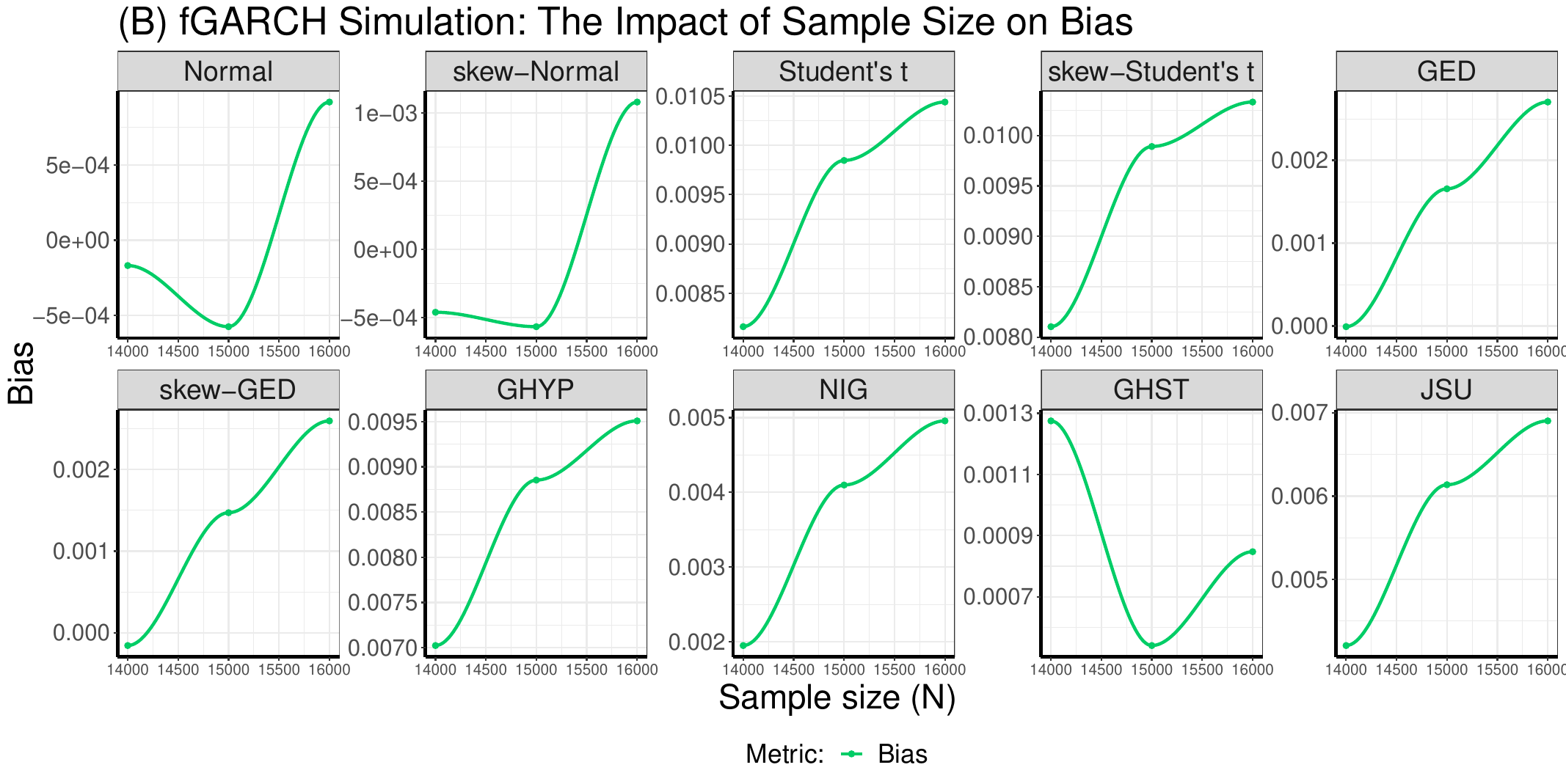}\hfil
  \end{adjustwidth}
\caption{Panels (A) and (B) display the impacts of sample size (N) on SE and bias, respectively, for MCS modelling involving the fGARCH(1,1)-Student's $t$. The SE shows $\sqrt{N}$ consistency, while bias does not dependent on $N$.\label{FirePure}}
\end{figure}

It is observed that the SE under each of the ten assumed innovations decreases as $N$ increases as presented in the table and also displayed in Panel A of Figure \ref{FirePure}. Such outcomes are consistent with the central limit theorem that suggests that the efficiency will improve as the sample size $N$ increases. In other words, the tabulated results show that the SEs of the $\hat{\alpha} + \hat{\beta}$ estimator display $\sqrt{N}$ consistency at recovering the true parameters under the ten innovations. It is further observed that the SEs of the $\hat{\alpha} + \hat{\beta}$ estimator for the assumed skew-Normal and the Normal innovations are the highest when they are compared with the SEs of the other eight innovation assumptions. This is a confirmation of the fact that the QMLE of the (f)GARCH-type model (with Normal error) is consistent but not efficient \cite{FengShi2017}. Panel B of Figure \ref{FirePure} shows that the bias is independent of $N$.

\begin{figure}[]
\begin{adjustwidth}{-\extralength}{0cm}
\centering
\includegraphics[height=2.5in, width=0.49\linewidth]{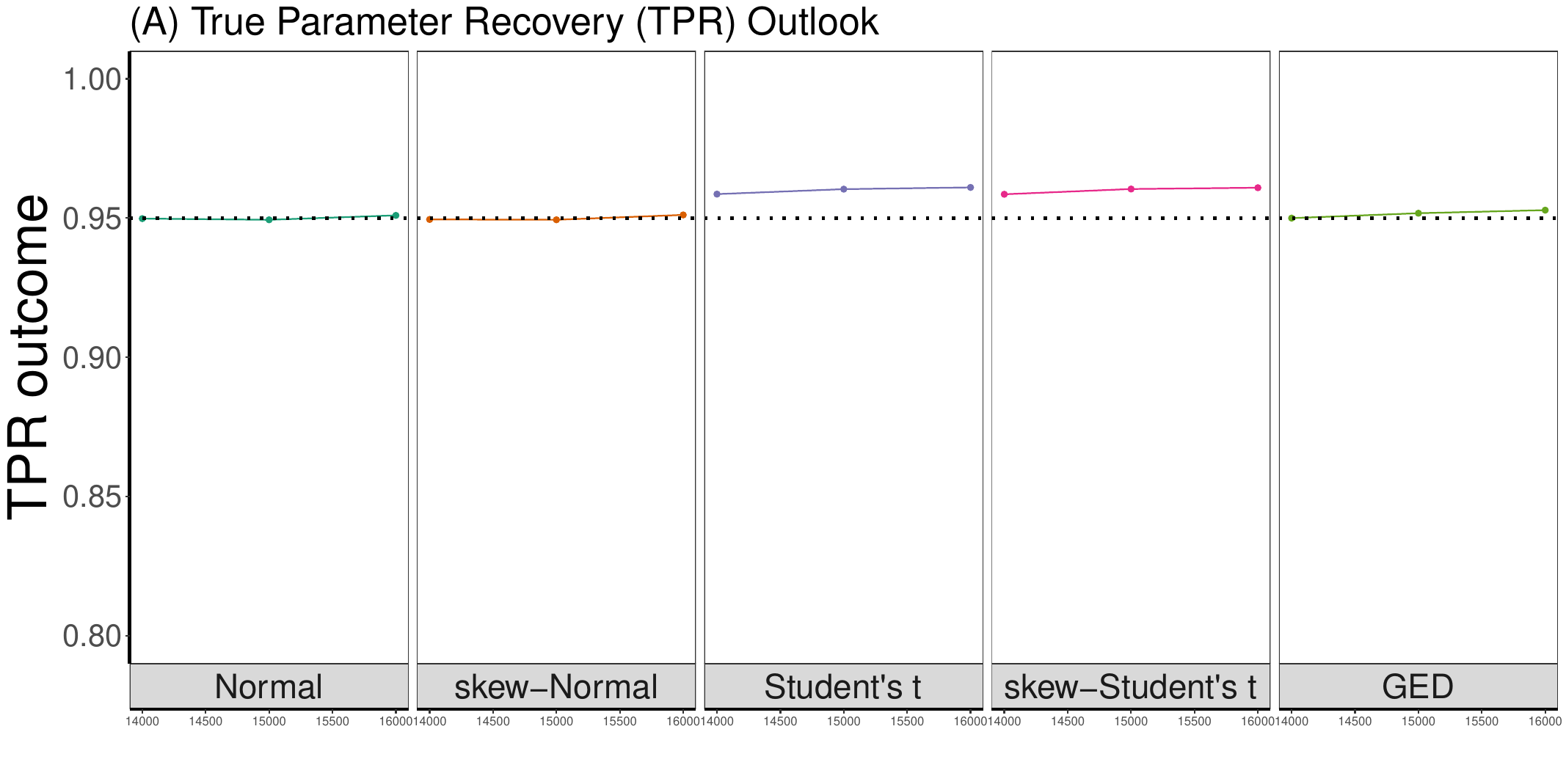}\hfil
   \includegraphics[height=2.5in, width=0.49\linewidth]{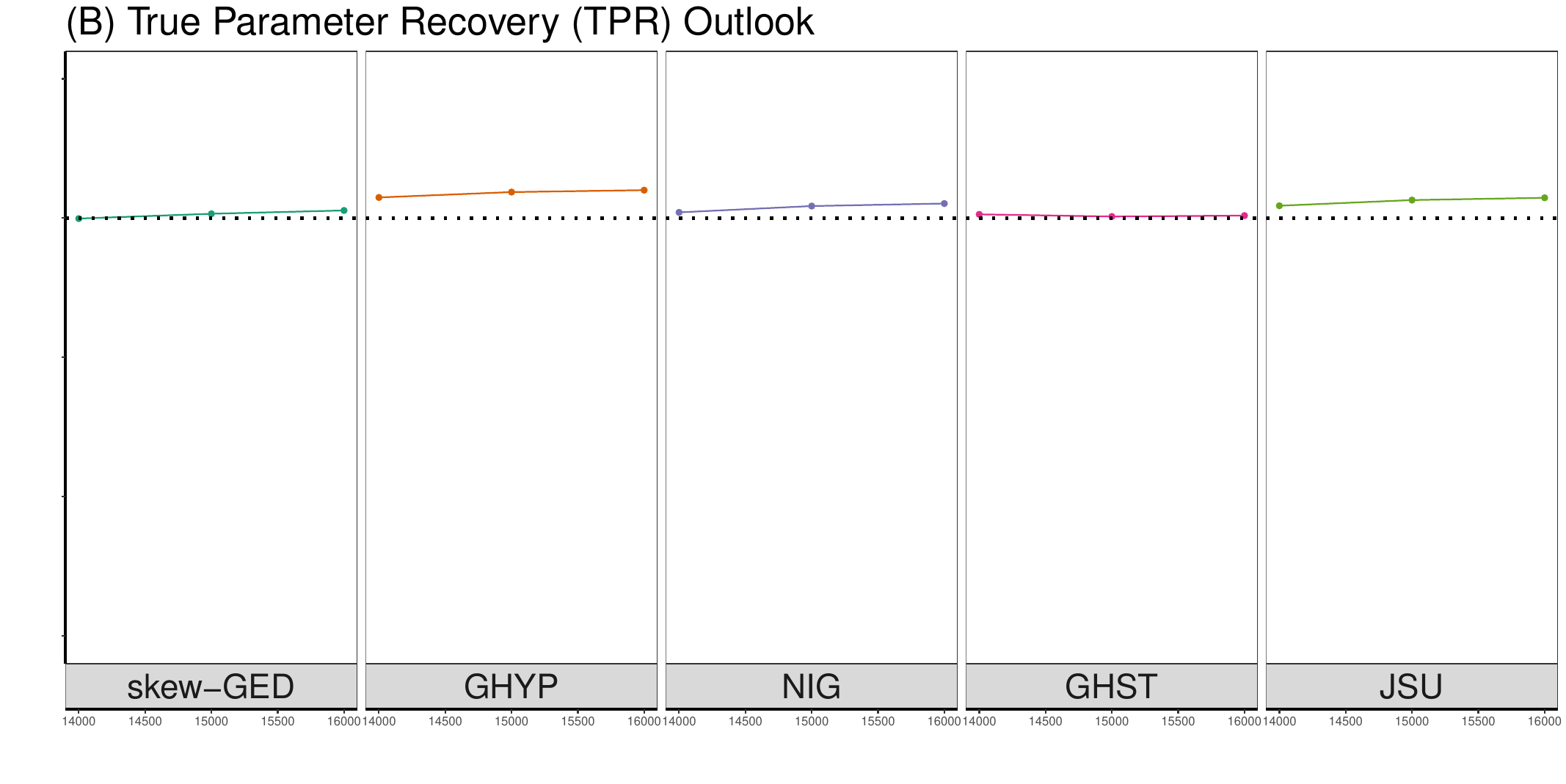}\hfil
\end{adjustwidth}
\caption{The TPR outcome (the coverage) in each assumed innovation for the true model fGARCH(1,1)-Student's $t$, where the dotted line denotes the 0.95 (or 95\%) nominal recovery level. The TPR outcomes for the Normal to the GED innovations are displayed in Panel (A), while those for the skew-GED to JSU are in Panel (B). \label{FirePeace}}
\end{figure}

To summarise, when the true innovation is Student's $t$, the NIG innovation assumption relatively outperforms the rest of the nine assumed innovations in precision and efficiency. It can also be seen from the table and as shown in Figure \ref{FirePeace} that the simulation estimates of the $\hat{\alpha} + \hat{\beta}$ estimator appreciably recover the true parameter value of 0.8999, with TPR outputs close to the 95\% (or 0.95) nominal recovery level under the ten innovations. This implies that the MCS experiments performed considerably well with suitably valid outcomes.

\subsection{Empirical Outcomes}\label{THELORD2}
Here, the result of the simulation modelling is verified empirically by using the actual returns from the S\&P Indian stock index. Among the selected ten innovations, the most suitable for the fGARCH model to describe the market's returns for volatility persistence estimation is investigated. To transform the daily closing price data to the log-returns, we take the log-difference of the value of the index as:
\begin{equation}\label{MSC22eniks}
r_t = \ln\left(\frac{P_{t+1}}{P_{t}}\right) \times 100,
\end{equation}
where $P_{t}$ is the daily closing equity price at time $t$, ln is natural logarithm and r$_{t}$ is the current returns.

\subsubsection{Exploratory Data Analysis}\label{THELORD1}
\begin{figure}[h!]
\begin{adjustwidth}{-\extralength}{0cm}
\centering
\includegraphics[height=3.85in, width=0.9\linewidth]{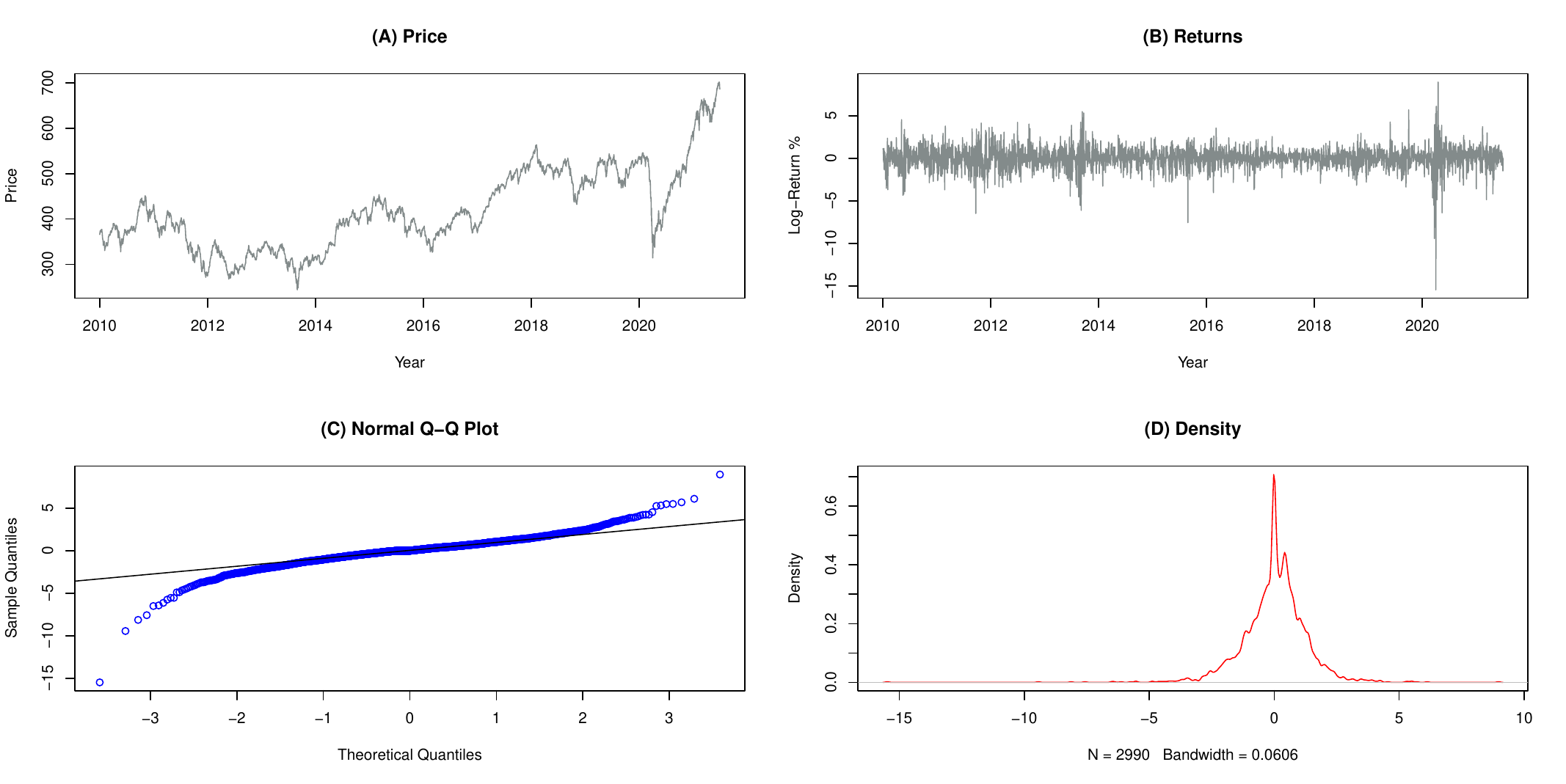}\hfil
    \includegraphics[height=3.85in, width=0.45\linewidth]{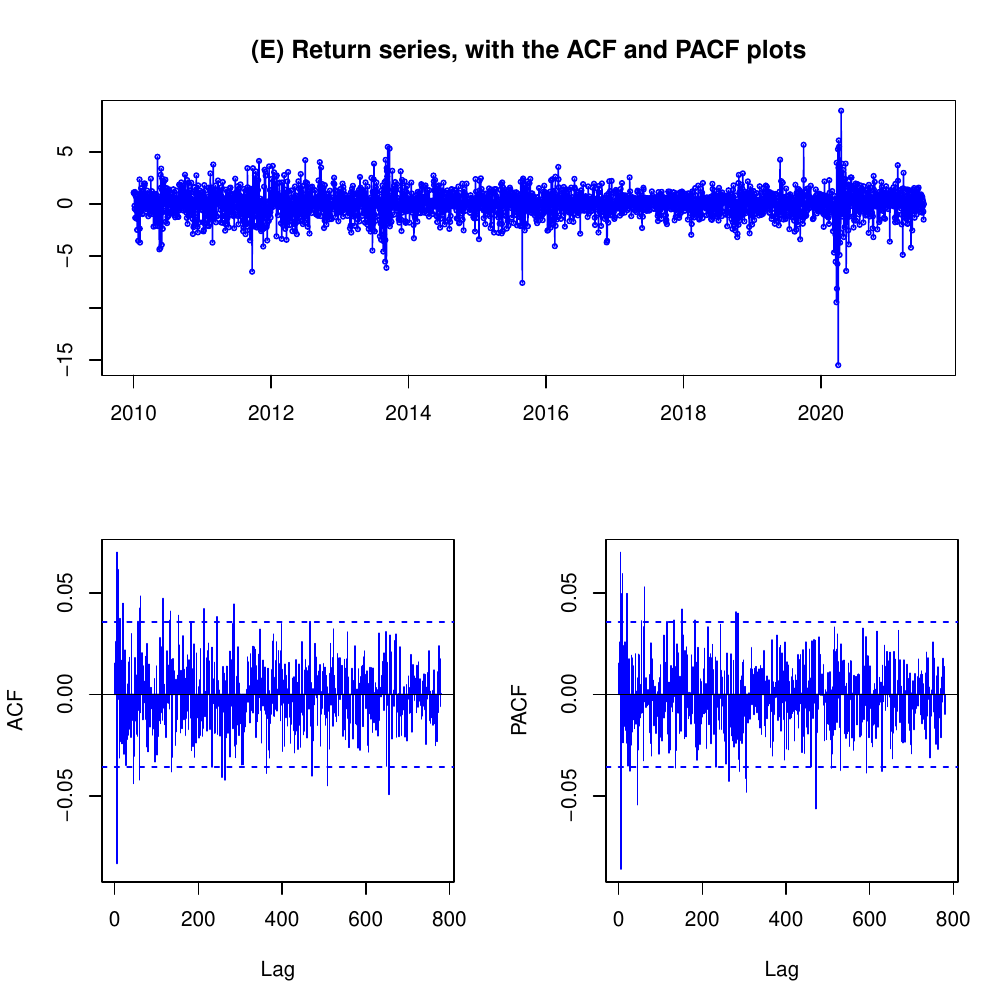}\hfil
    \includegraphics[height=3.85in, width=0.45\linewidth]{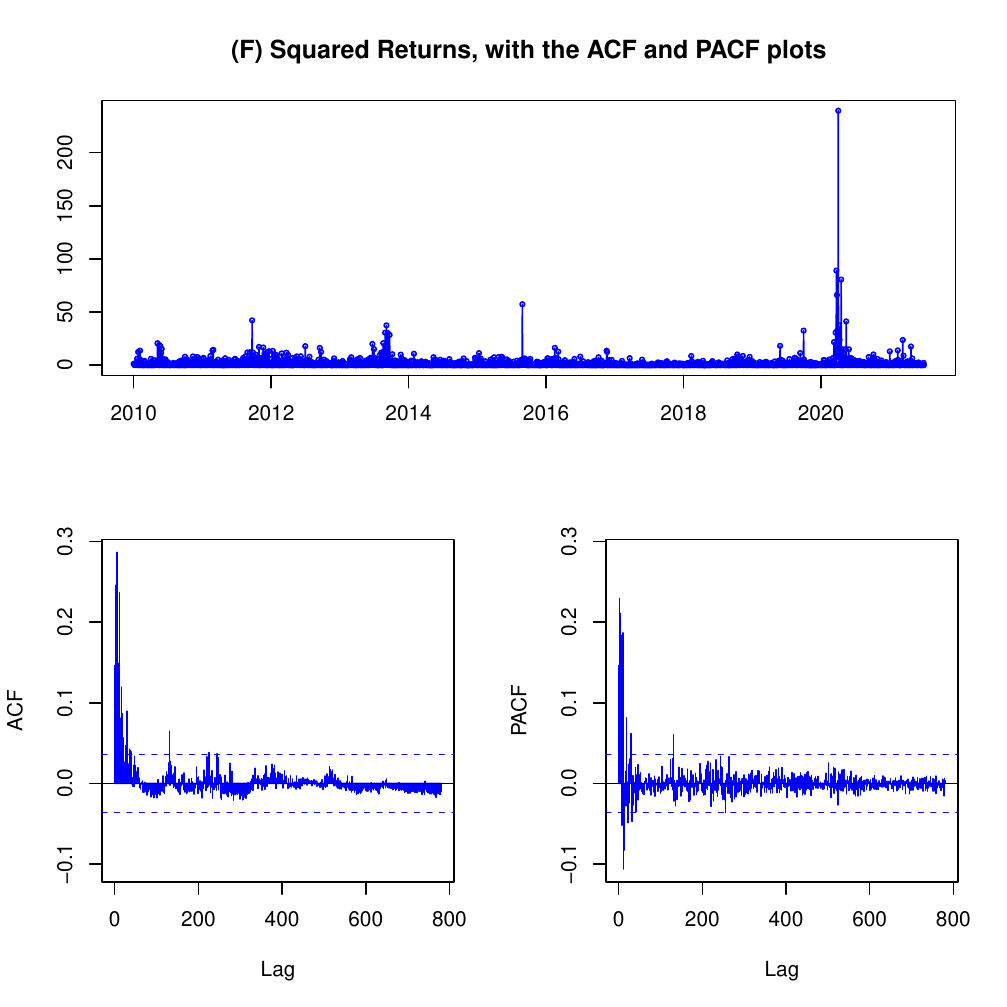}
\end{adjustwidth}
\caption{Panels (A) and (B) respectively show the plots of the price and return series, while Panels (C) and (D) display the quantile-quantile (Q-Q) and density plots, respectively, for the S\&P Indian Index. Panels (E) and (F) display the ACF and PACF plots of the returns and squared returns, respectively.\label{GloryPeace}}
\end{figure}

We begin with the visual inspections of the price level and the index returns over the sample period using exploratory data
analysis (EDA) as revealed in Figure \ref{GloryPeace}. The EDA gives relevant insights into the dataset to disclose vital information such as detecting possible outliers. The EDA shows that the market is characterised by time-varying volatility, with a steep plunge in volatility of price and return in 2020 due to the emergence of the global COVID-19 pandemic crisis. Volatility clustering is apparent in the return series, where large (small) changes tend to follow large (small) changes of either sign. The quantile-quantile (QQ) plot in Panel C shows that the returns are non-Normally distributed, while the density plot in Panel D denotes stability (stationarity) in the return series. The return series plot in Panel B also reveals stationarity in the returns. Panel E displays the correlogram of the returns, showing the Autocorrelation Function (ACF) and Partial Autocorrelation Function (PACF) plots. The plots indicate weak dependence in the mean of the series. Hence, we assume a constant conditional mean (see \cite{EnglePatton2001}). As for the squared returns, the correlogram as displayed in Panel F shows some moderate dependence at the initial lags, hence we fit an ARMA model as presented in the Section \ref{OlutayoDavid} to capture this.

\subsubsection{Descriptive Statistics}\label{MyComfort}
Panel A of Table \ref{DesEngleARCH} presents some descriptive statistics on the return data. The table shows that the mean of the daily return is close to zero. The daily variance is 1.6383, which implies an average annualised volatility of 20.32\% \footnote{Annualized volatility for 252 trading days (i.e., one year) $= \sqrt{252} \times \sqrt{\mbox{variance}} = 20.32\%$ (see \cite{EnglePatton2001}).}. The skewness coefficient shows that the distribution of returns is significantly negatively skewed, a familiar feature of stock returns. This outcome reveals the impact of information arrival in the market, and it shows that investors and other market participants tend to react more to bad news than they do to good news. The kurtosis coefficient is very high. Kurtosis is a measure of the thickness of the tails of the return distribution, and the result shows evidence of leptokurtosis with value greater than three. That is, it surpasses the kurtosis of the Normal distribution, which is three, and that suggests a fat-tailed distribution for describing the return series. Lastly, the outcome of the Shapiro-Wilk test rejects the assumption of normality with a $p$-value = 0. Hence, this implies that the assumption of a Normal distribution for modelling the volatility persistence of the returns is not realistic.

\begin{table}[h!]
\caption{\label{DesEngleARCH}}
\newcolumntype{C}{>{\centering\arraybackslash}X}
\begin{tabularx}{\textwidth}{LC|C|C|C|CC|CC}
 \hline
\multicolumn{3}{c}{$\mathbf{Panel~ A}$} & \multicolumn{1}{|c}{} & \multicolumn{5}{|c}{$\mathbf{Panel~ B}$}   \\
\multicolumn{3}{c}{Returns Descriptive Statistics} & \multicolumn{1}{|c}{} & \multicolumn{5}{|c}{Engle's ARCH Test Outcomes}   \\ \hline
 \multicolumn{2}{c}{} & \multicolumn{1}{c}{} &\multicolumn{1}{|c}{} & \multicolumn{1}{|c|}{} & \multicolumn{2}{|c|}{PQ test} & \multicolumn{2}{c}{LM test}  \\ \hline
   & & & &\textbf{Lag order} & \textbf{PQ} & \textbf{P-value} &  \textbf{LM} & \textbf{P-value} \\ \hline
 Mean& &0.0215 &&4 & 452 & 0 &  3446 & 0   \\
		Variance	& &1.6383	&& 8 & 1043 & 0 &  1130 &  0 \\
		Skewness	& &-0.9525	&& 12 & 1361 & 0 &  641 & 0   \\
		Kurtosis	& &11.6371	&& 16 & 1434 & 0 &  443 & 0   \\
		Shapiro-Wilk & &0.9222  &&       20 & 1487 & 0 &  343 & 0  \\
			$p$-value	& &0.0000 &&24 & 1502 & 0 &  283 &  0  \\ \hline
\end{tabularx}
\noindent{\footnotesize{Note: LM is the Lagrange-Multiplier statistic, while PQ is the Portmanteau-Q statistic in Panel (B).}}
\end{table}

\subsubsection{Tests for Autocorrelation and ARCH effects}\label{OlutayoDavid}
Next, autocorrelation and ARCH effects\footnote{ARCH effect is also known as heteroscedasticity.} are removed by fitting ARMA-fGARCH models with each of the ten innovations to the stationary returns. ARMA(1,1) model as shown in Equation \eqref{PCkxcs} is used as the best, among the candidates ARMA($p$,$q$) models, to capture autocorrelation in the residuals. Table \ref{PaperJARE} shows the results of the Weighted Ljung-Box test \cite{FisherGalla2012} on both the standardised residuals (denoted as WLB SRs) and the standardised squared residuals (WLB SSRs) from the fit of the ARMA(1,1) model. The table shows that for both instances, the $p$-values of the test at lag order 5 are greater than 5\% under the ten innovation distributions. Based on this, the null hypothesis of “no serial correlation” in the returns cannot be rejected. Hence, serial correlation is removed in the residuals.
\begin{equation}\label{PCkxcs}
r_t = \varsigma_0 + \varsigma_1 r_{t-1} + \psi_1 \varepsilon_{t-1} + \varepsilon_{t}
\end{equation}

After removing serial correlation in the returns, we carry out Engle's ARCH test \cite{Engle1982} using
the Portmanteau-Q (PQ) and Lagrange Multiplier (LM) tests to examine the existence of ARCH effects or heteroscedasticity in the residuals. Both tests are carried out based on the null hypothesis of homoscedasticity in the residuals of an ARIMA\footnote{ARIMA is Autoregressive Integrated Moving Average.} model. The results from the two tests reveal highly significant $p$-values of 0 from lag order 4 to 24 as shown in Panel B of Table \ref{DesEngleARCH}. Therefore, we reject the null hypothesis of ``no ARCH effects'' in the residuals, which implies the presence of volatility clustering in the returns. Based on this, we fit candidates fGARCH$(p,q)$ models, with each of the ten distributed errors, to capture the ARCH effects in the returns. Hence, following Engle and Patton \cite{EnglePatton2001}, we use the Bayesian Information Criterion (BIC) \cite{LiCoKiCh09} and found that the best candidate model, to capture heteroscedasticity in the returns, among the fGARCH$(p,q)$ class for $p$ $\in$ [1,5] and $q$ $\in$ [1,2] is the parsimonious fGARCH(1,1) (see Equation \eqref{LondyIyawoMiAtata}).

\begin{equation}\label{LondyIyawoMiAtata}
\sigma^{\gamma}_t = \omega + \alpha_1 \sigma^{\gamma}_{t-1}(|z_{t-1} - \zeta_{21}| - \zeta_{11}\{z_{t-1} - \zeta_{21}\})^{\delta} + \beta_1 \sigma^{\gamma}_{t-1}.
\end{equation}

Following the fit of the fGARCH(1,1) model to the returns, we carry out some checks using the weighted ARCH LM test to ascertain if heteroscedasticity has been removed. It can be seen from Table \ref{PaperJARE} that the $p$-value of the ``ARCH LM (5)'' statistic at lag order 5 exceeds 5\% under each of the ten assumed errors. Hence, this implies that ARCH effect is captured since we cannot reject the null hypothesis of ``no ARCH effects'' in the residuals.

\begin{table}[H]
\caption{The empirical outcomes of ARMA(1,1)-fGARCH(1,1) models.\label{PaperJARE}}
	\begin{adjustwidth}{-\extralength}{0cm}
		\newcolumntype{C}{>{\centering\arraybackslash}X}
          \begin{tabularx}{\fulllength}{LCCCCC}
\toprule
		\textbf{}	& \textbf{(A)}	& \textbf{(B)}	& \textbf{(C)} & \textbf{(D)} & \textbf{(E)}\\
  \textbf{}	& \textbf{Normal}	& \textbf{skew-Normal}	& \textbf{Student's $t$} & \textbf{skew-Student's $t$} & \textbf{GED}\\
		\midrule
$\hat{\varsigma_0}$  & 0.0045 & -0.0073 & 0.0462$^{**}$ & 0.0110 & 0.0389$^{**}$ \\
$\hat{\varsigma_1}$            & 0.5358$^{*}$& 0.5260$^{**}$& 0.3417& 0.3545& 0.2327$^{*}$\\
  $\hat{\psi_1}$            & -0.4668$^{**}$& -0.4612$^{***}$& -0.2724& -0.2872& -0.1767$^{*}$\\
$\hat{\omega}$	 & 0.0482$^{*}$ & 0.0453$^{*}$ & 0.0441$^{**}$ & 0.0447$^{*}$ & 0.0460$^{*}$\\
$\hat{\alpha}$ & 0.0769$^{*}$ & 0.0844$^{*}$ & 0.0871$^{*}$ & 0.0903$^{*}$ & 0.0812$^{*}$ \\
$\hat{\beta}$ & 0.8046$^{*}$ & 0.8168$^{*}$ & 0.8054$^{*}$ & 0.8075$^{*}$ &0.8093$^{*}$ \\
$\hat{\zeta}_{11}$ & -0.1721  & -0.1842$^{***}$  &  -0.1682  &  -0.1893  & -0.1725 \\
$\hat{\zeta}_{21}$  & 1.3867$^{*}$  & 1.3282$^{*}$  &  1.3281$^{*}$  &  1.3301$^{*}$  & 1.3288$^{*}$ \\
$\hat{\gamma}$ = $\hat{\delta}$  &  2.1253$^{*}$ &  1.9995$^{*}$ &  1.9723$^{*}$  &  2.0131$^{*}$  & 2.0521$^{*}$ \\
Persistence $(\hat{P})$ & 0.9726 & 0.9735 & 0.9728 & 0.9760 & 0.9711\\
WLB SRs (5)& 0.7405  & 0.6924 & 0.7014 & 0.8040 & 1.5361\\
$p$-value (5)  & (1.0000)& (1.0000) & (1.0000) & (1.0000) & (0.9973)\\
WLB SSRs (5)& 2.927 & 3.039 & 2.761 & 3.094 & 3.034 \\
$p$-value (5)  & (0.4205)& (0.4000) & (0.4524) & (0.3902) & (0.4009)\\
\multirow[m]{1}{*}{ARCH LM (5)}   & 3.9170 & 3.8687 & 3.4943 & 3.5942 & 3.7311\\
$p$-value (5)  & (0.1816)& (0.1862) & (0.2257) & (0.2144) & (0.1999)\\
AIC   & 3.0541 & 3.0426 & 3.0085 & 3.0006 & 3.0052\\
BIC & 3.0722 & 3.0627 & 3.0286 & 3.0227 & 3.0253\\
SIC & 3.0541 & 3.0426 & 3.0085 & 3.0006 & 3.0052\\
HQIC & 3.0606 & 3.0499 & 3.0157 & 3.0086 & 3.0124\\ \hline
		\textbf{}	& \textbf{(F)}	& \textbf{(G)}	& \textbf{(H)} & \textbf{(I)} & \textbf{(J)}\\
  \textbf{}	& \textbf{skew-GED}	& \textbf{GHYP}	& \textbf{NIG} & \textbf{GHST} & \textbf{JSU}\\
		\midrule
$\hat{\varsigma_0}$ &  0.0064& 0.0108 & $\mathbf{0.0106}$ & -0.0027 & 0.0097\\
$\hat{\varsigma_1}$            & 0.2684$^{**}$& 0.3145& $\mathbf{0.3208}$& 0.3712& 0.3312\\
  $\hat{\psi_1}$            & -0.2017$^{***}$& -0.2501& $\mathbf{-0.2556}$& -0.3052& -0.2651\\
$\hat{\omega}$ & 0.0458$^{*}$ & 0.0447$^{*}$ & 0.0443$^{*}$ & 0.0434$^{*}$ &0.0442$^{*}$ \\
$\hat{\alpha}$ & 0.0831$^{*}$ & 0.0878$^{*}$ & 0.0900$^{*}$ & 0.0963$^{*}$ & 0.0916$^{*}$\\
$\hat{\beta}$ & 0.8116$^{*}$ & 0.8104$^{*}$ & 0.8113$^{*}$ & 0.8187$^{*}$ &0.8112$^{*}$\\
$\hat{\zeta}_{11}$ &  -0.1796 & -0.1910  &  -0.1919  &  -0.1998$^{***}$  & -0.1918 \\
$\hat{\zeta}_{21}$  & 1.3298$^{*}$  &  1.3263$^{*}$ &  1.3191$^{*}$  &  1.2903$^{*}$  & 1.3151$^{*}$ \\
$\hat{\gamma}$ = $\hat{\delta}$ &  2.0496$^{*}$ &  2.0335$^{*}$ &  1.9993$^{*}$  &  1.8975$^{*}$  & 1.9821$^{*}$ \\
Persistence $(\hat{P})$ & 0.9735 & 0.9748 & 0.9749 & 0.9756 & 0.9754\\
WLB SRs (5)        & 1.1550 & 1.0025 & 0.9421 & 0.8057 & 0.8857\\
$p$-value (5)& (0.9999)& (1.0000) & (1.0000) & (1.0000) & (1.0000)\\
WLB SSRs (5)& 3.146 & 3.171 & 3.127 & 3.029 & 3.100 \\
$p$-value (5)  & (0.381)& (0.3767) & (0.3843) & (0.4017) & (0.3891)\\
\multirow[m]{1}{*}{ARCH LM (5)}   & 3.8738 & 3.6558 & 3.6329 & 3.7242 & 3.6114\\
$p$-value (5)  & (0.1857)& (0.2078) & (0.2102) & (0.2006) & (0.2125) \\
AIC   & 3.0008 & 2.9995 & 2.9990 & 3.0030 & 2.9992\\
BIC & 3.0229 & 3.0236 & 3.0210 & 3.0251 & 3.0213\\
SIC & 3.0008 & 2.9994 & 2.9989 & 3.0030 & 2.9992\\
HQIC & 3.0087 & 3.0081 & 3.0069 & 3.0110 & 3.0072\\
\bottomrule
		\end{tabularx}
	\end{adjustwidth}
	\noindent{\footnotesize{Note: WLB SRs (SSRs) denotes the Weighted Ljung-Box test for standardised residuals (standardised squared residuals), where "(5)" is lag order 5. The $p$-value at 5\% level of significance is presented in the round bracket for each error. The "$*$", "$**$" and "$***$" are 1\%, 5\% and 10\% significance levels, respectively. The information criteria are computed as AIC = $-2\textit{L}/N$ + $2p/N$, BIC = $-2\textit{L}/N$ + $p\log_e(N)/N$, HQIC = $-2\textit{L}/N$ + $2p\log_e(\log_e(N))/N$, and SIC = $-2\textit{L}/N$ + $\log_e((N + 2p)/N)$, where $N$ denotes the sample size, $L$ is the log-likelihood of the maximum likelihood of unknown parameter vector $L(\Theta)$, and $p$ denotes the number of estimated parameters\cite{LiCoKiCh09,Ghalanos2018}}.}
\end{table}

\subsubsection{Selection of an Optimal Error Distribution}
Next, selection of the most adequate innovation distribution (among the ten innovation assumptions) that can be used to describe
the returns for estimating the volatility persistence is carried out in Table \ref{PaperJARE}. Models’ comparison and selection are done using four information criteria that consist of the Bayesian Information Criterion (BIC), Akaike Information Criterion (AIC), Shibata Information Criterion (SIC), and Hannan-Quinn Information Criterion (HQIC) (see \cite{Ghalanos2018}). The assumed innovation with the minimum (or lowest) value of information criteria will be the most adequate innovation distribution required to estimate the persistence. To begin with, Table \ref{PaperJARE} shows that all, but one, of the fGARCH volatility parameter estimates ($\hat{\omega}$, $\hat{\alpha}$, $\hat{\beta}$,  $\hat{\gamma}$, $\hat{\zeta}_{11}$, and $\hat{\zeta}_{21}$) under the ten assumed errors are highly significant at 1\% level. To be precise, the only exclusions to the 1\% significance is the $\hat{\omega}$ that is significant at 5\% under the Student's $t$, and the $\hat{\zeta}_{11}$ that is nearly insignificant across the board. These highly significant outcomes in the parameters reveal the presence of volatility clustering in the conditional variance. Moreover, the strongly significant $\hat{\zeta}_{21}$ indicates that the market is driven more by short volatility shocks. In other words, the effects of short volatility shocks are more pronounced in the market than those of large shocks.

For model selection, Table \ref{PaperJARE} shows that the four information criteria have their lowest values under the NIG innovation distribution. Therefore, the NIG innovation (fitted with the fGARCH(1,1) model) is the most suitable to describe the market returns when the underlying error distribution is unknown for volatility modelling. This empirical result is consistent with the MC simulation outcomes. The result is evident to the claim that the NIG distribution is analytically tractable, and it can adequately model the skewness of financial market variables, such as equity prices, exchange rates, and interest rates \cite{AasHaff2006}. Moreover, it has been applied many times for financial applications both as the unconditional return distribution and as the conditional distribution of a GARCH model (see \cite{AasHaff2006}). Hence, this makes the distribution attractive for financial modelling applications.

\subsubsection{Persistence in Volatility and Mean Reversion}\label{STARSUN}
The estimated volatility persistence under this optimal NIG error distribution is  0.9749. This indicates that the volatility of the S\&P Indian equity market's returns is considerably highly persistent. To determine if a model has adequately captured all of the persistence present in the variance of returns, Engle and Patton \cite{EnglePatton2001} suggested that the standardised squared residuals (SSRs, henceforth) should be serially uncorrelated. Hence, from our outcomes in Table \ref{PaperJARE}, the $p$-value of the Weighted Ljung-Box test of the SSRs is greater than 5\% under the NIG innovation (and generally under all the remaining nine assumed innovation distributions)\footnote{These outcomes are observed up to the ninth lags for both SRs and SSRs under the ten error assumptions.}, implying that the SSRs are serially uncorrelated. This indicates that the fGARCH(1,1) model fitted with the NIG assumed innovation has adequately captured all of the persistence present in the variance of returns.

\begin{figure}[]
\begin{adjustwidth}{-\extralength}{0cm}
\centering
\includegraphics[height=4.5in, width=0.98\linewidth]{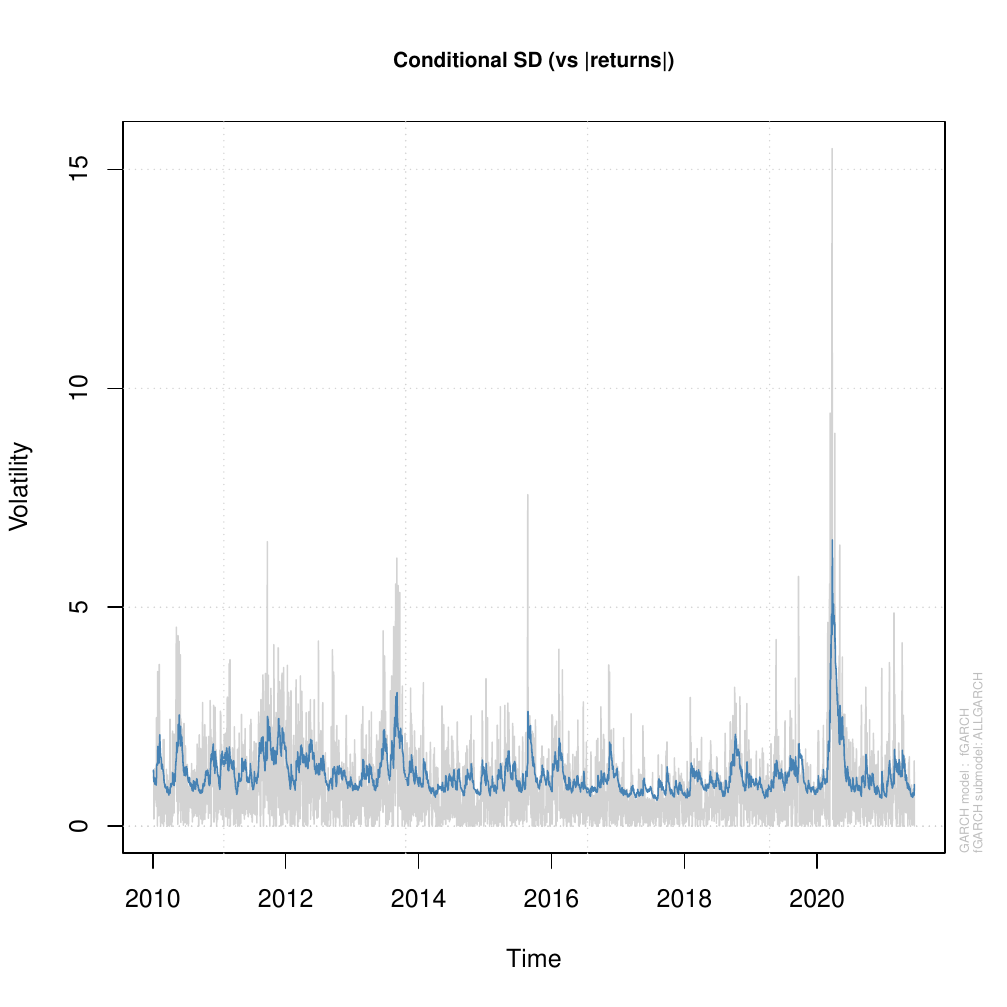}\hfil
\end{adjustwidth}
\caption{Conditional volatility of the fGARCH(1,1) model fitted with the NIG assumed error. \label{GloryPeacea}}
\end{figure}

Broadly speaking, the outcomes of our study show that the mean and variance equations are well specified, with no evidence of correlation in the standardised residuals (SRs) and SSRs of the model. The outcomes of the calculated ARCH LM test also show that ARCH effect is filtered out in the residuals since we cannot reject the null hypothesis of “no ARCH effects” based on the ARCH LM results. Thus, the estimated fGARCH(1,1) model in this study is adequately specified and correctly fitted, and the persistence is also adequately captured by the model. Figure \ref{GloryPeacea} shows the conditional volatility of the fGARCH(1,1) model fitted with the NIG assumed distribution, and it displays alternating phases of lower and higher volatility in the returns.

\begin{figure}[H]
\begin{adjustwidth}{-\extralength}{0cm}
\centering
\includegraphics[height=3.25in, width=0.98\linewidth]{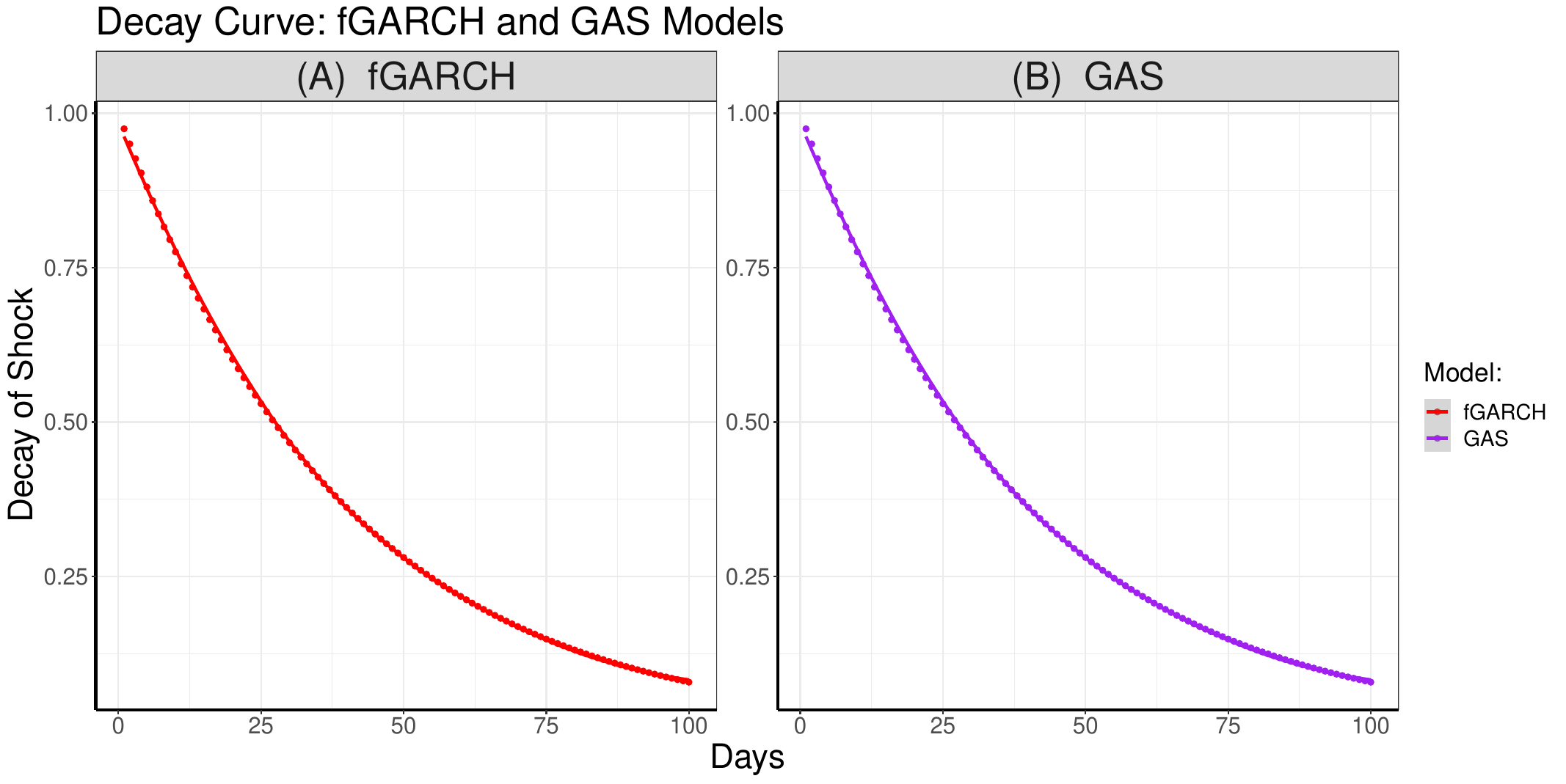}\hfil
   \end{adjustwidth}
\caption{Panels (A) and (B) display the decay of volatility persistence through the fGARCH(1,1)-NIG and GAS-AST1 models, respectively. The two decay curves follow the same trajectory because the persistence estimates from the two models are the same (i.e., 0.9749).\label{DecayGASFGARCH}}
\end{figure}

The obtained outcome of 0.9749 indicates that the market's returns volatility displays considerable persistence, with volatility half-life of about 27 days. This suggests the presence of considerable long memory in the returns volatility, but it is still mean reverting since the outcome is significantly less than one. This implies that even if it takes a while, the volatility process does go back to its mean (see \cite{EnglePatton2001}). With this decay rate of 0.9749, we plot the decay process using the $(0.9749)^{\mbox{day}}$ following the method of Chen and Shen \cite{ChenShen2004}, Chiang et al. \cite{ChiangETAL2009} and Chou \cite{Chou1988} for the first 100 days as shown in Figure \ref{DecayGASFGARCH}. Panel A of the plot displays the decay of shock impacts through the fGARCH(1,1)-NIG model, showing how the trajectory of the decay drops to half the intensity in about 27 days. The limit of the decay sequence is zero \cite{EnglePatton2001} as shown by the declines toward zero. This confirms that the volatility process is mean reverting.

The unconditional variance of the fGARCH(1,1)-NIG model for the S\&P Indian market over the sample period is 1.7659\footnote{The unconditional variance can be estimated using the "uncvariance()" function in R rugarch package \cite{Zivot2013} or directly computed using the unconditional variance formula as stated in Section \ref{thefGARCH777}.}, implying a mean annualised volatility of 21.10\%, which is quite close to the sample estimate of the unconditional volatility presented in Panel A of Table \ref{DesEngleARCH}.

From the outcomes of the ARMA(1,1)-fGARCH(1,1) models in Table \ref{PaperJARE}, the mean and variance equations of the fGARCH(1,1) model when fitted with the assumed NIG innovation can be stated as:

\begin{eqnarray*}
 \ r_t &=& \mu_t + \varepsilon_t \\
 &=& \varsigma_0 + \varsigma_1 r_{t-1} + \psi_1 \varepsilon_{t-1} + \varepsilon_{t} \\
 &=& 0.0106 + 0.3208 r_{t-1} - 0.2556 \varepsilon_{t-1} + \varepsilon_{t} \\
  \sigma^{\gamma}_t &=& \omega + \alpha_1 \sigma^{\gamma}_{t-1}(|z_{t-1} - \zeta_{21}| - \zeta_{11}\{z_{t-1} - \zeta_{21}\})^{\delta} + \beta_1 \sigma^{\gamma}_{t-1} \\
  \sigma^{1.9993}_t &=& 0.0443 + 0.0900\sigma^{1.9993}_{t-1}(|z_{t-1} - 1.3191| + 0.1919\{z_{t-1} - 1.3191\})^{1.9993} + 0.8113 \sigma^{1.9993}_{t-1}
\end{eqnarray*}

\subsection{Monte Carlo Simulation through the GAS Model}\label{PRECIOUSGOD}
In this section, we present the outcomes of the MC simulation study involving the GAS model by following the simulation design steps in Section \ref{LoveGOD}. Hence, we also use this MCS approach to determine the most suitable assumed innovation distribution, among seven selected assumed innovations, that is relevant for volatility persistence estimation through the GAS model. The seven assumed innovations are the Normal, skew-Normal, Student’s $t$, skew-Student’s $t$, asymmetric Student’s $t$ with two tail decay parameters (AST), asymmetric Student’s $t$ with one tail decay parameter (AST1), asymmetric Laplace distribution (ALD). Details on these error distributions can be found in \cite{CrealETAL2011,Crealetal2013a,Ardiaetal2019,Ghalanos2018,ZhuGalb2010,KotzETAL2001}.

The MCS study is carried out by generating simulated return observations from the GAS model as stated in Equation \eqref{JJ3gh/*hdf}, where we set $S_{t}$ = $\textbf{I}$ following Oh and Patton \cite{OhPatton2016}).

\begin{equation}\label{JJ3gh/*hdf}
   \vartheta_{t + 1} \equiv \kappa + \textbf{A} s_{t} + \textbf{B}\vartheta_{t},
\end{equation}

To keep a parsimonious model, the data are generated using the Student’s $t$ GAS model\footnote{Here, the GAS model uses the score of the Student’s $t$ conditional distribution to drive the dynamics of volatility \cite{Blasquesetal2014}.} with a constant shape parameter or degrees of freedom $\nu$ (following Oh and Patton \cite{OhPatton2016}), but a time-varying conditional mean (or location) and scale (or volatility) parameters. Hence, the Student’s $t$ is the true error distribution.

Next, the true model Student’s $t$ GAS is fitted to the actual return data to obtain the true parameter values for the simulation process. The true parameter values used for the data generation are $a_{\mu} = 0.0023$, $b_{\mu} = 0.9919$, $a_{\sigma} = 0.2061$, $b_{\sigma} = 0.9739$, $\mu^{\ast} = 0.0882$, $\sigma^{\ast} = 0.8867$ (and $\nu^{\ast}$ fixed to 4.1), where $a_{\mu}$, $b_{\mu}$ are location parameters and $a_{\sigma}$, $b_{\sigma}$ are the scale parameters, while $\mu^{\ast}$, $\sigma^{\ast}$ and $\nu^{\ast}$ are the unconditional location, scale and shape parameters\footnote{The unconditional expectations are referred to as the target value of the time-varying parameter $\vartheta_{t}$ in Ardia et al \cite{Ardiaetal2019}.}, respectively (see \cite{Ardiaetal2019}). Also, $a_{\mu}$ and $a_{\sigma}$ are the diagonal elements of $\textbf{A}$, while $b_{\mu}$ and $b_{\sigma}$ are the diagonal elements of $\textbf{B}$. The true parameter value $b_{\sigma} = 0.9739$ from the fit is set a priori for the simulation process as shown in Table \ref{GASNiwuraerSTD}. The focus here is on the scalar parameter $b_{\sigma}$, which denotes the volatility persistence (i.e., the persistence of the conditional variance) for the GAS model. Moreover, the persistence parameter $b_{\sigma}$ of the GAS model coincides with the persistence parameters $\alpha + \beta$ of the standard GARCH model\footnote{The GAS model with assumed Normal distribution coincides with the standard GARCH(1,1) model of Bollerslev \cite{Bollers1986} (see \cite{Blasquesetal2014,Crealetal2013a,OhPatton2016,Ardiaetal2019}). Hence, we empirically compared the estimate of the persistence $\hat{b}_{\sigma}$ from the GAS model fitted with a time-varying scale parameter, and the estimate $\hat{\alpha}_1+\hat{\beta}_1$ from GARCH(1,1) model. Both models were fitted to the real return data under the Normal error, and their outcomes yielded $\hat{b}_{\sigma} \equiv \hat{\alpha}_1+\hat{\beta}_1 \approx 0.97$.} of Bollerslev \cite{Blasquesetal2014}.

\begin{table}[H]
\caption{The GAS modelling simulation outcomes.\label{GASNiwuraerSTD}}
\newcolumntype{C}{>{\centering\arraybackslash}X}
\begin{tabularx}{\textwidth}{C|CCCCC}
 \hline
 \multicolumn{1}{l|}{} &  \multicolumn{5}{c}{True model: GAS-Student's $t$}  \\
  \multicolumn{1}{l|}{} &  \multicolumn{5}{c}{True parameter: $b_{\sigma}$ = 0.9739}  \\ \hline
\textbf{Assumed innovation}&\textbf{$N$}& \textbf{$\hat{b}_{\sigma}$}  &\textbf{SE$_{\hat{b}_{\sigma}}$} &\textbf{Bias$_{\hat{b}_{\sigma}}$} & \textbf{TPR$_{\hat{b}_{\sigma}}$ (95\%)}  \\ \hline
  & 23,000& 0.9327  &0.0052 & -0.0412  & 90.98\%  \\
   		  	\multirow[m]{1}{*}{Normal}& 24,000 & 0.9351  & 0.0049 & -0.0388 & 91.22\%  \\
				 & 25,000 & 0.9428  &0.0040 & -0.0311 & 91.97\%  \\ \hline
      & 23,000 &  0.9414   & 0.0045  & -0.0325  &  91.83\% \\
  \multirow[m]{1}{*}{skew}    & 24,000 & 0.9440    &  0.0041 & -0.0298 & 92.09\%  \\
   \multirow[m]{1}{*}{Normal}    & 25,000 &  0.9477  &  0.0037 & -0.0262 &  92.45\% \\	\hline
      & 23,000 & 0.9686    &  0.0032 & -0.0053 &  94.49\% \\
  \multirow[m]{1}{*}{Student's $t$}    & 24,000 & 0.9690   & 0.0031  &  -0.0048 & 94.53\%  \\
      & 25,000 &  0.9695   &  0.0030 &  -0.0043 &  94.58\% \\	\hline
      & 23,000 & 0.9686   & 0.0032  &  -0.0052 &  94.49\% \\
  \multirow[m]{1}{*}{skew}    & 24,000 & 0.9691    &  0.0031 &  -0.0048 &  94.53\% \\
   \multirow[m]{1}{*}{Student's $t$}    & 25,000 &  0.9695  &  0.0030 & -0.0043 &  94.58\% \\	\hline
      & 23,000 &  0.9686   &  0.0032 &  -0.0052 &  94.49\% \\
  \multirow[m]{1}{*}{AST}    & 24,000 & 0.9691   &  0.0031 &  -0.0048 &  94.53\% \\
      & 25,000 &  0.9696  &  0.0030 &  -0.0043 &  94.58\% \\	\hline
      & 23,000 &  0.9686  &  0.0032 &  -0.0052  & 94.49\%  \\
  \multirow[m]{1}{*}{AST1}    & 24,000 &  0.9691  & 0.0031  & -0.0048  &  94.53\% \\
       & 25,000 &  0.9695   & 0.0029  & -0.0044  &  94.57\% \\	\hline
      & 23,000 &  0.9662  & 0.0036  &  -0.0076 & 94.26\%  \\
  \multirow[m]{1}{*}{ALD}    & 24,000 & 0.9672   & 0.0034  &  -0.0066 &  94.35\% \\
      & 25,000 &  0.9680  &  0.0032 &  -0.0058 &  94.43\% \\	\hline
\end{tabularx}
\noindent{\footnotesize{NOTE: $\hat{b}_{\sigma}$ is the diagonal element of $\textbf{B}$ and it represents the persistence of the conditional variance.}}
\end{table}

Using the seed value 12345, with the true parameter values, we generate return datasets of sample size $N$ = 35,000 for the simulation process. However, after trimming down the data to avoid initial values effect, the last $N$ = \{23,000; 24,000; 25,000\} data points are used under each of the seven innovation assumptions as shown in Table \ref{GASNiwuraerSTD}. After generating the simulated returns, we fit the GAS model to each simulated return dataset under the seven assumed innovations. Data generation through the GAS process is generally designed to run only once. In their study through autoregressive process, Samuel et al. \cite{SamuelETAL2023} showed that the outcomes of MCS experiments using the (family) GARCH model are the same for datasets simulated with the same seed value, regardless of whether the data generating process (or simulation) is replicated multiple times or run only once. The authors showed that the estimate from the family GARCH simulation is the average of all the estimates from different replicated series. Hence, the outcome of a single run is the same as the average of multiple runs or replicates.

Next, we use the bias and SE meta-statistics to access the volatility persistence estimator $\hat{b}_{\sigma}$. The optimal (or the most adequate) innovation assumption that is required to estimate the persistence is obtained where the estimator yields the best efficiency and precision from comparisons of the meta-statistics. The comparisons are carried out under the seven selected assumed innovations. The bias and SE outcomes for the estimator are presented in Table \ref{GASNiwuraerSTD}.

We begin with comparing the bias for $\hat{b}_{\sigma}$. The outcomes from the table show that the true innovation Student’s $t$ and its three variants (i.e., the skew-Student’s $t$, AST and AST1) perform equally well with the lowest absolute values of biases than the remaining three assumed innovations as the sample size $N$ progresses from the beginning till the middle at 24,000. However, as $N$ tends to the peak at 25,000, the Student’s $t$, skew-Student’s $t$, and AST equally outperform the remaining four assumed innovations.

\begin{figure}[H]
\begin{adjustwidth}{-\extralength}{0cm}
\centering
\includegraphics[height=2.5in, width=0.90\linewidth]{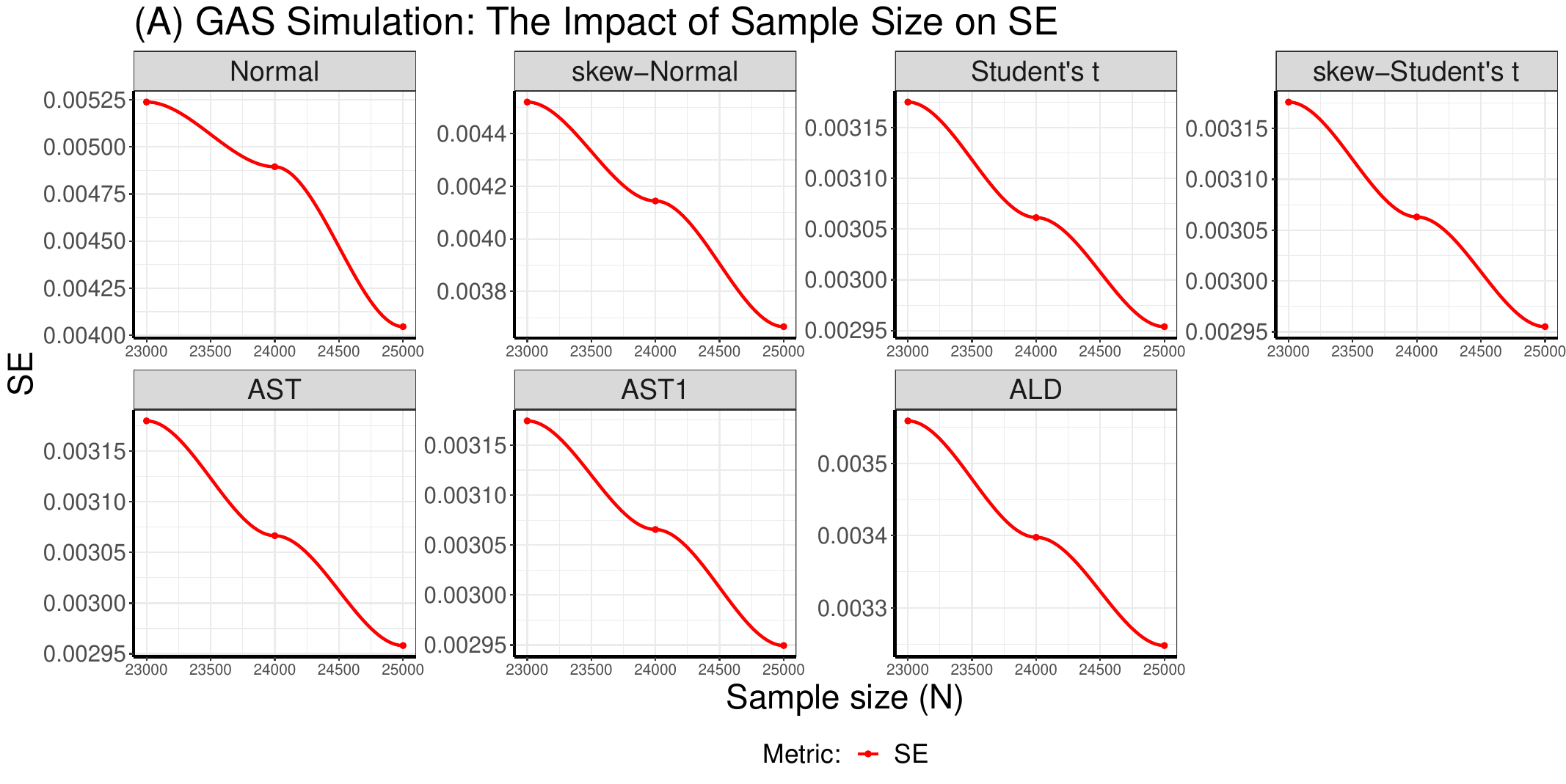}\hfil
  \includegraphics[height=2.5in, width=0.90\linewidth]{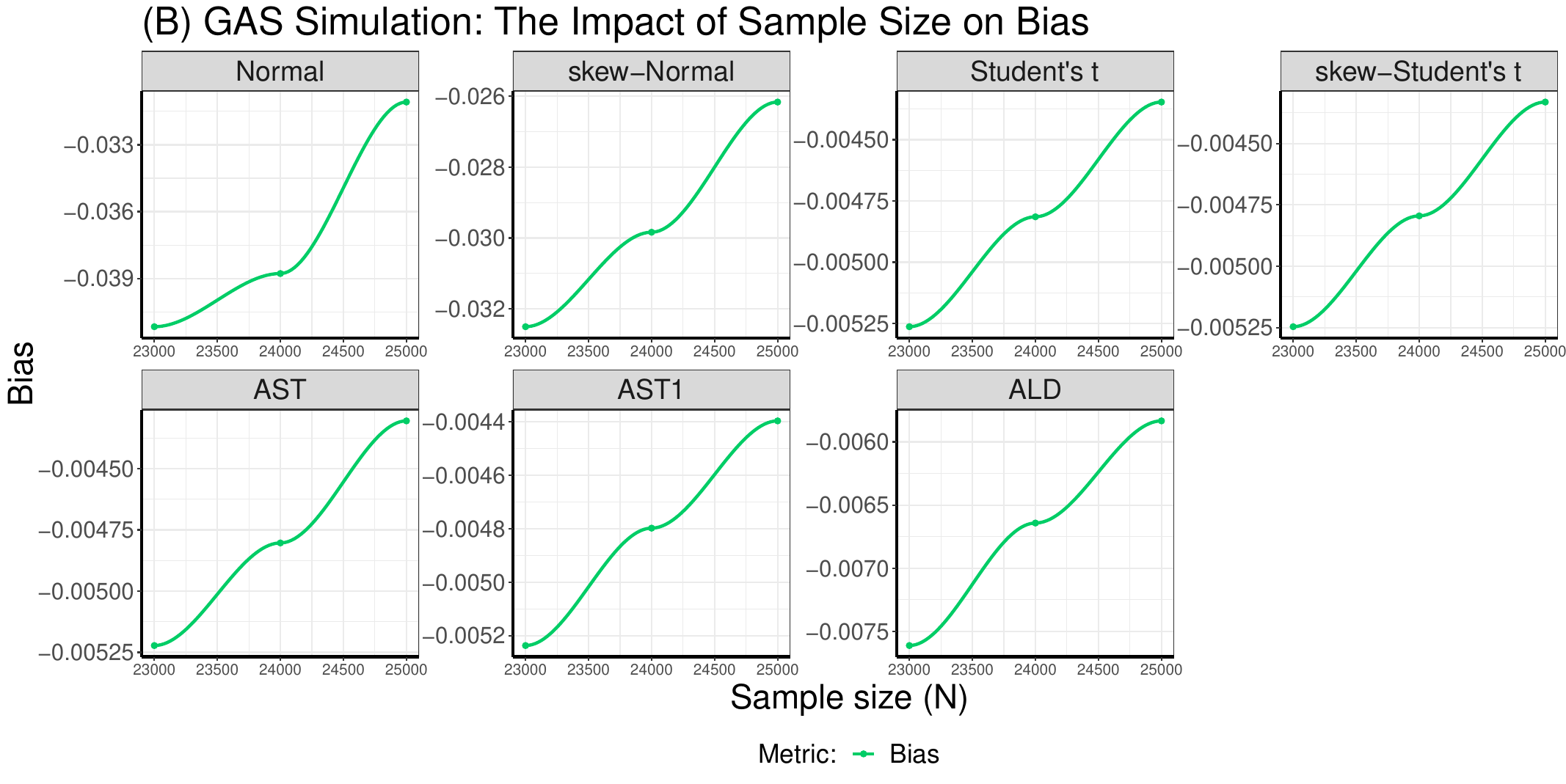}\hfil
  \end{adjustwidth}
\caption{Panels (A) and (B) display the impacts of sample size on SE and bias, respectively, for MCS process involving the GAS-Student's $t$. The SE displays $\sqrt{N}$ consistency. \label{Fire2Brand}}
\end{figure}

For comparisons through the main metric SE, the Student’s $t$ and its three variants also perform equally well in efficiency and precision with the least values than the remaining three assumed innovations from the beginning till the middle, but the AST1 slightly takes the lead by outperforming the rest of the six assumed errors as $N$ tends to the peak. It is observed from the table that the five non-Normal assumed innovations outperform the Normal and skew-Normal in efficiency. It is also observed that the SE under each of the seven assumed innovations decreases as $N$ increases as shown in the table and as displayed in Panel A of Figure \ref{Fire2Brand}. Figure \ref{Fire2Brand} shows that the SE performance measure showcases adequate $\sqrt{N}$ consistency in recovering the true parameter under the seven assumed errors. The bias is not a direct measure of consistency but it displays considerable $\sqrt{N}$ consistency in absolute values under this GAS process simulation as shown in the table. In summary, when the true innovation is Student's $t$, the AST1 assumed innovation distribution relatively outperforms the other six distributed innovation assumptions in efficiency and precision.

\begin{figure}[H]
\centering
\includegraphics[height=2.75in, width=14cm]{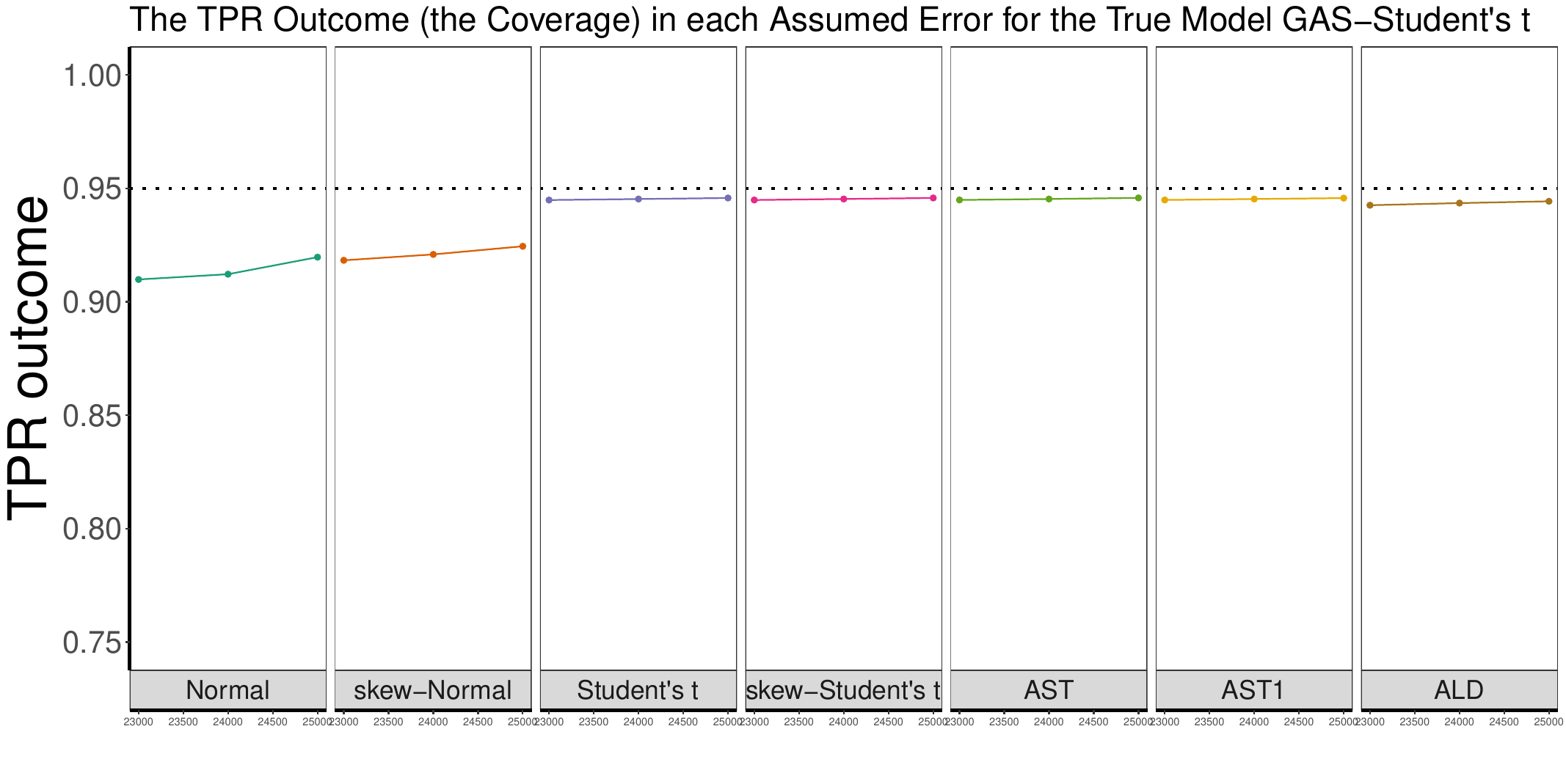}
\caption{The true parameter recovery (TPR) outlook, where the dotted line denotes the 95\% (or 0.95) nominal recovery level.}\label{STDTPRN}
\end{figure}

Furthermore, it is revealed in Table \ref{GASNiwuraerSTD} and visually displayed in Figure \ref{STDTPRN} that the MCS estimates of the $\hat{b}_{\sigma}$ estimator appreciably recover the true parameter value of 0.9739 with TPR outputs near the 95\% (or 0.95) nominal recovery level under each of the seven innovations. This implies that the MCS experiments performed well with considerably valid outcomes. The non-Normal assumed innovations however outperform the Normal and skew-Normal as shown in the plot.

\subsection{Empirical Study}\label{MYFATHERLORD}
Following the MCS experiment, we proceed with empirical verification using the actual returns from the S\&P Indian index. Among the seven innovations, the most adequate for the GAS model to estimate the market's return persistence of volatility is examined. Model selection and comparisons are carried out using two information criteria AIC and BIC as shown in Table \ref{GASStocksIndia}. It is observed from the table that the GAS parameter estimates $\hat{\kappa}_{\sigma}$, $\hat{a}_{\sigma}$ and $\hat{b}_{\sigma}$ are statistically significant at 1\% level under the seven innovation distributions, with the exception of the $\hat{\kappa}_{\sigma}$ that is insignificant under the Student $t$, and is 5\% significant under the skew-Student $t$. The table's results further show that both the AIC and BIC have their least values under the AST1 innovation. Hence, the volatility persistence in this market's returns can be most adequately described through the GAS model fitted with the AST1 assumed innovation. The outcome is consistent with the result of
the MCS investigation.

\begin{table}[H]
\caption{Empirical outcomes of the GAS modelling on real return data.\label{GASStocksIndia}}
\newcolumntype{C}{>{\centering\arraybackslash}X}
\begin{tabularx}{\textwidth}{L|CCCCCCC}
\hline
        & \textbf{Normal} & \textbf{skew-Normal}& \textbf{Student $t$} & \textbf{skew-Student $t$} & \textbf{AST}& \textbf{AST1} & \textbf{ALD} \\  \hline
$\hat{\kappa}_{\sigma}$  & 0.0080* & 0.0039* & -0.0030 & 0.0036** & 0.0227* & 0.0226* & 0.0048** \\
 $\hat{a}_{\sigma}$  & 0.1239*& 0.0305* & 0.2048* & 0.0499* & 0.0509* & 0.0508* & 0.0666* \\
   $\hat{b}_{\sigma}$  & 0.9689* & 0.9696* & 0.9743* & 0.9746* & 0.9747* & 0.9749* & 0.9724* \\
    AIC  & 9275.222 & 9259.497  &  9109.057 & 9103.281  &  9094.407 & 9092.654 & 9110.983 \\
    BIC   & 9299.234 &  9289.512 & 9139.072  & 9139.299  & 9136.429  & 9128.672 & 9140.998\\  \hline
\end{tabularx}
\noindent{\footnotesize{Note: The empirical outcomes of the fit of GAS model. The "*" and "**" denote 1\% and 5\% levels of significance, respectively.}}
\end{table}
The estimated volatility persistence $\hat{b}_{\sigma}$ under this optimal AST1 distributed error is 0.9749. The outcome indicates that the GAS parameter $\hat{b}_{\sigma}$ in $\textbf{B}$ is estimated close to unity with volatility half-life of about 27 days, which implies a considerable persistent dynamic process for $\vartheta_{t}$. This outcome coincides with the persistence outcome obtained from the best model fGARCH(1,1)-NIG in Section \ref{STARSUN}. The high persistence outcome is consistent with the findings of Pandey and Kumar \cite{PandeyKumar2017}, that used GARCH(1,1) model on the Indian S\&P CNX NIFTY 50 for sample period 1997 to 2012, and found high persistence in the volatility process. However, this study used a more robust approach that involved a comparative use of the omnibus fGARCH and GAS models to estimate the persistence of the return volatility. Panel B of Figure \ref{DecayGASFGARCH} shows that the impact of shocks through the GAS-AST1 model decays toward zero mean reversion, where the persistence impact dropped by half in about 27 days. That is, it follows the same trajectory as in fGARCH(1,1)-NIG model in Panel A since both models have the same persistence outcomes.

The outcome further suggests the existence of long-memory in the returns volatility. In addition, the estimated value of the unconditional scale (volatility or variance) under the GAS-AST1 model for the S\&P Indian market over the sample period is 2.4605, implying a mean annualised volatility of 24.90\%. To summarise, this study shows that when the underlying true innovation distribution is unknown, the fGARCH model fitted with the NIG assumed innovation and the GAS model fitted with the AST1 are the most suitable to describe the returns for volatility persistence estimation in the S\&P Indian market.

From the empirical outputs of the GAS-AST1 model in Table \ref{GASStocksIndia}, the equation of the GAS model for time-varying scale (or volatility) parameter can be stated as:

\begin{equation}\label{hj123gh/*hdf}
   \bm{\vartheta}_{t + 1} =  0.0226 + 0.0508 \bm{s}_{t} + 0.9749 \bm{\vartheta}_{t}.
\end{equation}

\subsection{Applications of the Extensions of the Two Models}\label{ETERNALGOD}
These estimates of the persistence from both the fGARCH and GAS models indicate that the returns volatility exhibits considerable long memory since the persistence tends towards (or, is close to) one. Moreover, their mean reversions (from the half-life computations) are somehow slow, which implies that the volatility of returns approaches the average or long-run volatility slowly. Based on this, we carry out further investigations on the long-memory behaviour and possible asymmetry in the process using the extended versions of the (f)GARCH and GAS model, and then compared their outcomes. To be precise, we used the Threshold GARCH (TGARCH) model, which is a sub-model of the fGARCH model, to investigate asymmetry (or leverage effect) in the returns. We further used the Component GARCH (CGARCH) model to investigate the long memory decomposition of volatility. That is, the CGARCH model is used to determine if the persistence of volatility can be decomposed into long-term and short-term processes. Following this, we then applied the dual functions of the Beta-Skew-$t$-EGARCH model to further estimate the leverage and long-memory behaviour of volatility. The Beta-Skew-$t$-EGARCH model is an unrestricted extension of the GAS model. The results of the leverage and long-memory process from these two extended models of the (f)GARCH (i.e., the TGARCH and CGARCH models) and the GAS model (through the one- and two-component Beta-Skew-$t$-EGARCH model) are then compared to determine a superior fit.

The robust fGARCH is a family model that nests the TGARCH model as a sub-model, hence we estimate asymmetry (or leverage) in the demeaned\footnote{The return is demeaned in the TGARCH and CGARCH model specifications by using ARMA(0,0) and by setting the “include.mean” argument as “FALSE” in the ugarchspec function of the R rugarch package. For the sake of comparison, the three models TGARCH, CGARCH and Beta-Skew-$t$-EGARCH are fitted to the demeaned returns.} returns using the TGARCH model fitted with the optimal NIG assumed innovation. We further estimate the long-memory decomposition of the returns volatility using the CGARCH model fitted with the NIG innovation assumption. However, in order to be able to compare the outcomes of the fits of the TGARCH and CGARCH models with those of the one- and two-component Beta-Skew-$t$-EGARCH model, we went further to fit all the four models with the skew Student’s $t$ assumed innovation. This is because the conditional error of the Beta-Skew-$t$-EGARCH model is distributed as a skew Student’s $t$ (see \cite{Sucarrat2013,HarveySuca2014}). Skewness $\eta$ in the four models is applied through the method proposed by \mbox{Fern$\acute{a}$ndez} and Steel \cite{FernSteel1998} (see \cite{Ghalanos2018,Sucarrat2013,HarveySuca2014}). Hence, $\eta < 1$($\eta > 1$) implies left (right) skewness.

\subsection{Application of the TGARCH Model}\label{tGETER}
To examine the presence of asymmetry in the returns volatility, we fit the TGARCH(1,1) model when the innovation is NIG. Following this, we proceed to further fit the model with the skew Student’s $t$ assumed innovation for comparison purposes. The outcomes of the fit are presented in Panel A of Table \ref{TGARCHCGARCH} for both assumed innovations (NIG and skew Student’s $t$). It is observed from the table that asymmetry is evident in the volatility since the asymmetric parameter estimates $\hat{\gamma}$ are non-zero and highly significant when the TGARCH model is fitted with both assumed innovations.  Moreover, the positive leverage coefficient ($\hat{\gamma}$ > 0) under both innovations indicates the existence of leverage effect, which is highly significant at 1\%. This implies that negative shocks or bad news will impact future volatility more than positive shocks or good news of the same size. For instance, under the NIG (skew Student’s $t$), good news has an impact of $\hat{\alpha}$ = 0.0665 ($\hat{\alpha}$ = 0.0671), while bad news has an impact of $\hat{\alpha} + \hat{\gamma}$ = 0.8644 ($\hat{\alpha} + \hat{\gamma}$ = 0.8709). Hence, in both cases, large negative returns are being followed by higher impact of volatility.

\begin{figure}[H]
\begin{adjustwidth}{-\extralength}{0cm}
\centering
\includegraphics[height=3.5in, width=0.45\linewidth]{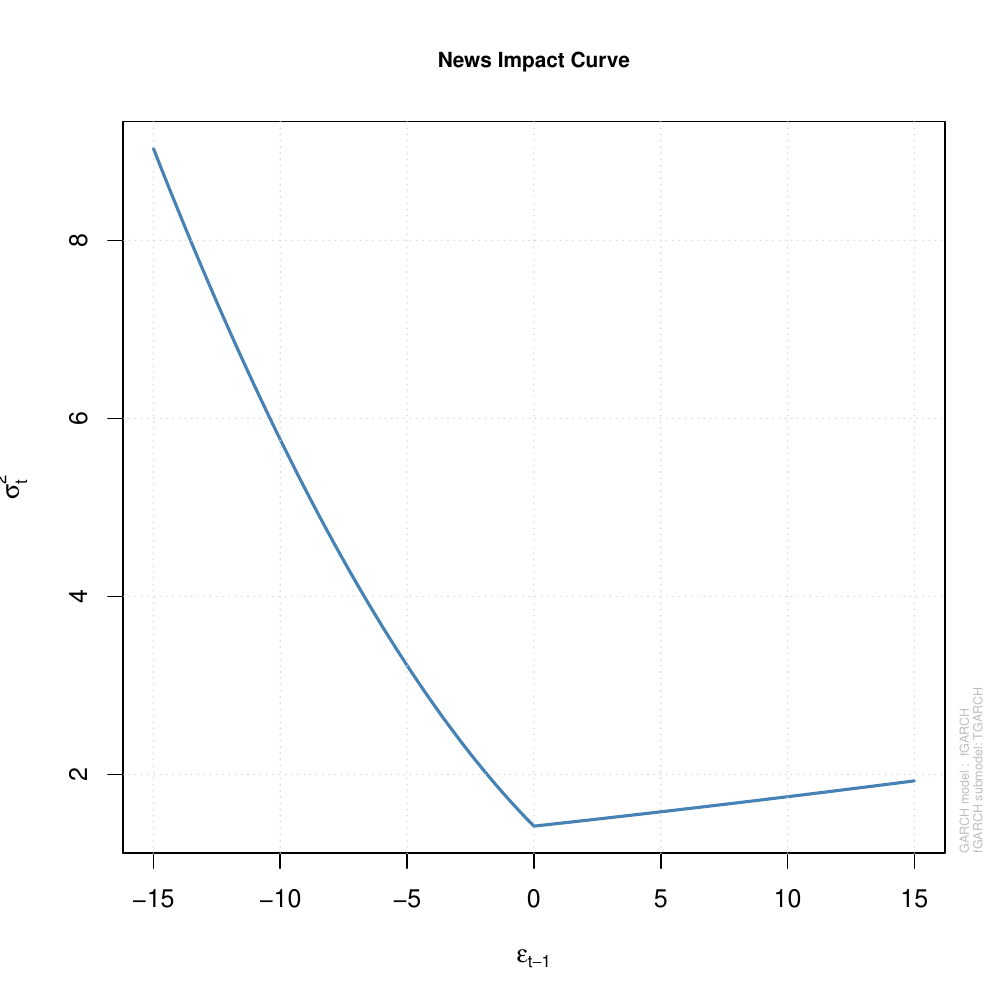}\hfil
    \includegraphics[height=3.5in, width=0.45\linewidth]{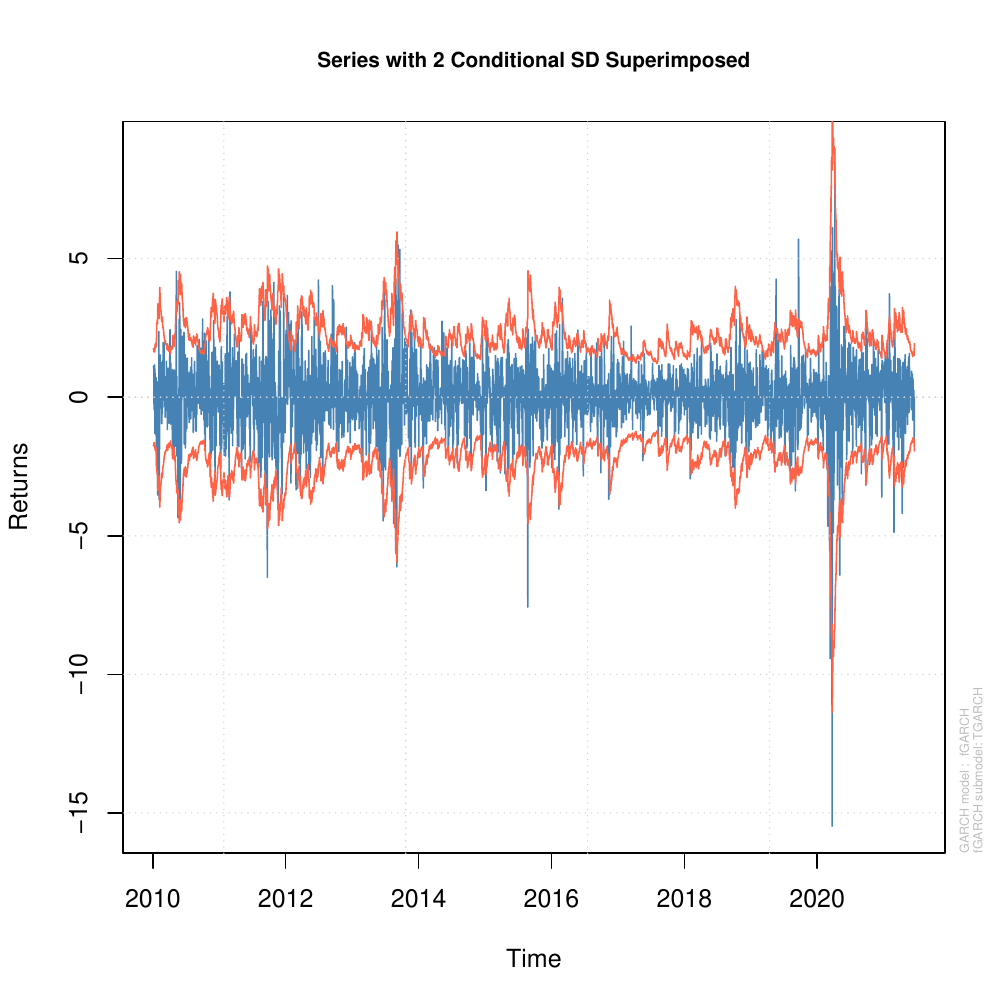}\par\medskip
  \end{adjustwidth}
\caption{The left panel displays the news impact curve, while the right panel showcase the skewness of the returns.\label{CACMercy}}
\end{figure}

\begin{table}[H]
\caption{Outcomes of Estimations Involving the TGARCH and CGARCH Models.\label{TGARCHCGARCH}}
\newcolumntype{C}{>{\centering\arraybackslash}X}
\begin{tabularx}{\textwidth}{L|CC|CC}
 \hline
\multicolumn{1}{c}{} & \multicolumn{2}{|c|}{Panel A} & \multicolumn{2}{c}{Panel B}  \\
\multicolumn{1}{c}{} & \multicolumn{2}{|c|}{ARMA(0,0)-TGARCH(1,1) Model} & \multicolumn{2}{c}{ARMA(0,0)-CGARCH(1,1) Model}  \\ \hline
 \textbf{} & \textbf{NIG} & \textbf{skew-Student $t$} &  \textbf{NIG} & \textbf{skew-Student $t$} \\ \hline
 				 $\hat{\omega}$ & 0.0359*  & 0.0361* &  0.0000* &  0.0001* \\
				 $\hat{\alpha}$ & 0.0665* & 0.0671* & 0.0852*  & 0.0845*  \\
				 $\hat{\beta}$ & 0.9211* & 0.9207* &  0.8799* & 0.8807*  \\
			         $\hat{\gamma}$ & 0.7979* & 0.8038* & $-$  & $-$ \\
                  $\hat{\eta}$  & -0.2094* & 0.8739* &  -0.2364* &  0.8601* \\
                  $\hat{\nu}$   & 1.7036* & 6.1582* &  1.4337* &  5.4048* \\
              BIC   & 3.0211 & 3.0232 &  3.0433 & 3.0462  \\ \hline
$\hat{\phi}$   &  $-$& $-$ &  0.0057* &  0.0056* \\
     $\hat{\rho}$     & $-$ &$-$ & 0.999991*    & 0.999959* \\
              $\hat{\alpha}+\hat{\beta}$   & $-$ & $-$ & 0.9651   &  0.9652\\ \hline
\end{tabularx}
\noindent{\footnotesize{Note: $\hat{\omega}$ estimates the long-run average volatility, estimator $\hat{\rho}$ ($\hat{\alpha} + \hat{\beta}$) is used to measure the persistence of the  permanent (transitory) component, and $\hat{\gamma}$ estimates the leverage or asymmetry. Furthermore, the $\hat{\eta}$ and $\hat{\nu}$ are estimates of the skewness and degrees of freedom, respectively, while $\hat{\phi}$ ($\hat{\alpha}$) estimates the effect of a shock to the permanent (transitory) component.}}
\end{table}
The impact can be visualised as shown in the "News Impact Curve" in the left panel of Figure \ref{CACMercy} (for the skew Student’s $t$ assumption). The curve shows an asymmetric effect with a stronger impact from negative shocks. In the event of asymmetric effects, Mishra \cite{Mishra2010} reported that investors' attentions become more short-term focused. The investors tend to constantly review their investment portfolios for liquidity and performance, even if such investments are purchased with a long-term view. This could have an adverse influence on economic growth and business investment spending because investors tend to move their funds to more liquid and less risky assets as a result of this. The skewness estimates are less than one ($\hat{\eta}$ < 1) under both innovations, indicating negative skewness in the returns volatility as shown in Panel A of Table \ref{TGARCHCGARCH}, and in the right panel of Figure \ref{CACMercy} when the innovation is skew Student’s $t$. Lastly, the degrees of freedom estimate $\hat{\nu}$ in the assumed skew Student’s $t$ innovation is 6.1582, implying a reasonably fat-tailed conditional Student’s $t$ density.

\subsection{Application of the CGARCH Model}\label{cGREAT}
The results in Panel B of Table \ref{TGARCHCGARCH} show that all the parameter estimates ($\hat{\omega}$, $\hat{\rho}$, $\hat{\phi}$, $\hat{\alpha}$ and $\hat{\beta}$) for the CGARCH model are highly significant at 1\% level, thereby suggesting the existence of a permanent-transitory component behaviour in the market. For the permanent component, the significance of the parameter estimates
$\hat{\omega}$ and $\hat{\rho}$ indicates that permanent conditional volatility displays long memory.
The persistence of a shock to the permanent component as measured by the estimate $\hat{\rho}$ is very
high, in excess of 0.99, under the NIG and skew Student's $t$ innovations. This implies that the long-run
volatility $(m_t)$ approaches the mean $(\omega)$ very slowly such that shocks to volatility persist for a long time.

As for the transitory component, the sum of the estimates $\hat{\alpha}$ and $\hat{\beta}$ is used to measure the
persistence of shocks to the volatility. The results under both NIG and skew Student's $t$ in Panel B of Table  \ref{TGARCHCGARCH} show that both parameter estimates are significant at 1\% level, implying that the persistence of shocks to volatility is reasonably high, in excess of 0.9650 under both innovations. However, the effect of shocks to volatility is much more pronounced in the transitory component (with $\hat{\alpha} = 0.0845$) than it is in the permanent component with $\hat{\phi} = 0.0056$.
In summary, the general outcomes indicate that shocks to volatility persist much more in the permanent component than in the transitory component, but the impact of the shocks are more felt in the transitory component than in the permanent component. Due to the very high persistence in the permanent component, we carry out further investigation on the long-memory behaviour of volatility decomposition using the Beta-Skew-$t$-EGARCH model and then compare the outcomes with those of the CGARCH model.

\subsection{Application of the Beta-Skew-$t$-EGARCH Model}\label{BetaSkewT}
\begin{figure}[H]
\begin{adjustwidth}{-\extralength}{0cm}
\centering
\includegraphics[height=3in, width=0.45\linewidth]{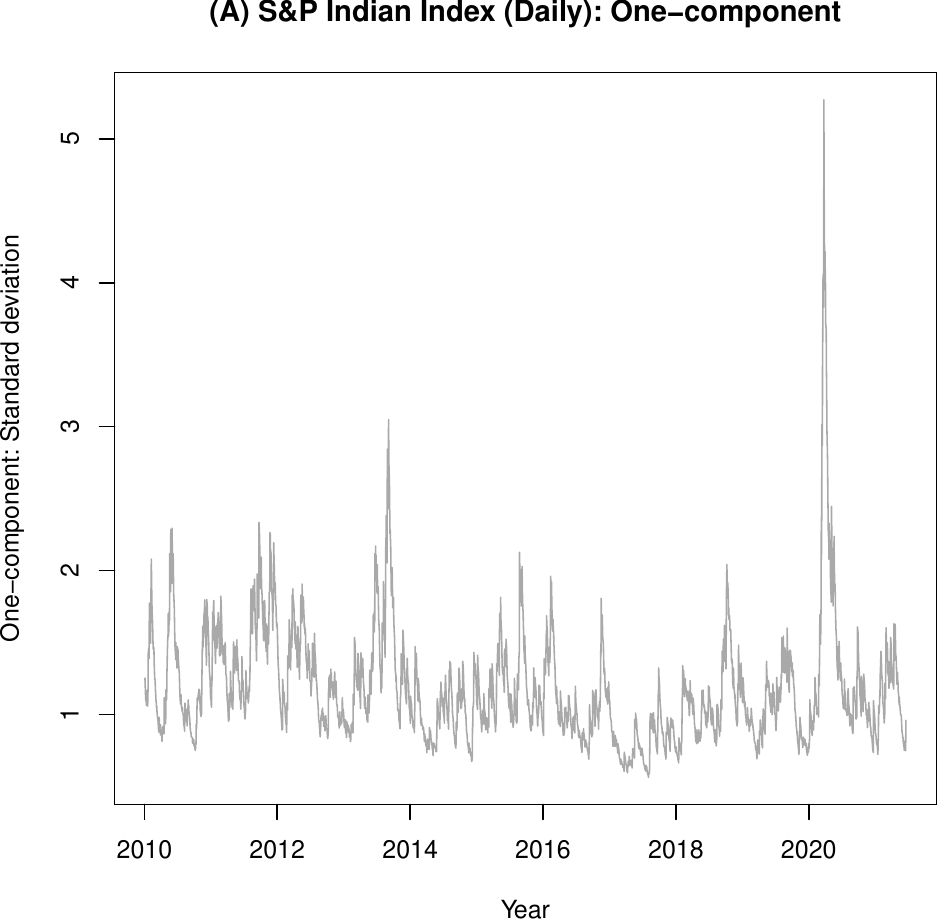}\hfil
    \includegraphics[height=3in, width=0.45\linewidth]{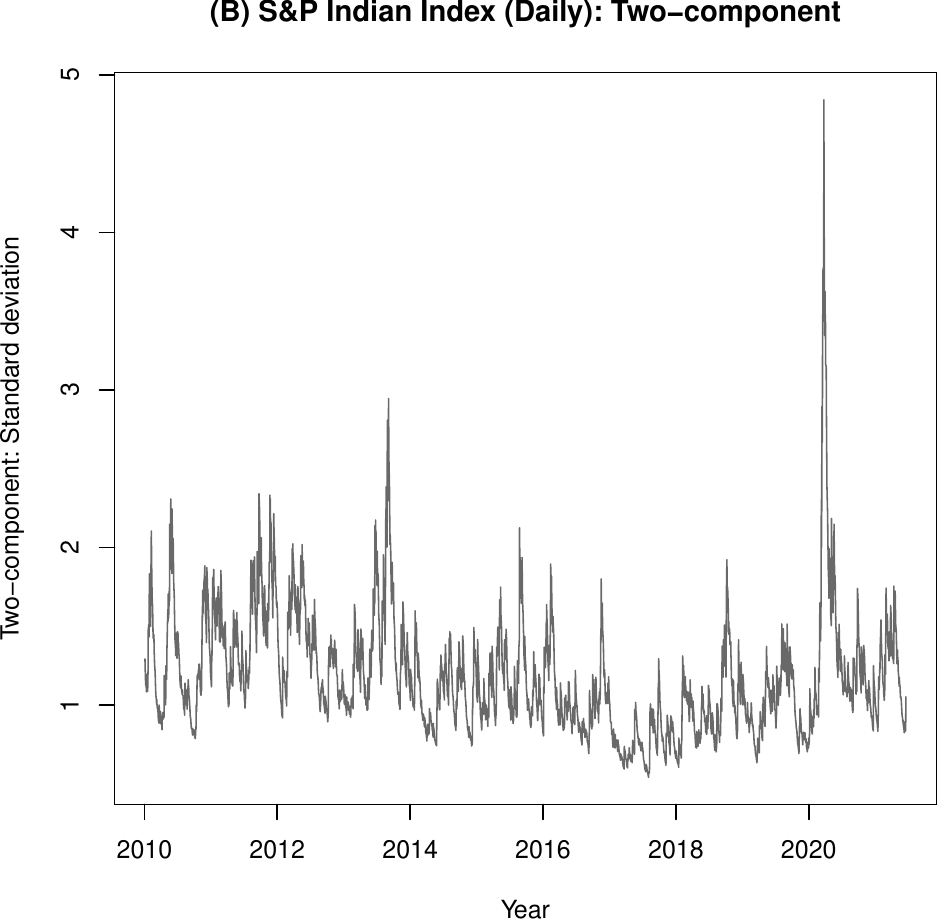}\par\medskip
  \end{adjustwidth}
\caption{Panels (A) and (B) present the fitted conditional standard deviations of the one- and two-component models, respectively.\label{1n2COMPS}}
\end{figure}

We begin with the one-component modelling. Panel A of Figure \ref{1n2COMPS} displays the plot of the fitted conditional standard deviations for the one-component model. The plot shows that the return series is characterised by time-varying volatility with a strong volatility spike caused by the global COVID-19 pandemic crisis in 2020. The outcomes of estimation through the one-component Beta-Skew-$t$-EGARCH model are presented in Panel A of Table \ref{BETESKEWT}. The estimated degrees of freedom $\hat{\nu}$ in the skew Student’s $t$ innovation is 6.2102, which is a reasonably fat-tailed conditional Student’s $t$ density. The skewness estimate $\hat{\eta}$ is about 0.8709, which relates to pronounced negative skewness ($\eta < 1$) in the residuals $z_t$. The leverage effect estimate $\hat{\kappa}^{\ast}$ is positive, which indicates that large negative returns are being followed by higher volatility.

Next is the two-component modelling. Panel B of Figure \ref{1n2COMPS} displays the plot of the fitted conditional standard deviations for the two-component model. Panel A of Table \ref{BETESKEWT} presents the outcomes of the estimations involving the two-component model. From the table, the degrees of freedom estimate in the skew Student’s $t$ innovation is 6.3534, suggesting fat-fails in the conditional Student’s $t$ density. The estimated skewness is about 0.8742, which indicates a pronounced negative skewness. This implies that the risk of a large negative demeaned stock return is greater than that of a large positive demeaned stock return. The persistence of shocks in the long-run component  $\hat{\phi}_{1}$ is very high at 0.9988, with a mean-reversion half-life ($h2l_{{\phi}_{1}}$) of about 577 days (i.e., about a year and seven months). This long-run half-life outcome suggests that the long run effect of the volatility, which was partly caused by the 2020/2021 global COVID-19 pandemic crisis, would persist for as long as about a year and seven months before returning halfway back to the normal state. This indicates that even if the volatility of returns appears to have quite a long memory, it will still mean revert since the persistence estimate is less than one \cite{EnglePatton2001,ChiangETAL2009}. This implies that even though it takes a long time to revert, the volatility process does go back to its mean \cite{EnglePatton2001,ChiangETAL2009}. Due to the persistence caused by the pandemic outbreak, the Indian economy contracted 6.6 percent during fiscal year 2021, but staged a mild recovery in fiscal year 2022 when it grew 8.7 percent \cite{CRISIL2022}.

\begin{figure}[H]
\begin{adjustwidth}{-\extralength}{0cm}
\centering
\includegraphics[height=2.75in, width=0.45\linewidth]{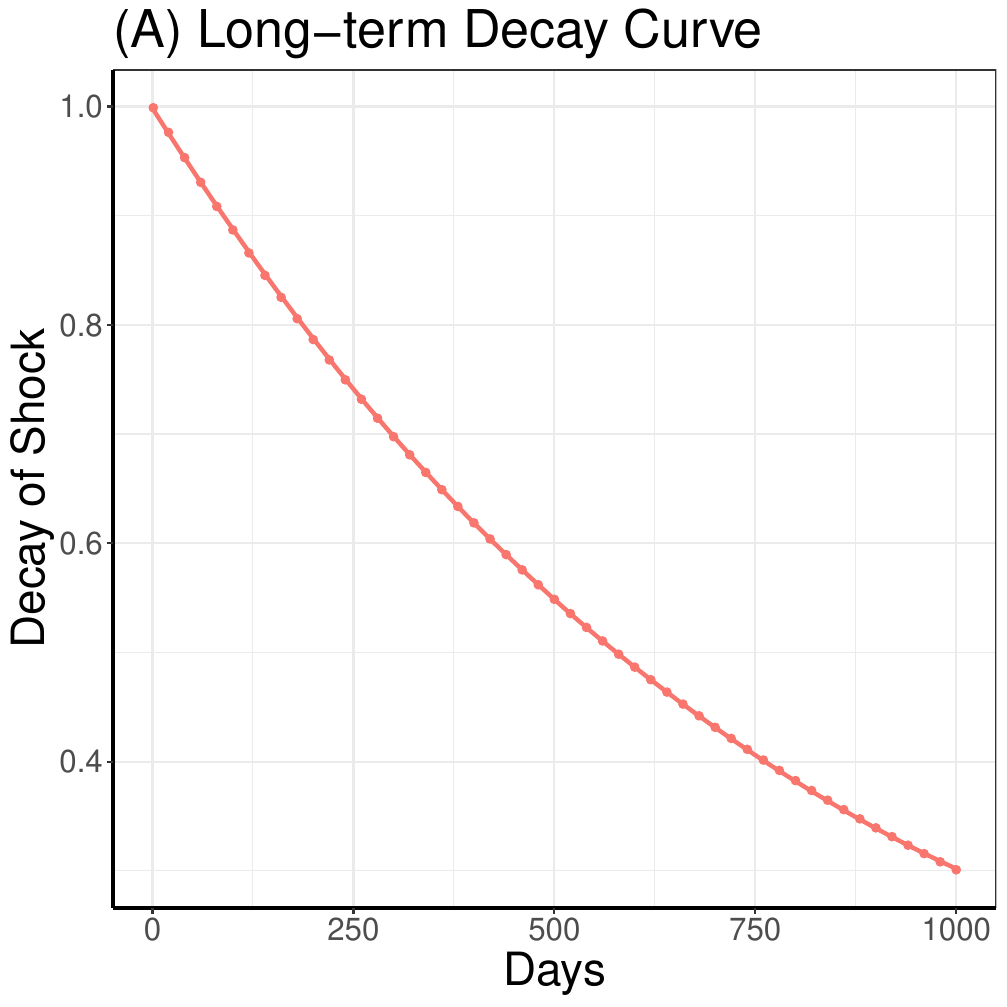}\hfil
    \includegraphics[height=2.75in, width=0.45\linewidth]{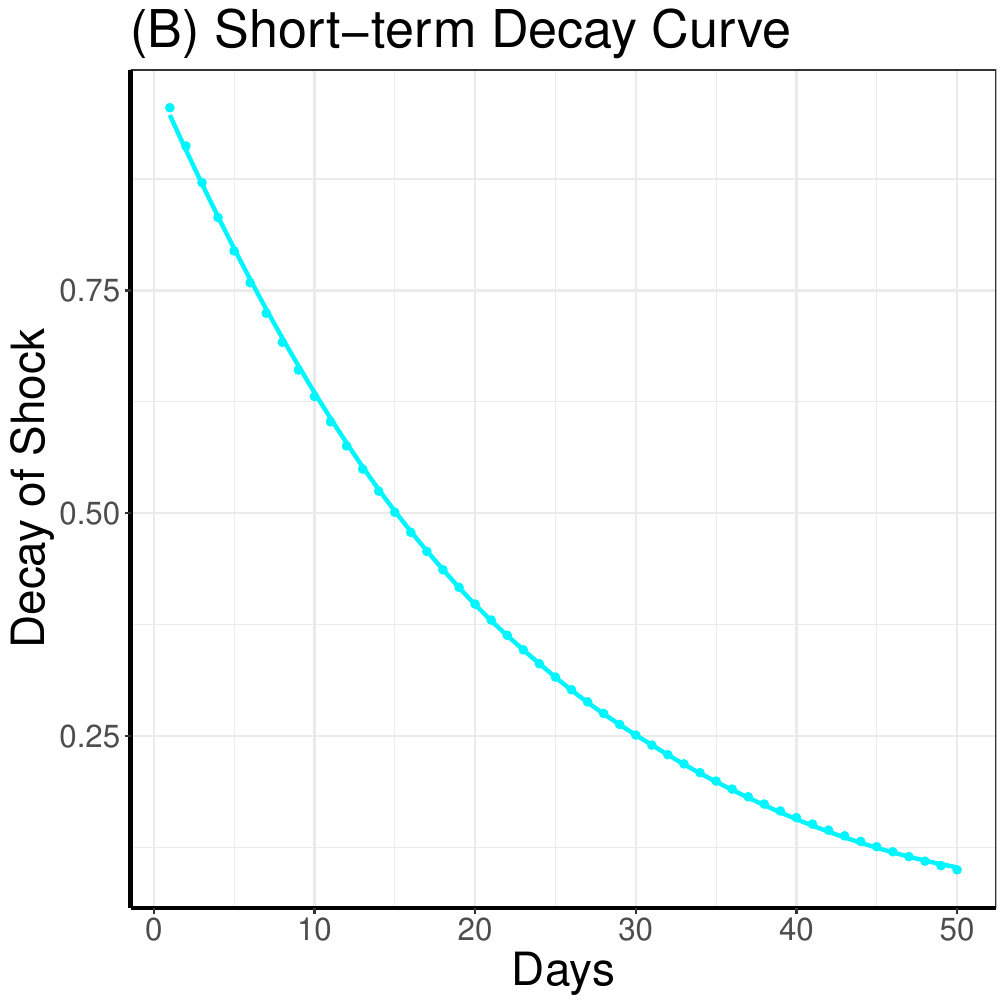}\par\medskip
  \end{adjustwidth}
\caption{Panels (A) and (B) present the long-term and short-term decay curves, respectively.\label{LSDecayCurves}}
\end{figure}

\begin{table}[H]
\caption{Empirical and Simulation Outcomes of Beta-Skew-$t$-EGARCH Model.\label{BETESKEWT}}
\newcolumntype{C}{>{\centering\arraybackslash}X}
\begin{tabularx}{\textwidth}{L|CC|CC}
 \hline
 \multicolumn{1}{c}{} & \multicolumn{2}{|c|}{\textbf{Panel A}} & \multicolumn{2}{c}{\textbf{Panel B}}  \\ \hline
\multicolumn{1}{c}{} & \multicolumn{2}{|c|}{Empirical Outcomes} & \multicolumn{2}{c}{Simulation Outcomes}  \\
\multicolumn{1}{c}{} & \multicolumn{2}{|c|}{Beta-Skew-$t$-EGARCH Model} & \multicolumn{2}{c}{Beta-Skew-$t$-EGARCH Model}  \\ \hline
 \textbf{} & \textbf{One-Component} & \textbf{Two-Component} &  \textbf{One-Component} & \textbf{Two-Component} \\ \hline
 				 $\hat{\omega}$ [se] & 0.0142 [0.0482]  & 0.0529 [0.1293] &  0.0212 [0.0430] &  0.0461 [0.1106] \\
				 $\hat{\phi}_{1}$ [se] & 0.9721 [0.0056]& 0.9988 [0.0015] & 0.9771 [0.0034]  & 0.9979 [0.0012]  \\
                 $h2l_{{\phi}_{1}}$(Days) &  & 577 &   &  \\
				 $\hat{\phi}_{2}$ [se] & $-$ & 0.9550 [0.0092] &  $-$ &  0.9529 [0.0078] \\
                      $h2l_{{\phi}_{2}}$(Days)  &  & 15 &   &  \\
			         $\hat{\kappa}_1$ [se] & 0.0402 [0.0054] & 0.0076 [0.0030] & 0.0413 [0.0041]  & 0.0151 [0.0038] \\
                  $\hat{\kappa}_2$ [se]  & $-$ &  0.0301 [0.0061] &  $-$ &  0.0238 [0.0062] \\
                  $\hat{\kappa}^{\ast}$ [se]   & 0.0337 [0.0039] & 0.0379 [0.0044] &  0.0324 [0.0030] &  0.0414 [0.0035] \\
              $\hat{\eta}$ [se]   & 0.8709 [0.0207] & 0.8742 [0.0208] &  0.8611 [0.0163] &  0.8613 [0.0162] \\
      $\hat{\nu}$ [se]     & 6.2102 [0.6855] & 6.3534 [0.7171]& 6.0746 [0.4639]    & 6.3577 [0.5032] \\
      BIC   & 3.0255 & 3.0266 &  &   \\ \hline
\end{tabularx}
\noindent{\footnotesize{Note: $\hat{\omega}$ estimates the long-term log-volatility, $\hat{\phi}_1$ and $\hat{\phi}_2$ are estimators for the long-term and short-term persistence parameters, respectively, and $\hat{\kappa}_1$ ($\hat{\kappa}_2$) estimates the long-run (short-run) response to shocks. $h2l_{{\phi}_{1}}$ and $h2l_{{\phi}_{2}}$ are measures of the half-life for $\hat{\phi}_1$ and $\hat{\phi}_2$, respectively. $\hat{\kappa}^{\ast}$ estimates the leverage parameter, while $\hat{\eta}$ and $\hat{\nu}$ are estimates of the skewness and degrees of freedom, respectively. [se] in square bracket is the standard error of the estimated parameter.}}
\end{table}
The persistence of shocks in the short-run component $\hat{\phi}_{2}$ is 0.9550, with a half-life ($h2l_{{\phi}_{2}}$) outcome of about 15 days. Hence, the short run component decays much faster than the long run component. In other words, it can be seen from the persistence outcomes that $\hat{\phi}_{2} < \hat{\phi}_{1} < 1$ and from the half-life results that $h2l_{{\phi}_{2}} < h2l_{{\phi}_{1}}$, hence the short-run component decays more quickly than the long-run component that dominates the volatility persistence process. Panel A (Panel B) of Figure \ref{LSDecayCurves} shows that the decay of persistence impact in the long-term (short-term) reached half in about 577 days (15 days) and it continues towards zero mean reversion.

The parameter estimate $\hat{\kappa}_1$ that indicates the long-run response of volatility to shocks is 0.0076, while the estimate $\hat{\kappa}_2$ for the short-run response to shocks is 0.0301. These estimates show a significant discrepancy, where the outcomes reveal that the unexpected arrival of news influences the short-run component considerably more than the long-run component. In other words, the response to the effect of shocks in the short-run is higher than in the long-run volatility. More precisely, the long-run component displays smaller proportional effects of about 20 percent response to volatility shocks as compared to 80 percent from the short-run component\footnote{The total displayed response to shocks via $\hat{\kappa}_1$ and $\hat{\kappa}_2$ from the model as shown in Table \ref{BETESKEWT} is 0.0377 (i.e., 0.0076 + 0.0301). Hence, there are 20.16\% $\approx$ 20\% and 79.84\% $\approx$ 80\% responses to volatility shocks by the long-term and short-term components, respectively.}. However, even though the short-run component has a stronger shock effect, it is short lived (see \cite{ChiangETAL2009,ChenShen2004,EngleLee1999} for related outcomes).

To summarise, this study finds the existence of both short-term and long-term volatility in the persistence process, where the response to the effect of shocks in the short-run is much higher than in the long-run volatility. This infers that higher volatility in the process is mostly due to the short-run volatility increase. However, the short run volatility fluctuation is brief, while the long run mean-reversion of volatility persistence will dominate thereafter. Precisely, the short run effect is big but short-lived. Although the long-run component displays a pronounced longer persistence into the future, its response to volatility shocks is much lower than that of the transient short-run component. This means that investment and other market risks in the long-term seem to be considerably under control in the market.

\subsection{Simulation Study}\label{BetatSimStd}
Following the empirical outcomes, we run a set of Monte Carlo experiments using the true parameter outcomes from the Beta-Skew-$t$-EGARCH specification to further ascertain the validity of both the one- and two-component Beta-Skew-$t$-EGARCH model’s results.

For the one-component simulation, we used $N$ = 5000 simulated returns, generated from the true parameters $\omega = 0.0142$, $\phi_{1} = 0.9721$, $\kappa_1 = 0.0402$, $\kappa^{\ast} = 0.0337$, $\eta = 0.8709$, and $\nu = 6.2102$. These true parameter values are empirical outcomes (i.e., MLE estimates) from fitting the first order one-component Beta-Skew-$t$-EGARCH model to the real Indian return data. We used seed value 12345 for the simulation. Next, we fit the one-component Beta-Skew-$t$-EGARCH model to the simulated dataset and obtained the outcomes as presented in Panel B of Table \ref{BETESKEWT} under the "One-Component" model. From the table, the estimated degrees of freedom $\hat{\nu}$ in the skew Student’s $t$ is 6.0746, the leverage parameter estimate $\hat{\kappa}^{\ast}$ is 0.0324, while the skewness estimate $\hat{\eta}$ is about 0.8611. These outcomes are quite close to the empirical results in Panel A of the table.

For the two-component simulation, we follow the same steps by using $N$ = 5000 simulated returns, generated from the true parameters $\omega = 0.0529$, $\phi_{1} = 0.9988$, $\phi_{2} = 0.9550$, $\kappa_1 = 0.0076$, $\kappa_2 = 0.0301$, $\kappa^{\ast} = 0.0379$, $\eta = 0.8742$, and $\nu = 6.3534$. We used seed value 12345 for the simulation. The true parameter values are obtained from fitting the first order two-component Beta-Skew-$t$-EGARCH model to the real Indian return data. Next, we fit the two-component Beta-Skew-$t$-EGARCH model to the simulated dataset and obtained the outcomes as presented in Panel B of Table \ref{BETESKEWT} under the "Two-Component" model. The estimates of the long-run $\hat{\phi}_{1}$ and short-run $\hat{\phi}_{2}$ are 0.9979 and 0.9529, respectively, and they are quite close and consistent with the empirical outcomes in Panel A of the table.

\begin{figure}[H]
\begin{adjustwidth}{-\extralength}{0cm}
\centering
    \includegraphics[height=2.85in, width=0.45\linewidth]{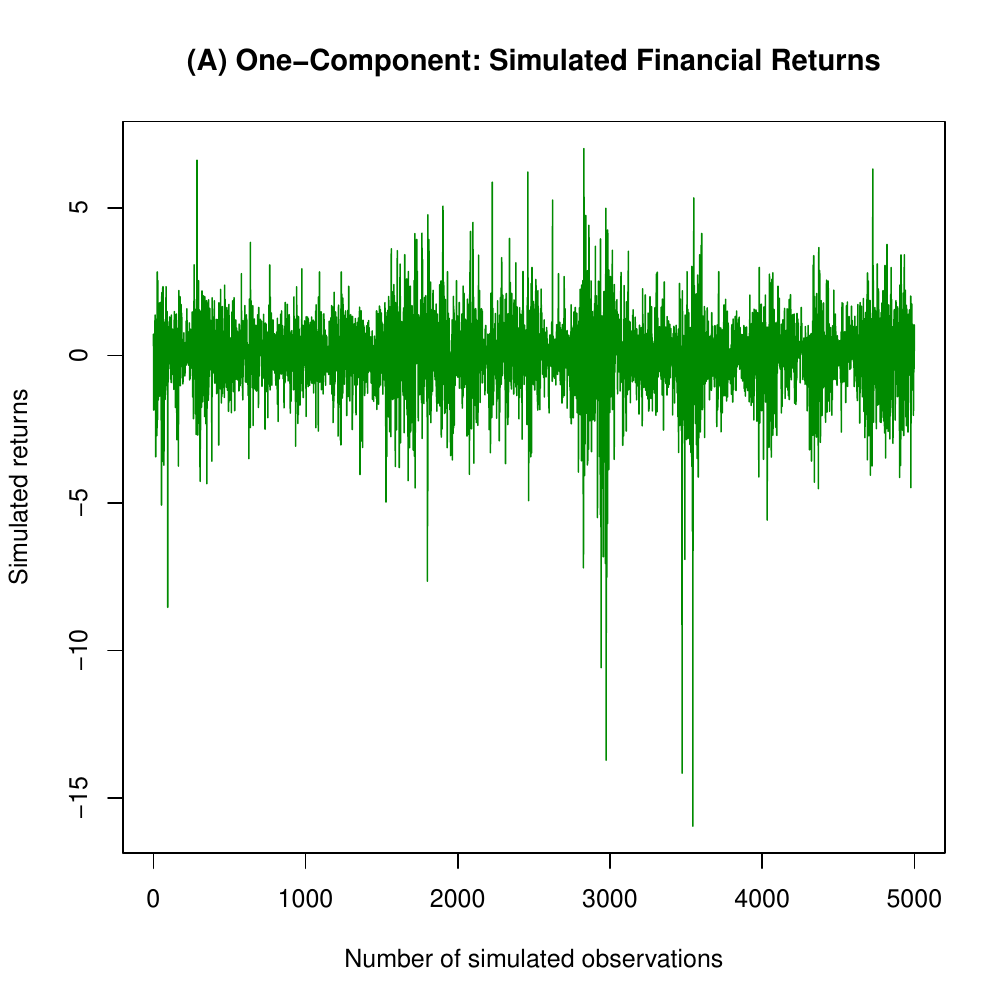}\hfil
    \includegraphics[height=2.85in, width=0.45\linewidth]{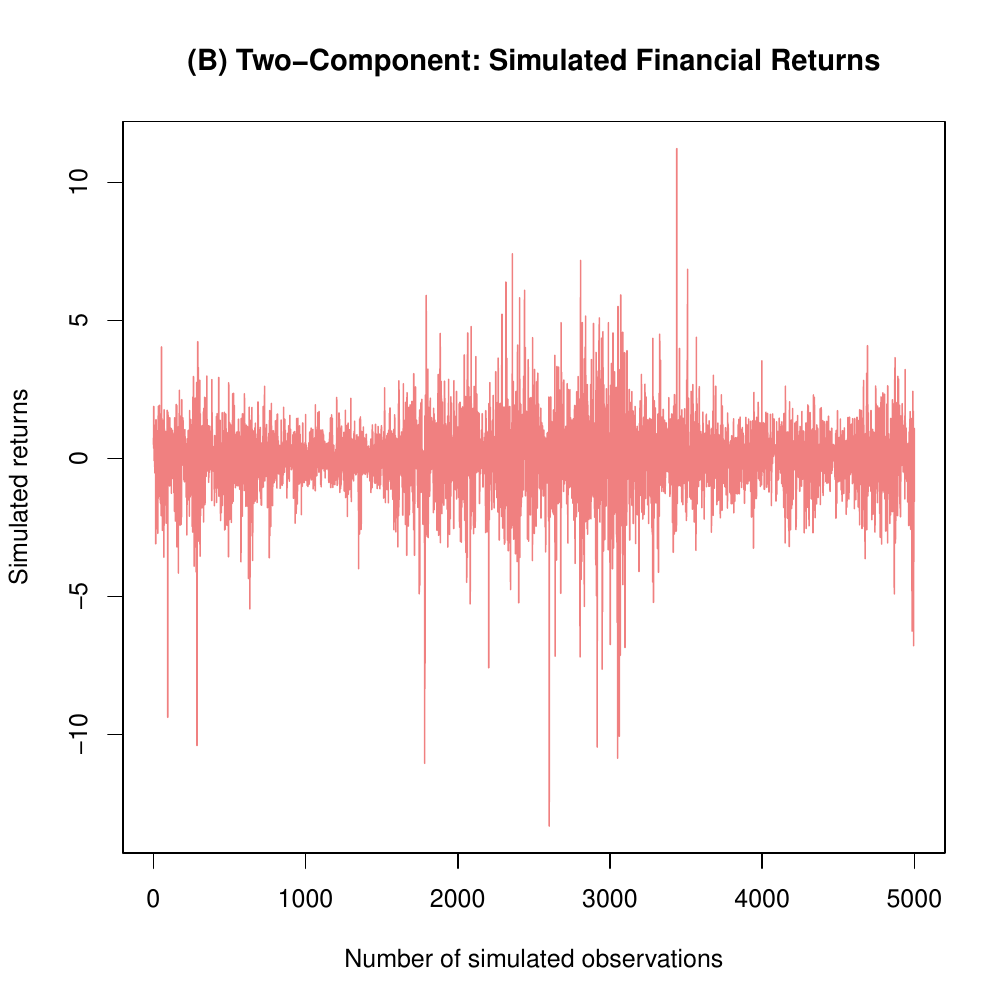}
\end{adjustwidth}
\caption{Panels (A) and (B) display the plots for one-component simulated returns and two-component simulated returns, respectively.\label{FireAlpha}}
\end{figure}

Moreover, given a 95\% nominal recovery level, the TPR outcome for the leverage parameter estimate $\hat{\kappa}^{\ast}$ in the one-component model is 91.34\%, while the TPR outcome for the long-run (short-run) estimate is 94.91\% (94.79\%) in the two-component model. These outcomes indicate a good performance of the simulation experiments. Hence, leverage effect is evident, with long-memory decomposition into the long-run and short run components. The response to volatility shocks is more pronounced in the short-lived short-run component than it is for long-run component. Panels A and B of Figure \ref{FireAlpha} display the plots for one- and two-component simulated returns, respectively. Both plots show that the simulation return series is characterised
by time-varying volatility.
\subsection{Model Comparison}\label{ModComp}
Next, we compare the results from the one- and two-component Beta-Skew-$t$-EGARCH model with those of the TGARCH and CGARCH models where the innovations are skew Student’s $t$ in Table \ref{TGARCHCGARCH}. The degrees of freedom and skewness estimates of the one- and two-component models are relatively similar to those of the TGARCH and CGARCH models. Comparison of the CGARCH model with the Beta-Skew-$t$-EGARCH model through the BIC\footnote{We used BIC for model selection because it shows consistency as the sample size increases, such that the criterion will select a true model of finite dimension if it is included among the candidate models \cite{Christensen2018}.} values suggests that the latter provides a better fit. This outcome provides evidence in favour of the Beta-Skew-$t$-EGARCH model for modelling the long-memory behaviour of the financial stock market returns. The outcome is consistent with the findings of Sucarrat \cite{Sucarrat2013}, and Harvey and Sucarrat \cite{HarveySuca2014} (see details in \cite{Sucarrat2013,HarveySuca2014}). However, when it comes to the leverage effect estimation, based on the BIC value, the TGARCH model gives a better fit than the Beta-Skew-$t$-EGARCH specification for both one- and two-component.

Lastly, according to the BIC comparison, the one-component model outperforms the two-component model. This outcome is consistent with that of Harvey and Sucarrat \cite{HarveySuca2014}, where the one-component model also performed better than the two-component model; but the outcome is in contrast with the findings of Sucarrat \cite{Sucarrat2013}, where the two-component model outperformed the one-component model. Moreover, as reported by Harvey and Sucarrat \cite{HarveySuca2014}, the use of the two-component model does not always give a better fit.

\section{Discussion}\label{CCC444}
In this study, we used five autoregressive models comprising of the fGARCH, GAS, TGARCH, CGARCH and Beta-Skew-$t$-EGARCH models to estimate six features of return volatility that are relevant for a robust risk management in the S\&P Indian market index. These features are stylised facts that characterise the market and they include volatility persistence, mean-reversion, asymmetry (or leverage effect), skewness, fat-tails and the long-memory behaviour of volatility decomposition into long-term and short-term components. The ability of these models to capture these stylised facts provided a high degree of robustness for volatility modelling in the market’s returns. Both simulations and empirical evidence were used to show the accuracy of the estimations.

To begin with, the study comparably used the robust fGARCH and GAS models to estimate the magnitude and dynamics of the persistence in the conditional volatility using the returns from the market index. The study began by using simulation with empirical verifications to obtain the NIG and AST1 assumed errors as the most adequate (or optimal) assumed error distributions to use with the fGARCH and GAS models, respectively, for volatility modelling when the underlying error distribution is unknown in the returns. Hence, the NIG (AST1) distribution may be widely used with the fGARCH (GAS) model to improve the accuracy of volatility modelling for risk measures in finance and other areas. Appropriate risk management with proper economic policy implementation could create a channel for profit maximization by financial institutions and individual investors \cite{Dralle2011}.

Through fitting each of these optimal assumed innovations to their respective model, the study found considerably high volatility persistence in the market returns. The high persistence apparently suggests the presence of volatility clustering in the returns. Knowledge about the clustering of volatility allows market agents to adopt dynamic and flexible trading strategies that are suitable either for high volatility or low volatility regimes \cite{PandeyKumar2017}. The study further showed that the fGARCH and GAS models performed equally well in the volatility persistence (and mean-reversion) estimation when fitted with their respective optimal assumed error.

Next, we used the extensions of these two models to estimate other features of the return volatility that include leverage effect or asymmetry, skewness, fat-tails, and long-memory behaviour. Specifically, we used the extensions of the (f)GARCH model consisting of the TGARCH and CGARCH models, and the extension of the GAS model involving the one- and two-components Beta-Skew-$t$-EGARCH model to estimate the stated volatility features, and we compared the outcomes from the two model extensions. Through the TGARCH and Beta-Skew-$t$-EGARCH models, our findings show that negative skewness and leverage effects are pronounced, with considerable fat-tails in the conditional density. The leverage estimate is positive, which indicates that large negative returns are being followed by higher volatility. The pronounced negative skewness estimate implies that the risk of a large negative demeaned stock return is greater than that of a large positive demeaned stock return.

Through the CGARCH and Beta-Skew-$t$-EGARCH models, the study finds the existence of both permanent (or long-run) and transitory (or short-run) components of volatility in the persistence process, but the response to the effect of shocks in the short-run is higher than in the long-run volatility. This response to shock effects is also part of the findings of the fGARCH modelling. This implies that higher volatility in the process is mostly due to the short-run volatility increase. Further findings through the Beta-Skew-$t$-EGARCH model using the half-life estimation showed that the short run volatility fluctuation reverts much faster to the mean or normal volatility state than the long run volatility persistence. Consequently, these results show that with the arrival of news in the stock market, the long-run component displays much lower response to the effect of shocks (or change to volatility) than the transient short-run component. In summary, the short run effect is big but short-lived, while the long-run effect is much lower but persists into the future. This means that investment and other market risks in the long term seem to be considerably under control in the market.

Next, comparison of the CGARCH model with the Beta-Skew-$t$-EGARCH model suggests that the latter provided a better fit. This outcome provides evidence in favour of the Beta-Skew-$t$-EGARCH model for modelling the long-memory behaviour of the market returns. However, when it comes to the leverage effect, skewness, and fat-tails estimations, our findings showed that the TGARCH model outperformed the Beta-Skew-$t$-EGARCH model. Lastly, comparison of the two versions of the Beta-Skew-$t$-EGARCH model showed that the one-component model outperformed the two-component model. These discussed outcomes summarily answered all the four research questions, and the study shows that the market returns are characterised by the six stated volatility features.

\section{Conclusion}\label{CCC555}
In conclusion, this study largely contributes to the literature by comparing the fit of two robust models involving the fGARCH and GAS (and their extensions as explained in the discussion). The fGARCH model uses the dynamics of the residuals to drive the conditional volatility while the GAS model uses the dynamics of the conditional score to drive the time-varying parameters. To the best of the authors' knowledge, the combination of these two robust models have not been used comparatively in previous studies for modelling the dynamics of volatility. Risk management systems are highly dependent on the underlying assumed distribution, and the identification of a distribution that adequately captures every aspect of the given financial data may be of great benefit to investors and risk managers \cite{ChinhamuETAL2017}. Hence, our findings showed that to obtain a reliable volatility (risk) modelling in a financial time series with unknown underlying error distribution, the NIG (AST1) assumed error could be recommended for use with the fGARCH (GAS) model. The study further revealed considerably high volatility persistence, negative skewness, leverage effect and fat-tails in the S\&P Indian financial market returns.

Lastly, our findings from the long-memory behaviour revealed that although the response to shocks is greater in the short-term component, it is however short-lived. On the contrary, despite a high degree of persistence in the long-term component, market information or unexpected news arrival only has a low long-run impact on the market. Based on this, the long-run investment risks within the Indian stock market seems to be under control. Hence, our findings suggest that rational investors should try to stay calm with the arrival of unexpected news or unforeseen events in the Indian stock market because the long run effect of such news will not be severe, and the market will eventually return to the normal state. With the presence of short-term and long-term components and their impacts on the market, this study also suggests that market managers and government, in particular, should make efforts to understand the implications of changes in their system of trading and policies implemented. Such moves (or actions) will facilitate improvement in the market activities and further enable them to better control risks in the market.

For future studies, the authors intend exploring the functionalities of other extensions of the GARCH and GAS models for volatility modelling and forecasting. Specifically, the authors intend using the apARCH model \cite{Dingetal1993} as an extension of the GARCH model, and the GAS extensions involving the two‑component Beta‑$t$‑QVAR‑M‑lev model proposed by Haddad et al  \cite{HaddadETAL2023}, and the Beta-$t$-EGARCH model with random shifts (RS-Beta-$t$-EGARCH model) developed by Alanya-Beltran \cite{AlanyaBeltran2022}.

\authorcontributions{Conceptualization, R.T.A.S., C.C. and C.S.; methodology, R.T.A.S., C.C. and C.S.; software, R.T.A.S., C.C. and C.S.; validation, R.T.A.S., C.C. and C.S.; formal analysis, R.T.A.S.; investigation, R.T.A.S.; resources, R.T.A.S., C.C. and C.S.; data curation, R.T.A.S.; writing---original draft preparation, R.T.A.S.; writing---review and editing, R.T.A.S., C.C. and C.S.; visualization, R.T.A.S., C.C. and C.S.; supervision, C.C. and C.S.; project administration, C.C., C.S. and R.T.A.S.; funding acquisition, R.T.A.S., C.C. and C.S. All authors have read and agreed to the published version of the manuscript.}

\funding{This research received no external funding.}

\institutionalreview{Not applicable.}

\dataavailability{Restrictions apply to the availability of these data. Data was obtained from Thomson Reuters Datastream and are available from the authors with the permission of Thomson Reuters Datastream.}

\acknowledgments{The authors are grateful to Thomson Reuters Datastream for providing the data, and to the University of the Witwaterstrand and the University of Venda for their resources.}

\conflictsofinterest{The authors declare no conflict of interest.}

%

\appendixtitles{yes} 
\appendixstart
\appendix
\begin{adjustwidth}{-\extralength}{0cm}

\section*{References}
\bibliography{MendeleyRef}

\begin{thebibliography}{999}

\bibitem[Harvey and Sucarrat(2014)]{HarveySuca2014}
Harvey, A.; Sucarrat, G.
\newblock {EGARCH models with fat tails, skewness and leverage}.
\newblock {\em Computational Statistics and Data Analysis} {\bf 2014}, {\em
  26},~320--338.

\bibitem[Sucarrat(2013)]{Sucarrat2013}
Sucarrat, G.
\newblock {betategarch: Simulation, estimation and forecasting of first-order
  beta-skew-t-EGARCH models}.
\newblock {\em The R Journal} {\bf 2013}, {\em 5},~137--147.

\bibitem[Engle and Patton(2001)]{EnglePatton2001}
Engle, R.F.; Patton, A.J.
\newblock {What good is a volatility model?}
\newblock {\em Quantitative Finance} {\bf 2001}, {\em 1},~237--245.

\bibitem[Porterba and Summers(1986)]{PortSumm1986}
Porterba, J.; Summers, L.
\newblock {The persistence of volatility and stock market fluctuations}.
\newblock {\em American Economic Review} {\bf 1986}, {\em 76},~1143--1151.

\bibitem[Ardia et~al.(2019)Ardia, Boudt, and Catania]{Ardiaetal2019}
Ardia, D.; Boudt, K.; Catania, L.
\newblock {Generalized autoregressive score models in R: The GAS package}.
\newblock {\em Journal of Statistical Software} {\bf 2019}, {\em 88},~1--28.
\newblock {\url{https://doi.org/10.18637/jss.v088.i06}}.

\bibitem[Chou(1988)]{Chou1988}
Chou, R.Y.
\newblock {Volatility persistence and stock valuations : Some empirical
  evidence using GARCH}.
\newblock {\em Journal of Applied Econometrics} {\bf 1988}, {\em 3},~279--294.

\bibitem[Takaishi(2018)]{TakaishiT2018}
Takaishi, T.
\newblock {Volatility estimation using a rational GARCH model}.
\newblock {\em Quantitative Finance and Economics} {\bf 2018}, {\em
  2},~127--136.

\bibitem[Feng and Shi(2017)]{FengShi2017}
Feng, L.; Shi, Y.
\newblock {A simulation study on the distributions of disturbances in the GARCH
  model}.
\newblock {\em Cogent Economics and Finance} {\bf 2017}, {\em 5},~1355503.
\newblock {\url{https://doi.org/10.1080/23322039.2017.1355503}}.

\bibitem[Hentschel(1995)]{Hentschel1995}
Hentschel, L.
\newblock {All in the family nesting symmetric and asymmetric GARCH models}.
\newblock {\em Journal of Financial Economics} {\bf 1995}, {\em 39},~71--104.
\newblock {\url{https://doi.org/10.1016/0304-405X(94)00821-H}}.

\bibitem[Pandey and Kumar(2017)]{PandeyKumar2017}
Pandey, R.; Kumar, A.
\newblock {Modelling persistence in conditional volatility of asset returns}.
\newblock {\em Afro-Asian J. Finance and Accounting} {\bf 2017}, {\em
  7},~16--34.

\bibitem[Tsay(2005)]{Tsay2005}
Tsay, R.S.
\newblock {\em {Analysis of Financial Time Series}}; John Wiley and Sons:
  London,  2005; pp. 154--205.

\bibitem[Oh and Patton(2016)]{OhPatton2016}
Oh, D.; Patton, A.
\newblock {The time-varying systemic risk: Evidence from a dynamic copula model
  of CDS spreads}.
\newblock {\em Journal of Business and Economic Statistics} {\bf 2016}, {\em
  36},~181--195.
\newblock {\url{https://doi.org/10.1080/07350015.2016.1177535}}.

\bibitem[Creal et~al.(2014)Creal, Schwaab, Koopman, and Lucas]{Crealetal2014}
Creal, D.; Schwaab, B.; Koopman, S.; Lucas, A.
\newblock {Observation-driven mixed-measurement dynamic factor models with an
  application to credit risk}.
\newblock {\em Rev. Econ. Stat.} {\bf 2014}, {\em 96},~898--915.
\newblock {\url{https://doi.org/https://doi. org/10.1162/REST_a_00393}}.

\bibitem[Blasques et~al.(2014)Blasques, Koopman, and Lucas]{BlasqKOOPetal2014}
Blasques, F.; Koopman, S.; Lucas, A.
\newblock {Maximum likelihood estimation for generalized autoregressive score
  models}.
\newblock {\em TI 2014-029/III, Tinbergen Institute Discussion Paper} {\bf
  2014}, pp. 1--65.
\newblock
  {\url{https://doi.org/https://personal.vu.nl/f.blasques/Paper_Blasques2014b.pdf.}}

\bibitem[Catania and Bill$\acute{e}$(2017)]{CataniaBill2017}
Catania, L.; Bill$\acute{e}$, A.
\newblock {Dynamic spatial autoregressive models with autoregressive and
  heteroskedastic disturbances}.
\newblock {\em J. Appl. Econom.} {\bf 2017}, {\em 32},~1178--1196.
\newblock {\url{https://doi.org/https://doi.org/10.1002/jae.2565}}.

\bibitem[Opschoor et~al.(2018)Opschoor, Janus, Lucas, and {Van
  Dijk}]{Opschooretal2018}
Opschoor, A.; Janus, P.; Lucas, A.; {Van Dijk}, D.
\newblock {New HEAVY models for fat-tailed realized covariances and returns}.
\newblock {\em Journal of Business and Economic Statistics} {\bf 2018}, {\em
  36},~643--657.
\newblock {\url{https://doi.org/10.1080/07350015.2016.1245622}}.

\bibitem[Gorgi et~al.(2019)Gorgi, Hansen, Janus, and Koopman]{Gorgietal2019}
Gorgi, P.; Hansen, P.; Janus, P.; Koopman, S.
\newblock {Realized Wishart-GARCH: a score-driven multi-asset volatility
  model}.
\newblock {\em J. Financ. Economet.} {\bf 2019}, {\em 17},~1--32.
\newblock {\url{https://doi.org/https://doi.org/10.1093/jjfinec/nby007}}.

\bibitem[Haddad et~al.(2023)Haddad, Blazsek, Arestis, Fuerst, and
  Sheng]{HaddadETAL2023}
Haddad, M.F.C.; Blazsek, S.; Arestis, P.; Fuerst, F.; Sheng, H.H.
\newblock {The two‑component Beta‑t‑QVAR‑M‑lev: a new forecasting
  model}.
\newblock {\em Financial Markets and Portfolio Management} {\bf 2023}, {\em
  37},~379--401.
\newblock {\url{https://doi.org/https://doi.org/10.1007/s11408-023-00431-4}}.

\bibitem[Chen and Shen(2000)]{ChenShen2004}
Chen, S.; Shen, C.
\newblock {GARCH, jumps, and permanent and transitory components of volatility:
  The case of the Taiwan exchange rate}.
\newblock {\em Mathematics and Computers in Simulation} {\bf 2000}, {\em
  67},~201--216.

\bibitem[Chiang et~al.(2006)Chiang, Yeh, and Chiu]{ChiangETAL2009}
Chiang, S.M.; Yeh, C.P.; Chiu, C.L.
\newblock {Permanent and transitory components in the Chinese stock market: The
  ARJI-Trend model}.
\newblock {\em Taylor and Francis} {\bf 2006}, {\em 45},~35--55.

\bibitem[Ane(2006)]{Ane2006}
Ane, T.
\newblock {Short- and long-term components of volatility in Hong Kong stock
  returns}.
\newblock {\em Applied Financial Economics} {\bf 2006}, {\em 16},~439--460.

\bibitem[Engle and Lee(1999)]{EngleLee1999}
Engle, R.; Lee, G.
\newblock {A permanent and transitory component model of stock return
  volatility, in R.F. Engle and H. White (eds.), cointegration, causality, and
  forecasting: A festschrift in honor of Clive W.J. Granger}.
\newblock {\em Oxford University Press, New York} {\bf 1999}, pp. 475--497.

\bibitem[Datastream(2021)]{Datastream2021}
Datastream.
\newblock {Thomson reuters datastream. [Online]. Available at: Subscription
  Service (Accessed: June 2021)},  2021.

\bibitem[Hertog et~al.(2022)Hertog, Gerland, and Wilmoth]{HertogETAL2022}
Hertog, S.; Gerland, P.; Wilmoth, J.
\newblock {India overtakes China as the world’s most populous country}.
\newblock {\em UN DESA Population Division, World Population Prospects: Summary
  of Results} {\bf 2022}, pp. 1--5.

\bibitem[Bollerslev(1986)]{Bollers1986}
Bollerslev, T.
\newblock {Generalized autoregressive conditional heteroskedastic}.
\newblock {\em Journal of Econometrics} {\bf 1986}, {\em 31},~307--327.

\bibitem[Engle(1982)]{Engle1982}
Engle, R.F.
\newblock {Autoregressive conditional heteroscedacity with estimates of
  variance of United Kingdom inflation}.
\newblock {\em Econometrica} {\bf 1982}, {\em 50},~987--1008.

\bibitem[Harvey(2013)]{Harvey2013}
Harvey, A.C.
\newblock {\em {Dynamic models for volatility and heavy tails: With
  applications to financial and economic time series}}; Cambridge University
  Press: New York, USA,  2013; pp. 1--261.
\newblock {\url{https://doi.org/https://doi.org/10.1017/CBO9781139540933}}.

\bibitem[Ghalanos(2018)]{Ghalanos2018}
Ghalanos, A.
\newblock {Introduction to the rugarch package. (Version 1.3-8)},  2018.

\bibitem[Zakoian(1994)]{Zakoian1994}
Zakoian, J.M.
\newblock {Threshold heteroscedastic models}.
\newblock {\em Journal of Economic Dynamics and Control} {\bf 1994}, {\em
  18},~931--955.

\bibitem[Engle and Ng(1993)]{EngleNg1993}
Engle, R.F.; Ng, V.K.
\newblock {Measuring and testing the impact of news on volatility}.
\newblock {\em The journal of Finance} {\bf 1993}, {\em 48},~17749--1778.

\bibitem[Taylor(1986)]{Taylor1986}
Taylor, S.J.
\newblock {\em {Modelling financial time series}}, second ed.; World Scientific
  Publishing Co. Pte. Ltd.,  1986; pp. 1--297.

\bibitem[Schwert(1990)]{Schwert1990}
Schwert, G.W.
\newblock {Stock volatility and the crash of '87}.
\newblock {\em The Review of Financial Studies} {\bf 1990}, {\em 3},~77--102.

\bibitem[Higgins and Bera(1992)]{HigginsBera1992}
Higgins, M.L.; Bera, A.K.
\newblock {A class of nonlinear Arch models}.
\newblock {\em International Economic Review,} {\bf 1992}, {\em 33},~137--158.

\bibitem[Nelson(1991)]{Nelson1991}
Nelson, B.D.
\newblock {Conditional heteroskedasticity in asset returns : A new approach}.
\newblock {\em Econometrica} {\bf 1991}, {\em 59},~347--370.

\bibitem[Glosten et~al.(1993)Glosten, Jagannathan, and Runkle]{GlostenETAL1993}
Glosten, L.R.; Jagannathan, R.; Runkle, D.E.
\newblock {On the relation between the expected value and the volatility of the
  nominal excess return on stocks}.
\newblock {\em The Journal of Finance} {\bf 1993}, {\em 48},~1779--1801.
\newblock {\url{https://doi.org/10.1111/j.1540-6261.1993.tb05128.x}}.

\bibitem[Ding et~al.(1993)Ding, Granger, and Engle]{Dingetal1993}
Ding, Z.; Granger, C.W.; Engle, R.F.
\newblock {A long memory property of stock market returns and a new model}.
\newblock {\em Journal of Empirical Finance} {\bf 1993}, {\em 1},~83--106.

\bibitem[Ghalanos(2022)]{Ghalanos2022}
Ghalanos, A.
\newblock {rugarch: Univariate GARCH models. R package version 1.4-7},  2022.

\bibitem[Glosten et~al.(1993)Glosten, Jagannathan, and Runkle]{GJR1993}
Glosten, L.; Jagannathan, R.; Runkle, D.
\newblock {On the relation between expected value and the volatility of the
  nominal excess return on stocks}.
\newblock {\em Journal of Finance} {\bf 1993}, {\em 48},~1779--1801.

\bibitem[Creal et~al.(2013)Creal, Koopman, and Lucas]{Crealetal2013a}
Creal, D.; Koopman, S.J.; Lucas, A.
\newblock {Generalized autoregressive score models with applications}.
\newblock {\em Journal of Applied Econometrics} {\bf 2013}, {\em 28},~777--795.
\newblock {\url{https://doi.org/10.1002/jae.1279}}.

\bibitem[Catania et~al.(2020)Catania, Ardia, and Boudt]{Cataniaetal2020}
Catania, L.; Ardia, D.; Boudt, K.
\newblock {Generalized autoregressive score models. Package ‘GAS' Version
  0.3.3},  2020.
\newblock {\url{https://doi.org/10.18637/jss.v088.i06}}.

\bibitem[Alanya-Beltran(2022)]{AlanyaBeltran2022}
Alanya-Beltran, W.
\newblock {Modelling stock returns volatility with dynamic conditional score
  models and random shifts}.
\newblock {\em Finance Research Letters} {\bf 2022}, {\em 45},~102121.

\bibitem[Creal et~al.(2012)Creal, Koopman, and Lucas]{Crealetal2012}
Creal, D.; Koopman, S.J.; Lucas, A.
\newblock {Generalized autoregressive score models}.
\newblock {\em Aenorm} {\bf 2012}, {\em 20},~37--41.

\bibitem[Koopman et~al.(2017)Koopman, Lucas, and Zamojski]{Koopmanetal2017}
Koopman, S.J.; Lucas, A.; Zamojski, M.
\newblock {Dynamic term structure models with score driven time varying
  parameters: estimation and forecasting. NBP Working Paper No. 258}.

\bibitem[Blasques et~al.(2014)Blasques, Creal, Janus, Koopman, Lucas, Scharth,
  and Schwaab]{Blasquesetal2014}
Blasques, F.; Creal, D.D.; Janus, P.; Koopman, S.J.; Lucas, A.; Scharth, M.;
  Schwaab, B.
\newblock {Generalized autoregressive score models for time-varying parameters
  : new models and applications},  2014.

\bibitem[Creal et~al.(2011)Creal, Koopman, and Lucas]{CrealETAL2011}
Creal, D.; Koopman, S.J.; Lucas, A.
\newblock {A dynamic multivariate heavy-tailed model for time-varying
  volatilities and correlations}.
\newblock {\em Journal of Business and Economic Statistics} {\bf 2011}, {\em
  29},~552--563.
\newblock {\url{https://doi.org/10.1198/jbes.2011.10070}}.

\bibitem[Harvey and Chakravarty(2008)]{HarvChakr2008}
Harvey, A.; Chakravarty, T.
\newblock {Beta-t-EGARCH. Cambridge working papers in economics 0840}.
\newblock {\em Faculty of Economics, University of Cambridge} {\bf 2008}, p.
  p137.

\bibitem[Creal et~al.(2008)Creal, Koopman, and Lucas]{CrealETAL2008}
Creal, D.; Koopman, S.; Lucas, A.
\newblock {A general framework for observation driven time-varying parameter
  models}.
\newblock {\em Tinbergen Institute Discussion Paper, Amsterdam, TI} {\bf 2008},
  pp. 2008--108/4.

\bibitem[\mbox{Fern$\acute{a}$ndez} and Steel(1998)]{FernSteel1998}
\mbox{Fern$\acute{a}$ndez}, C.; Steel, M.
\newblock {On Bayesian modelling of fat tails and skewness}.
\newblock {\em Journal of the American Statistical Association} {\bf 1998},
  {\em 93},~359--371.

\bibitem[Alizadeh et~al.(2002)Alizadeh, Brandt, and Diebold]{AlizadETAL2002}
Alizadeh, S.; Brandt, M.; Diebold, F.
\newblock {Range-based estimation of stochastic volatility models}.
\newblock {\em Journal of Finance} {\bf 2002}, {\em 57},~1047--1092.

\bibitem[Chalmers(2019)]{Chalmers2019}
Chalmers, P.
\newblock {Introduction to Monte Carlo simulations with applications in R using
  the SimDesign package} {\bf 2019}.
\newblock pp. 1--46.

\bibitem[Samuel et~al.(2023)Samuel, Chimedza, and Sigauke]{SamuelETAL2023}
Samuel, T.; Chimedza, C.; Sigauke, C.
\newblock {Simulation framework to determine suitable innovations for
  volatility persistence estimation: The GARCH approach}.
\newblock {\em J. Risk Financial Manag.} {\bf 2023}, {\em 16},~392.
\newblock {\url{https://doi.org/https://doi.org/10.3390/jrfm16090392}}.

\bibitem[Hilary(2002)]{Hilary2002}
Hilary, T.
\newblock {Descriptive statistics for research},  2002.

\bibitem[Yuan et~al.(2015)Yuan, Tong, and Zhang]{Yuanetal2015}
Yuan, K.H.; Tong, X.; Zhang, Z.
\newblock {Bias and efficiency for SEM with missing data and auxiliary
  variables: Two-stage robust method versus two-stage ML}.
\newblock {\em Structural Equation Modeling: A Multidisciplinary Journal} {\bf
  2015}, {\em 22},~178--192.
\newblock {\url{https://doi.org/10.1080/10705511.2014.935750}}.

\bibitem[Morris et~al.(2019)Morris, White, and Crowther]{Morrisetal2019}
Morris, T.P.; White, I.R.; Crowther, M.J.
\newblock {Using simulation studies to evaluate statistical methods}.
\newblock {\em Statistics in Medicine} {\bf 2019}, {\em 38},~2074--2102.
\newblock {\url{https://doi.org/https://doi.org/10.1002/sim.8086}}.

\bibitem[Patton(2011)]{Patton2011}
Patton, A.J.
\newblock {Volatility forecast comparison using imperfect volatility proxies}.
\newblock {\em Journal of Econometrics} {\bf 2011}, {\em 160},~246--256.

\bibitem[Wang and Yang(2018)]{WangYang2018}
Wang, J.X.; Yang, M.
\newblock {Conditional volatility persistence} {\bf 2018}.
\newblock p. 60 Pages.
\newblock {\url{https://doi.org/10.2139/ssrn.3080693}}.

\bibitem[Chalmers and Adkins(2020)]{ChalmerAndAdkins2020}
Chalmers, R.P.; Adkins, M.C.
\newblock {Writing effective and reliable Monte Carlo simulations with the
  SimDesign package}.
\newblock {\em The Quantitative Methods for Psychology} {\bf 2020}, {\em
  16},~248--280.
\newblock {\url{https://doi.org/10.20982/tqmp.16.4.p248}}.

\bibitem[Sigal and Chalmers(2016)]{SigalChalm2016}
Sigal, M.J.; Chalmers, P.R.
\newblock {Play it again: Teaching statistics with Monte Carlo simulation}.
\newblock {\em Journal of Statistics Education} {\bf 2016}, {\em 24},~136--156.
\newblock {\url{https://doi.org/10.1080/10691898.2016.1246953}}.

\bibitem[Bollerslev(1987)]{Bollerslev1987}
Bollerslev, T.
\newblock {Conditionally heteroskedasticity time series model for speculative
  prices and rates of returns}.
\newblock {\em The Review of Economic and Statistics} {\bf 1987}, {\em
  69},~542--547.

\bibitem[Bollerslev and Wooldridge(1992)]{BollerWoold1992}
Bollerslev, T.; Wooldridge, J.M.
\newblock {Quasi-maximum likelihood estimation and inference in dynamic models
  with time-varying covariances}.
\newblock {\em Econometric Reviews} {\bf 1992}, {\em 11},~143--172.
\newblock {\url{https://doi.org/10.1080/07474939208800229}}.

\bibitem[Zivot(2013)]{Zivot2013}
Zivot, E.
\newblock {Univariate GARCH},  2013.

\bibitem[White(1982)]{White1982}
White, H.
\newblock {Maximum likelihood estimation of misspecified models}.
\newblock {\em Econometrica} {\bf 1982}, {\em 50},~1--25.

\bibitem[Qi et~al.(2010)Qi, Xiu, and Fan]{Qietal2010}
Qi, L.; Xiu, D.; Fan, J.
\newblock {Non-gaussian quasi maximum likelihood estimation of GARCH models},
  2010,  \href{http://xxx.lanl.gov/abs/1001.3895}{{\normalfont [1001.3895]}}.
\newblock {\url{https://doi.org/https://doi.org/10.48550/arXiv.1001.3895}}.

\bibitem[Eling(2014)]{Eling2014}
Eling, M.
\newblock {Fitting asset returns to skewed distributions: Are the skew-normal
  and skew-student good models?}
\newblock {\em Insurance: Mathematics and Economics} {\bf 2014}, {\em
  59},~45--56.
\newblock {\url{https://doi.org/10.1016/j.insmatheco.2014.08.004}}.

\bibitem[Barndorff-Nielsen et~al.(2013)Barndorff-Nielsen, Mikosch, and
  Resnick]{BarndorNieletal2013}
Barndorff-Nielsen, O.E.; Mikosch, T.; Resnick, S..
\newblock {\em {Levy processes: Theory and applications}}; Birkhauser Science,
  Springer Nature, Global publisher: London,  2013.
\newblock {\url{https://doi.org/10.1016/b978-0-12-407795-9.00011-6}}.

\bibitem[Lee and Pai(2010)]{LeePai2010}
Lee, Y.H.; Pai, T.Y.
\newblock {REIT volatility prediction for skew-GED distribution of the GARCH
  model}.
\newblock {\em Expert Syst. Appl.} {\bf 2010}, {\em 37},~4737--4741.

\bibitem[Ashour and Abdel-hameed(2010)]{AshAbdham2010}
Ashour, S.K.; Abdel-hameed, M.A.
\newblock {Approximate skew normal distribution}.
\newblock {\em Journal of Advanced Research} {\bf 2010}, {\em 1},~341--350.
\newblock {\url{https://doi.org/10.1016/j.jare.2010.06.004}}.

\bibitem[Pourahmadi(2007)]{Pourahmadi2007}
Pourahmadi, M.
\newblock {Construction of skew-normal random variables: Are they linear
  combinations of normal and half-normal?}
\newblock {\em J. Stat. Theory Appl.} {\bf 2007}, {\em 3},~314--328.

\bibitem[Azzalini and Capitanio(2003)]{AzzaliniCapitanio2003}
Azzalini, A.; Capitanio, A.
\newblock {Distributions generated by perturbation of symmetry with emphasis on
  a multivariate skew t-distribution}.
\newblock {\em Journal of the Royal Statistical Society. Series B: Statistical
  Methodology} {\bf 2003}, {\em 65},~367--389,
  \href{http://xxx.lanl.gov/abs/0911.2342}{{\normalfont [0911.2342]}}.
\newblock {\url{https://doi.org/10.1111/1467-9868.00391}}.

\bibitem[Branco and Dey(2001)]{BrancoDey2001}
Branco, M.D.; Dey, D.K.
\newblock {A general class of multivariate skew-elliptical distributions}.
\newblock {\em Journal of Multivariate Analysis} {\bf 2001}, {\em 79},~99--113.
\newblock {\url{https://doi.org/10.1006/jmva.2000.1960}}.

\bibitem[Azzalini(1985)]{Azzalini1985}
Azzalini, A.
\newblock {A class of distributions which includes the normal ones}.
\newblock {\em Scandinavian Journal of Statistics} {\bf 1985}, {\em
  12},~171--178.

\bibitem[Javed and Mantalos(2013)]{JavedMant2013}
Javed, F.; Mantalos, P.
\newblock {GARCH-type models and performance of information criteria}.
\newblock {\em Communications in Statistics: Simulation and Computation} {\bf
  2013}, {\em 42},~1917--1933.
\newblock {\url{https://doi.org/10.1080/03610918.2012.683924}}.

\bibitem[Francq and Thieu(2019)]{FrancqThieu2019}
Francq, C.; Thieu, L.Q.
\newblock {QML inference for volatility models with covariates}.
\newblock {\em Econometric Theory} {\bf 2019}, {\em 35},~37--72.
\newblock {\url{https://doi.org/https://doi.org/10.1017/S0266466617000512}}.

\bibitem[Heracleous(2007)]{Heracleous2007}
Heracleous, M.S.
\newblock {Sample kurtosis, GARCH-t and the degrees of freedom issue}.
\newblock {\em EUR Working Papers} {\bf 2007}, pp. 1--22.

\bibitem[Chib(2015)]{Chib2015}
Chib, S.
\newblock {Monte Carlo methods and Bayesian computation: Overview},  2015.
\newblock {\url{https://doi.org/10.1016/B978-0-08-097086-8.42149-5}}.

\bibitem[Chalmers(2020)]{ChalmersP2020}
Chalmers, R.
\newblock {SimDesign: structure for organizing Monte Carlo simulation designs.
  R package version 2.1},  2020.

\bibitem[Chalmers and Adkins(2020)]{ChalmersAdkin2020}
Chalmers, R.P.; Adkins, M.C.
\newblock {Writing effective and reliable Monte Carlo simulations with the
  SimDesign package}.
\newblock {\em The Quantitative Methods for Psychology} {\bf 2020}, {\em
  16},~248--280.
\newblock {\url{https://doi.org/10.20982/tqmp.16.4.p248}}.

\bibitem[Fisher and Gallagher(2012)]{FisherGalla2012}
Fisher, T.J.; Gallagher, C.M.
\newblock {New weighted portmanteau statistics for time series goodness of fit
  testing}.
\newblock {\em Journal of the American Statistical Association} {\bf 2012},
  {\em 107},~777--787.
\newblock {\url{https://doi.org/10.1080/01621459.2012.688465}}.

\bibitem[{Li, F. and Cohen, A. S. and Kim, S. and Cho, S.}(2009)]{LiCoKiCh09}
{Li, F. and Cohen, A. S. and Kim, S. and Cho, S.}.
\newblock {Model selection methods for mixture dichotomous IRT models}.
\newblock {\em Applied Psychological Measurement} {\bf 2009}, {\em
  33},~353--373.

\bibitem[Aas and Haff(2006)]{AasHaff2006}
Aas, K.; Haff, I.H.
\newblock {The generalized hyperbolic skew student's t-distribution}.
\newblock {\em Journal of Financial Econometrics} {\bf 2006}, {\em
  4},~275--309.

\bibitem[Zhu and Galbraith(2010)]{ZhuGalb2010}
Zhu, D.; Galbraith, J.
\newblock {A generalized asymmetric Student-t distribution with application to
  financial econometrics}.
\newblock {\em Journal of Econometrics} {\bf 2010}, {\em 157},~297--305.

\bibitem[Kotz et~al.(2001)Kotz, Kozubowski, and Podgorski]{KotzETAL2001}
Kotz, S.; Kozubowski, T.; Podgorski, K.
\newblock {\em {The laplace distribution and generalizations}};
  Springer-Verlag: New York,  2001; pp. 1--441.
\newblock {\url{https://doi.org/doi:10.1007/978-1-4612-0173-1.}}

\bibitem[Mishra(2010)]{Mishra2010}
Mishra, P.K.
\newblock {A GARCH model approach to capital market volatility: The case of
  India}.
\newblock {\em Indian Journal of Economics \& Business} {\bf 2010}, {\em
  9},~631--641.

\bibitem[CRISIL(2022)]{CRISIL2022}
CRISIL.
\newblock {Indian Economy: States of aftermath}.
\newblock {\em CRISIL: An S\&P Global Economy} {\bf 2022}.

\bibitem[Christensen()]{Christensen2018}
Christensen, W.
\newblock {Model selection using information criteria (Made Easy in SAS®)}.
\newblock {\em University of California, Los Angeles, wchristensen@ucla.edu}.

\bibitem[Dralle(2011)]{Dralle2011}
Dralle, B.
\newblock {Modelling volatility in financial time series}.
\newblock Master of science in statistics, University of KwaZulu-Natal,  2011.

\bibitem[Chinhamu et~al.(2017)Chinhamu, Huang, and Chikobvu]{ChinhamuETAL2017}
Chinhamu, K.; Huang, C.K.; Chikobvu, D.
\newblock {Evaluating risk in gold prices with generalized hyperbolic and
  stable distributions}.
\newblock {\em Proceedings of the 57th Annual Conference of SASA} {\bf 2017},
  pp. 17--24.

\end{thebibliography}
\end{adjustwidth}
\end{document}